\documentclass[manuscript]{aastex}
\usepackage{epstopdf}
\usepackage{subfigure}
\usepackage{amsmath}

\usepackage{longtable}

\shorttitle{Isoenergetic Beams}
\shortauthors{Reep, Bradshaw, \& Alexander}

\begin{document}

\title{Optimal Electron Energies for Driving Chromospheric Evaporation in Solar Flares}

\author{J. W. Reep}
\affil{Department of Applied Mathematics and Theoretical Physics, University of Cambridge, Wilberforce Road, Cambridge CB3 0WA, UK}
\email{jr665@cam.ac.uk}\and
\author{S. J. Bradshaw}
\affil{Department of Physics and Astronomy, Rice University, Houston, TX 77005, USA}
\email{stephen.bradshaw@rice.edu}\and
\author{D. Alexander}
\affil{Department of Physics and Astronomy, Rice University, Houston, TX 77005, USA}
\email{dalex@rice.edu}

\begin{abstract}
In the standard model of solar flares, energy deposition by a beam of electrons drives strong chromospheric evaporation leading to a significantly denser corona and much brighter emission across the spectrum.  Chromospheric evaporation was examined in great detail by \citet{fisher1985a,fisher1985b,fisher1985c}, who described a distinction between two different regimes, termed explosive and gentle evaporation.  In this work, we examine the importance of electron energy and stopping depths on the two regimes and on the atmospheric response.  We find that with explosive evaporation, the atmospheric response does not depend strongly on electron energy.  In the case of gentle evaporation, lower energy electrons are significantly more efficient at heating the atmosphere and driving up-flows sooner than higher energy electrons.  We also find that the threshold between explosive and gentle evaporation is not fixed at a given beam energy flux, but also depends strongly on the electron energy and duration of heating.  Further, at low electron energies, a much weaker beam flux is required to drive explosive evaporation. 

\end{abstract}


\section{Introduction}

To understand the physics underpinning the evolution of solar flares, it is necessary to understand the transport of mass, momentum, and energy through the solar atmosphere.  The focus of the work presented here is to determine which component of the electron beam, assuming the thick-target model \citep{brown1971}, is most important to powering the chromospheric evaporation that fills the corona with hot, soft X-ray emitting plasma. We do this by separating the electron beam into a set of isoenergetic beams and examining their relative efficiency at driving the flare. One outcome of this work is to show where future instrumentation should be most sensitive in order for progress to be made in understanding the coupling between the beam and the flare dynamics, particularly the threshold between gentle and explosive evaporation.   

At the beginning of a solar flare, energy released from the magnetic field is partitioned between accelerated particles, {\it in situ} heating, and bulk motions in the plasma.  The accelerated electrons stream down magnetic field lines, depositing their energy into the dense chromosphere via collisions with the ambient plasma.  This energy deposition in turn drives an increase in the pressure of the chromosphere, causing an ablation of material and energy up along the field lines (commonly called chromospheric evaporation, \citealt{antiochos1978}), creating arcades of hot, bright coronal loops.

Under the thick-target model, all of the energy in an accelerated electron beam is deposited within the magnetic loop \citep{brown1971}.  The location of the energy deposition changes as the loop begins to fill due to chromospheric evaporation and the material front advances into the corona, thus decreasing the mean-free path of the streaming electrons.  \citet{nagai1984} showed that an electron with energy $E$ will be stopped at a column density of around $\approx 10^{17} [E$ (keV)$]^{2}$ cm$^{-2}$.  Since this depth changes in time as the density of the loop changes, and since the electron beam properties evolve ({\it e.g.} \citealt{holman2003}), the location of energy deposition can help to elucidate the dynamics of flaring loops.  This model necessarily assumes that the flare is compact and occurring on a single loop, whereas reconnection may drive the formation of many loops (see \citealt{warren2006}).  

The speed at which the material travels has been found to differ significantly depending on the strength of the beam.  \citet{fisher1985a,fisher1985b,fisher1985c} found that above a beam flux of about $10^{10}$ erg sec$^{-1}$ cm$^{-2}$, material from the chromosphere explosively evaporates, {\it i.e.} material is driven up into the corona at a few hundred km sec$^{-1}$.  This explosive evaporation also drives a material front in the opposite direction, traveling more slowly but with much greater mass (termed a chromospheric condensation).  Below the explosive threshold, the material slowly ($\approx$ 30 km sec$^{-1}$) expands upward, raising the density of the corona slightly.  

However, \citet{fisher1985a,fisher1985b,fisher1985c} make a few assumptions that we now inspect more closely.  They assume that the beams last for 5 seconds, with a fixed energy flux, a fixed spectral index (4), and a fixed low energy cut-off (20 keV), and they use a sharp cut-off distribution for the beam.  It is not clear how consistent their results are for different values of the cut-off energy or for time-dependent values.  One striking feature of Fisher et al.'s choice of values is that there is an abundance of high energy electrons, with none below 20 keV.  In many observed flares, however, there are accelerated electrons at energies as low as a few keV ({\it e.g.} \citealt{oflannagain2013}), so the results of \citet{fisher1985b} would not apply.  Due to the sharpness of accelerated electron spectra, electrons below 20 keV dominate the energy flux, so the nature of explosive evaporation requires further investigation.

Many observations, with many different satellites, have confirmed the existence of explosive evaporation and significantly blue-shifted material in flares: Solar Maximum Mission (SMM; \citealt{antonucci1982}; \citealt{antonucci1983}); Naval Research Laboratory Solar Flares X-ray experiment (SOLFLEX; \citealt{doschek1979}; \citealt{doschek1990}); Yohkoh Bragg Crystal Spectrometer (Yohkoh BCS \citealt{doschek2005}); Hinode Extreme Ultraviolet Imaging Spectrometer (Hinode-EIS; \citealt{delzanna2008}; \citealt{milligan2009}; \citealt{milligan2011}; \citealt{doschek2013}); and recently the Interface Region Imaging Spectrograph (IRIS; \citealt{polito2015}).  Further, observations have shown that the hottest plasma travels faster than cooler plasma, with a clear relation between velocity and temperature ({\it e.g.} \citealt{delzanna2008}; \citealt{milligan2009}).  

With the advent of the RHESSI satellite \citep{lin2002}, it has become routine to use observed bremsstrahlung emissions to derive the non-thermal electron distribution function from observations of solar flares \citep{brown2006}.  These electron spectra ({\it e.g.} \citealt{holman2003}) can range in energy from a few keV to well over a few hundred keV.  Because they are described by a power-law in general (with a negative slope), there are significantly more low-energy electrons than high-energy electrons in the distribution.  

The shape of the electron distribution in observed solar flares is discussed in depth in \citet{holman2011}.  Beneath a certain energy, referred to as the low-energy cut-off, or more simply the cut-off energy, the thermal emissions from the hot, dense plasma mask the non-thermal signatures of the electron beam.  Therefore, beneath that energy, the shape of the electron beam is uncertain.  Many different shapes for the electron beam have been assumed, including isoenergetic ({\it e.g.} \citealt{haug1985,melnik1999,stepan2007}; this paper), sharp cut-offs \citep{emslie1978,macneice1984}, and a low-energy power-law \citep{mariska1989,warren2006,reep2013}.  More exotic, {\it i.e.} non-Maxwellian, particle distributions have been suggested for solar flares that do not require unphysical assumptions about the distribution of the low energy electrons, such as the kappa distribution \citep{owocki1983, bian2014, dzifcakova2015,dudik2015}.

It is interesting to consider the relative importance of high energy electrons versus low energy electrons in the heating process.  In the extremely high energy limit, the stopping depth can extend deep into the solar atmosphere, where the energy deposited will be radiated away almost immediately, or, due to the large heat capacity, the electrons will make only a negligible contribution to the total thermal energy of the plasma.  \citet{reep2013}, for example, show that having a cut-off energy of greater than 100 keV leads to essentially no rise in temperature in a flare.  This implies that sufficiently high energy electrons contribute little to flare heating, and that lower energy electrons dominate the energy deposition.  The stopping depth of electrons is a straight-forward function of energy \citep{emslie1978}, so that by knowing the energy range of electrons that dominate the heating, the primary location of energy deposition, and thus the source of mass up-flows, can be determined.  In this work, the importance of electron energy on the dynamics of the flaring atmosphere is examined directly.  

In order to develop an understanding of the hydrodynamic response to specific particle energies, a number of studies have adopted an isoenergetic beam to examine the impact of accelerated electrons on the atmospheric response.  In particular, \citet{haug1985, bakaya1997} studied the evolution of the energy and angle distributions of an initially isoenergetic electron beam in order to calculate the resultant bremsstrahlung spectra.  \citet{brown1998, karlicky2000, brown2000} adopted an isoenergetic neutralized ion beam, in order to study hard X-ray production, radio bursts, and the generation of Langmuir waves, while limiting extraneous assumptions from their model.  \citet{melnik1999}, under the observational consideration of centimeter-wavelength emissions, used an isoenergetic electron beam to consider large-scale effects without solving the Fokker-Planck equation for the distribution function in orderto study the production of radio emissions in solar flares.  \cite{stepan2007, karlicky2009} utilized an isoenergetic electron beam to simplify their model, physically and numerically, in order to study the hydrogen Balmer line and return current formation, noting that the total current of the beam is more important than the shape of the distribution function.  

We similarly adopt heating due to an isoenergetic electron beam here.  The idea is straightforward: if all the electrons in a beam are approximately the same energy, their stopping depths will be approximately equal.  Any changes in the atmospheric response can be attributed directly to electrons of that energy, and then compared/contrasted with beams of different electron energy.  We wish to determine directly what range of electron energy is most effective at driving a solar flare.

In Section \ref{sec:isoenergetic}, the conditions for an isoenergetic beam are derived, which then form the basis of the numerical experiments.  In Section \ref{sec:isosims}, simulations are performed to examine the atmospheric response to different isoenergetic beams.  In Section \ref{sec:samenumber}, the importance of electron number flux is briefly examined.  In Section \ref{sec:evap}, the threshold of explosive evaporation is examined at different electron energies.  The main results are summarized and discussed in Section \ref{sec:isoconclusions}.

\section{Isoenergetic Electron Beams}
\label{sec:isoenergetic}

We wish to determine the efficiency of electrons at various energies in driving the atmospheric response to heating by a beam.  Under the isoenergetic beam assumption, we can limit extraneous assumptions and focus on the key aspects that drive chromospheric evaporation.  We couple this to a state-of-the-art hydrodynamic model in order to calculate key parameters: velocity flows, temperatures, and densities, all of which are readily observable.  Although electron spectra measured in observed flares are generally of the form of a sharp power-law, due to the steepness of the observed power-laws, most of the energy in a beam is concentrated close to the low-energy cut-off.  Thus, the isoenergetic assumption simplifies the problem and allows us to directly and thoroughly examine chromospheric evaporation.  

Consider the case where an electron beam consists of electrons at nearly the same energy $E_{\ast}$.  Then, the vast majority of the energy will be deposited at the same location (same column depth).  By employing isoenergetic beams we can assess the contributions made to the flare dynamics from the different energy deposition regions, free of confusion from coupled time dependence, spatial convolution, and velocity dispersion effects.

For these isoenergetic beams, the following distribution is assumed:
\begin{equation}
\mathfrak{F}(E_{0},t) = \frac{F_{0}(t)}{E_{\ast}^{2}}\ \frac{(\delta + 2)\ (\delta - 2)}{2\delta}
 \begin{cases}
	\Big(\frac{E_{0}}{E_{\ast}}\Big)^{\delta} & \text{if } E_{0} < E_{\ast} \\
	\Big(\frac{E_{0}}{E_{\ast}}\Big)^{- \delta} & \text{if } E_{0} \geq E_{\ast} \\
 \end{cases}
\end{equation}

\noindent where $E_{0}$ is the initial electron energy, $F_{0}(t)$ is the beam energy flux, and $\delta$ is the spectral index.    The distribution function has a maximum at $E_{0} = E_{\ast}$, dropping off as a power-law to either side of the maximum.  Assuming that $x$\% of the electrons are within $\pm \Delta E$ of the maximum of the distribution function, the condition would be:
\begin{equation}
\Bigg(\frac{E_{\ast} + \Delta E}{E_{\ast}}\Bigg)^{-\delta} = \frac{100 - x}{100}
\end{equation}

\noindent Suppose the desired tolerance is that 99\% of electrons are within $\pm \Delta E$ of $E_{\ast}$.  Then, the following condition must hold:
\begin{equation}
\Bigg(\frac{E_{\ast}}{E_{\ast} + \Delta E}\Bigg)^{\delta} = \frac{1}{100}
\end{equation}

\noindent Solving, the spectral index $\delta$ must meet the following condition for the beam to be approximately isoenergetic (within a 99\% tolerance for $\Delta E$):
\begin{equation}
\delta = \frac{-2}{\log_{10}(\frac{E_{\ast}}{E_{\ast} + \Delta E})}
\label{isoenergeticindex}
\end{equation}

\noindent For example, for an electron energy of 5 keV, a spectral index $\delta = 13.7$ is required for 99\% of electrons to be within $\pm$ 2 keV of $E_{\ast}$.  

Note that these spectral indices are articially large, {\it i.e.} observed beams have a larger spread in electron energy.  However, in the case of a sharp cut-off at energy of 20 keV, assuming a spectral index of 5, more than 50\% of electrons are within 3 keV of the cut-off \Big($(\frac{E_{0}}{E_{c}})^{-5} = (\frac{23}{20})^{-5} = 0.497$\Big) carrying around one third of the total energy.  73\% of the electrons, carrying more than 50\% of the total energy, are within 6 keV of the cut-off.  Further, the mean electron energy is $\langle E \rangle =  \Big(\frac{\delta - 1 }{\delta - 2}\Big)\ E_{c} = 26.7$ keV \citep{reep2013}.  So, although the isoenergetic case is extreme, even with a more modest spectral index, most of the electrons are concentrated near one energy.

These equations are then combined with previously derived heating functions and bremsstrahlung emission calculations (see \citealt{reep2013}).  Numerical experiments can then be run, and the dynamical response of the solar atmosphere and its radiative emission can be examined in detail.  

\section{Isoenergetic Beam Simulations}
\label{sec:isosims}

Numerical experiments have been performed with the HYDRAD code \citep{bradshaw2013} to examine in detail the response of the atmosphere to isoenergetic beams, to determine the importance of different energy components of the beam to driving a flare.  Table \ref{isoenertable} shows the details of 24 simulations, with maximal beam fluxes below ($10^{9}$ erg sec$^{-1}$ cm$^{-2}$), at ($10^{10}$ erg sec$^{-1}$ cm$^{-2}$), and above ($10^{11}$ erg sec$^{-1}$ cm$^{-2}$) the canonical explosive evaporation threshold of \citet{fisher1985b}.  Each simulation assumes an electron energy $E_{\ast}$ of [5, 10, 15, 20, 25, 30, 40, 50] keV, with spectral indices derived from Equation \ref{isoenergeticindex} with a width of $\pm 2$ keV in all cases.  The simulations were performed on loops of length $2L = 50$ Mm, with cross-sectional areas $A = 7.8 \times 10^{16}$ cm$^{2}$.  The beams lasted for 300 seconds, assuming a symmetric triangular temporal envelope.  

\LTcapwidth=\textwidth

\begin{longtable}{c c c c c c c c c c}
\caption{The results of 24 simulations of isoenergetic electron beams.  The first 8 simulations were performed with a maximal energy flux $F_{0}$ value of $10^{9}$ erg sec$^{-1}$ cm$^{-2}$, so that they are beneath the canonical explosive evaporation threshold, the second 8 are at the threshold, while the final 8 simulations are well above the threshold.  The GOES intensities are listed (W m$^{-2}$), along with the the maximal bulk flow velocity (km sec$^{-1}$), maximal electron temperature (MK), and maximal apex electron density (cm$^{-3}$).    \label{isoenertable}}  \\

\hline
Run \# & $E_{\ast}$ & $F_{0,max}$ & $\delta$ & GOES Class & $v_{\mbox{max}}$ & $T_{\mbox{max}}$ & $n_{\mbox{apex,max}}$ \\
 & (keV) & (erg sec$^{-1}$ cm$^{-2}$) &  & (1-8 \AA) & (km sec$^{-1}$) & (MK) & (cm$^{-3}$) \\
\hline
1 & 5 & $1.00 \times 10^{9}$ & 13.7 & B1.7 & 338.1 & 13.3 & $9.2 \times 10^{9}$  \\
2 & 10 & $1.00 \times 10^{9}$ & 25.3 & C2.4 & 457.8 & 11.5 & $1.0 \times 10^{10}$  \\
3 & 15 & $1.00 \times 10^{9}$ & 36.8 & M1.0 & 157.7 & 4.5 & $2.1 \times 10^{9}$  \\
4 & 20 & $1.00 \times 10^{9}$ & 48.3 &M2.4 & 104.9 & 3.3 & $1.2 \times 10^{9}$  \\
5 & 25 & $1.00 \times 10^{9}$ & 59.8 & M3.4 & 72.1 & 2.6 & $9.1 \times 10^{8}$  \\
6 & 30 & $1.00 \times 10^{9}$ & 71.4 & M7.0 & 52.4 & 2.1 & $7.4 \times 10^{8}$  \\
7 & 40 & $1.00 \times 10^{9}$ & 94.4 & X1.4 & 38.0 & 1.6 & $6.1 \times 10^{8}$  \\
8 & 50 & $1.00 \times 10^{9}$ & 117.4 & X2.4 & 27.8 & 1.3 & $5.4 \times 10^{8}$  \\
\hline
9 & 5 & $1.00 \times 10^{10}$ & 13.7 & C1.7 & 576.3 & 26.2 & $5.6 \times 10^{10}$  \\
10 & 10 & $1.00 \times 10^{10}$ & 25.3 & M2.2 & 729.5 & 25.8 & $5.7 \times 10^{10}$  \\
11 & 15 & $1.00 \times 10^{10}$ & 36.8 & M8.0 & 837.2 & 24.7 & $6.1 \times 10^{10}$  \\
12 & 20 & $1.00 \times 10^{10}$ & 48.3 & X1.9 & 753.3 & 22.3 & $5.0 \times 10^{10}$  \\
13 & 25 & $1.00 \times 10^{10}$ & 59.8 & X3.9 & 601.4 & 17.8 & $5.2 \times 10^{10}$  \\
14 & 30 & $1.00 \times 10^{10}$ & 71.4 & X5.7 & 341.1 & 12.4 & $3.8 \times 10^{10}$  \\
15 & 40 & $1.00 \times 10^{10}$ & 94.4 & X13 & 195.2 & 5.6 & $3.2 \times 10^{9}$  \\
16 & 50 & $1.00 \times 10^{10}$ & 117.4 & X21 & 146.0 & 4.2 & $2.0 \times 10^{9}$  \\
\hline
17 & 5 & $1.00 \times 10^{11}$ & 13.7 & M2.1 & 965.3 & 47.6 & $1.6 \times 10^{11}$  \\
18 & 10 & $1.00 \times 10^{11}$ & 25.3 & X1.8 & 1057 & 52.6 & $2.8 \times 10^{11}$  \\
19 & 15 & $1.00 \times 10^{11}$ & 36.8 & X6.5 & 1121 & 51.8 & $3.2 \times 10^{11}$  \\
20 & 20 & $1.00 \times 10^{11}$ & 48.3 & X15 & 1204 & 51.1 & $3.1 \times 10^{11}$  \\
21 & 25 & $1.00 \times 10^{11}$ & 59.8 & X27 & 968.2 & 50.0 & $3.2 \times 10^{11}$  \\
22 & 30 & $1.00 \times 10^{11}$ & 71.4 & X42 & 1043 & 48.9 & $3.2 \times 10^{11}$  \\
23 & 40 & $1.00 \times 10^{11}$ & 94.4 & X84 & 880.5 & 47.0 & $3.2 \times 10^{11}$  \\
24 & 50 & $1.00 \times 10^{11}$ & 117.4 & X140 & 872.7 & 44.5 & $3.3 \times 10^{11}$  \\
\end{longtable}

\begin{figure}
\centering
\begin{minipage}[b]{0.32\linewidth}
\centering
\includegraphics[width=2.2in]{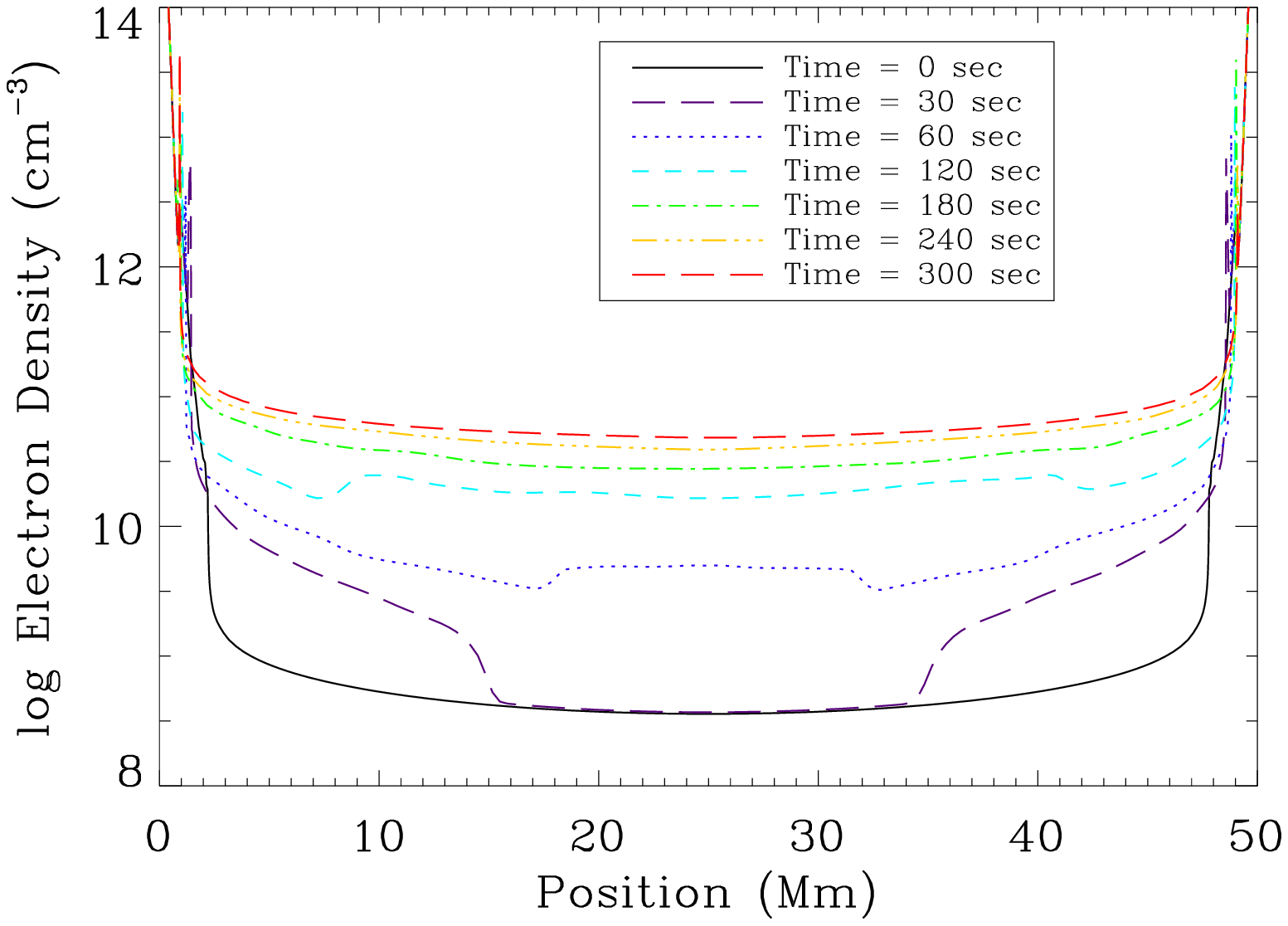}
\end{minipage}
\begin{minipage}[b]{0.32\linewidth}
\centering
\includegraphics[width=2.2in]{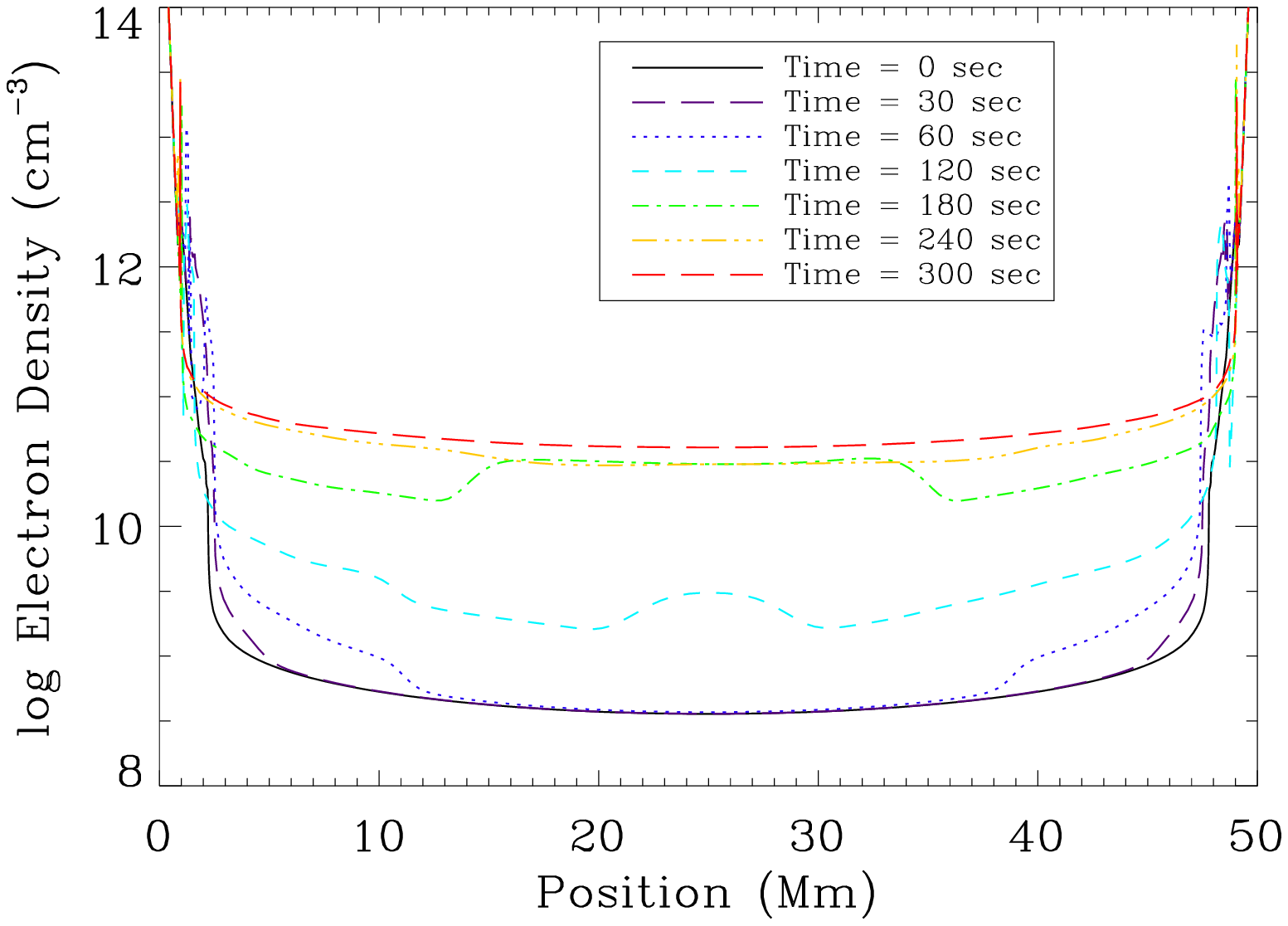}
\end{minipage}
\begin{minipage}[b]{0.32\linewidth}
\centering
\includegraphics[width=2.2in]{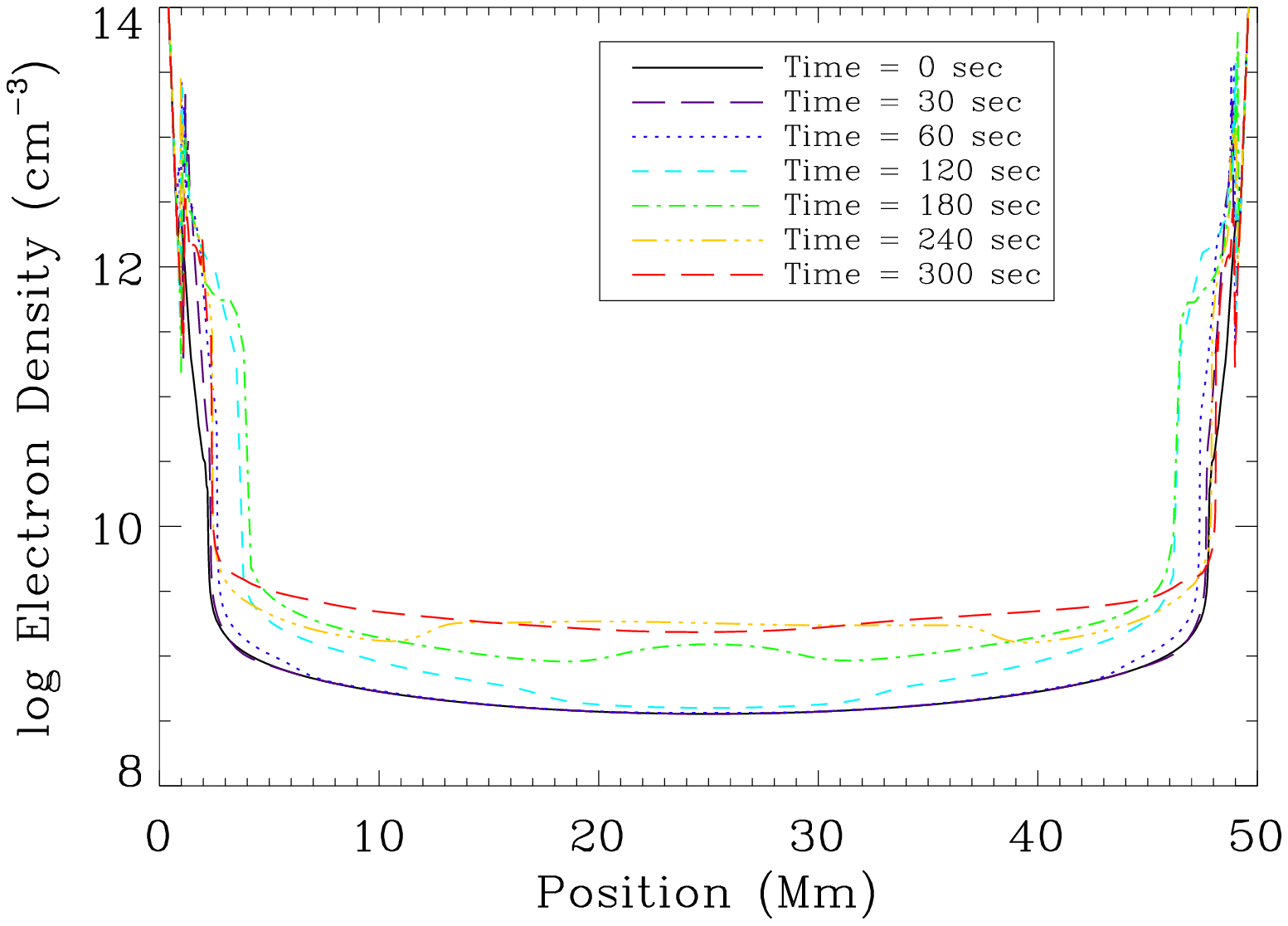}
\end{minipage}
\begin{minipage}[b]{0.32\linewidth}
\centering
\includegraphics[width=2.2in]{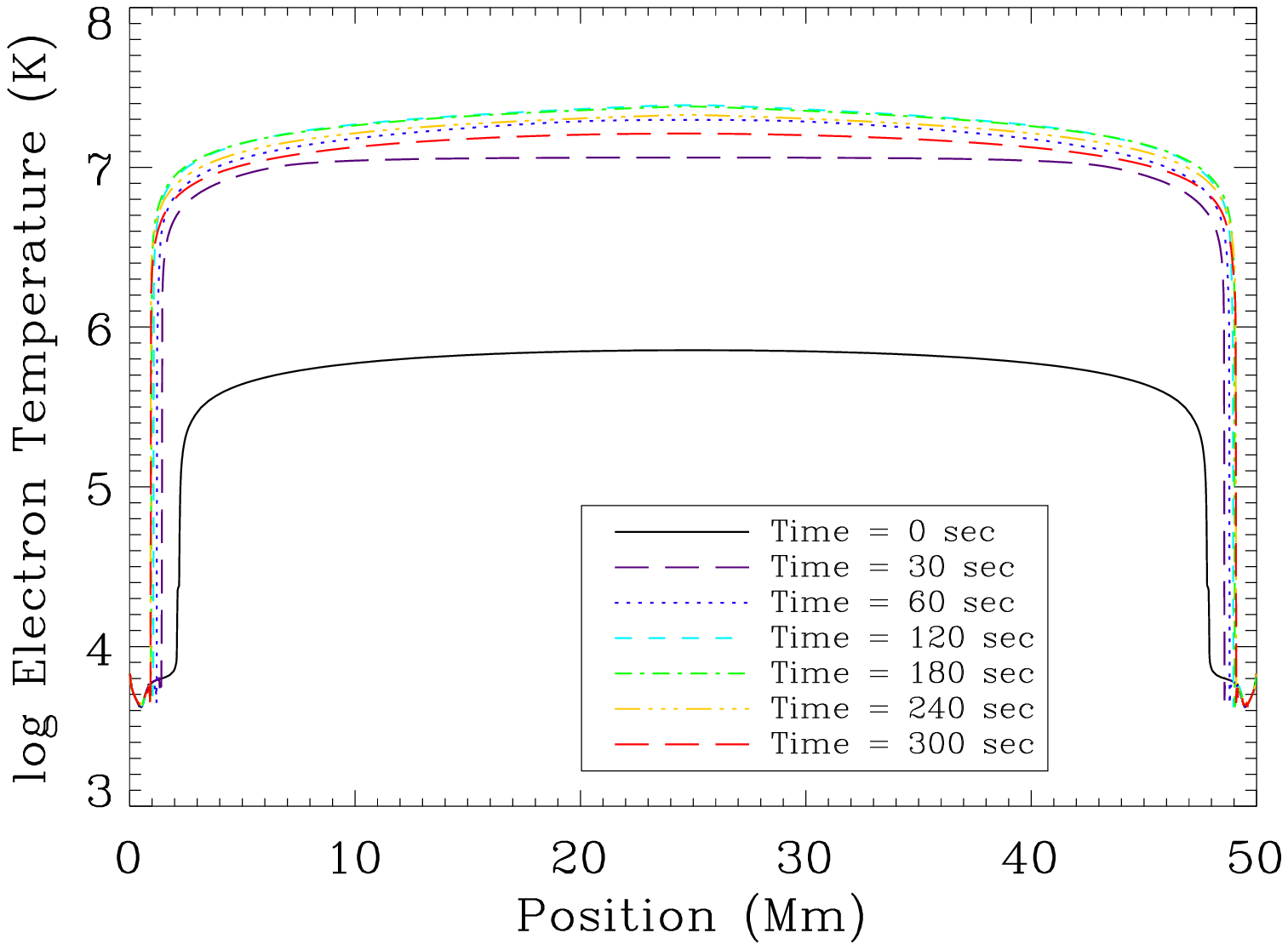}
\end{minipage}
\begin{minipage}[b]{0.32\linewidth}
\centering
\includegraphics[width=2.2in]{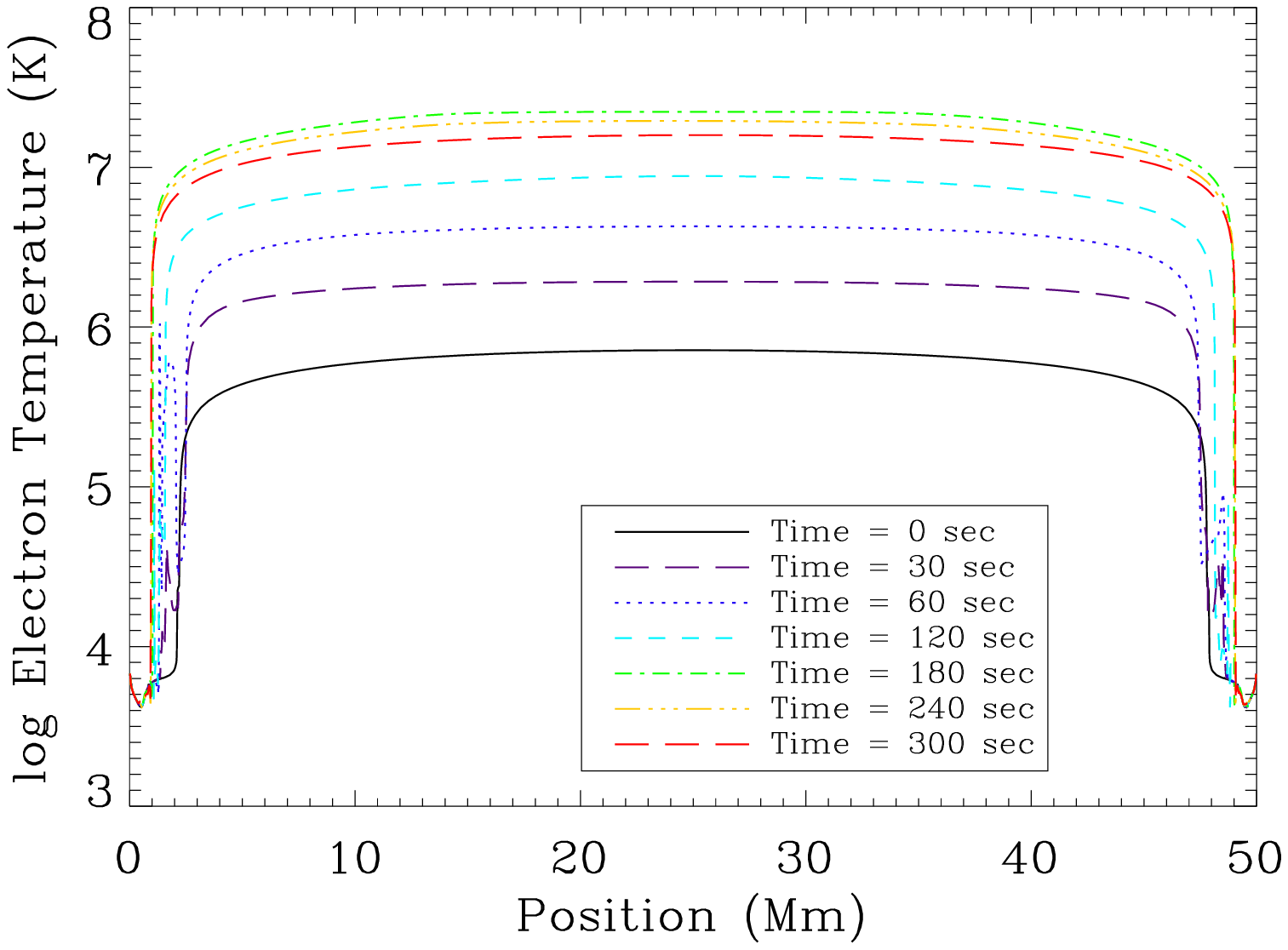}
\end{minipage}
\begin{minipage}[b]{0.32\linewidth}
\centering
\includegraphics[width=2.2in]{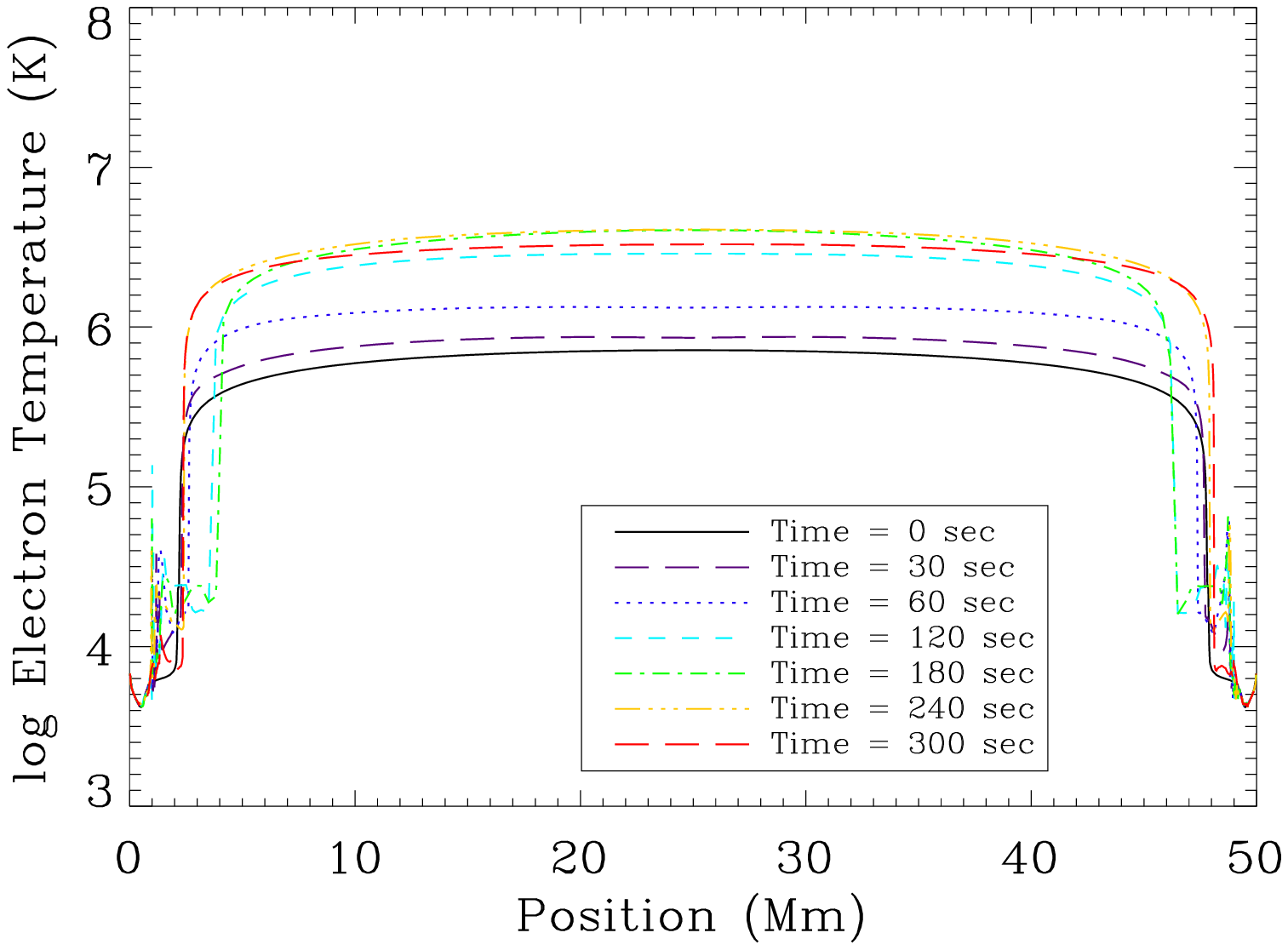}
\end{minipage}
\begin{minipage}[b]{0.32\linewidth}
\centering
\includegraphics[width=2.2in]{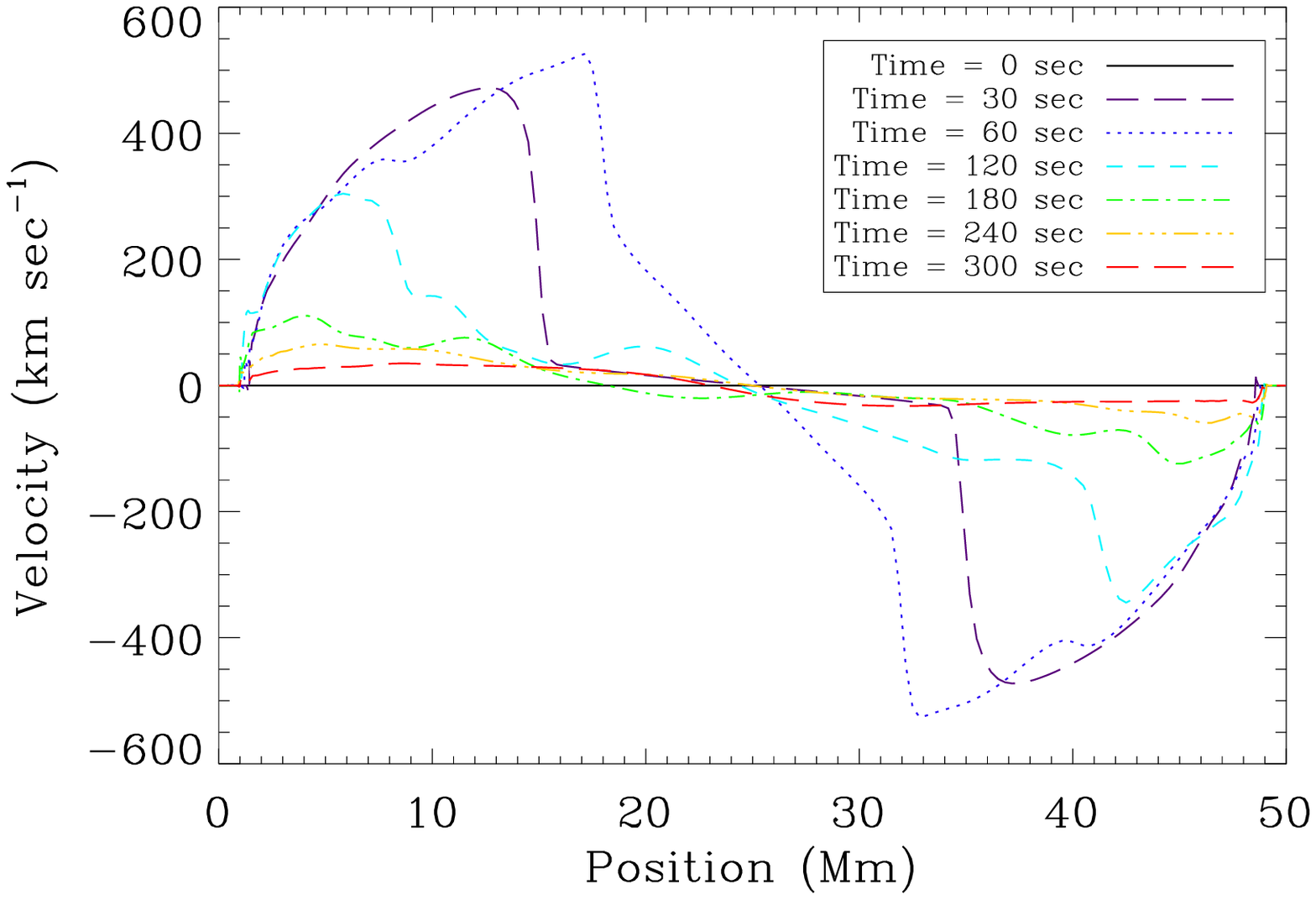}
\end{minipage}
\begin{minipage}[b]{0.32\linewidth}
\centering
\includegraphics[width=2.2in]{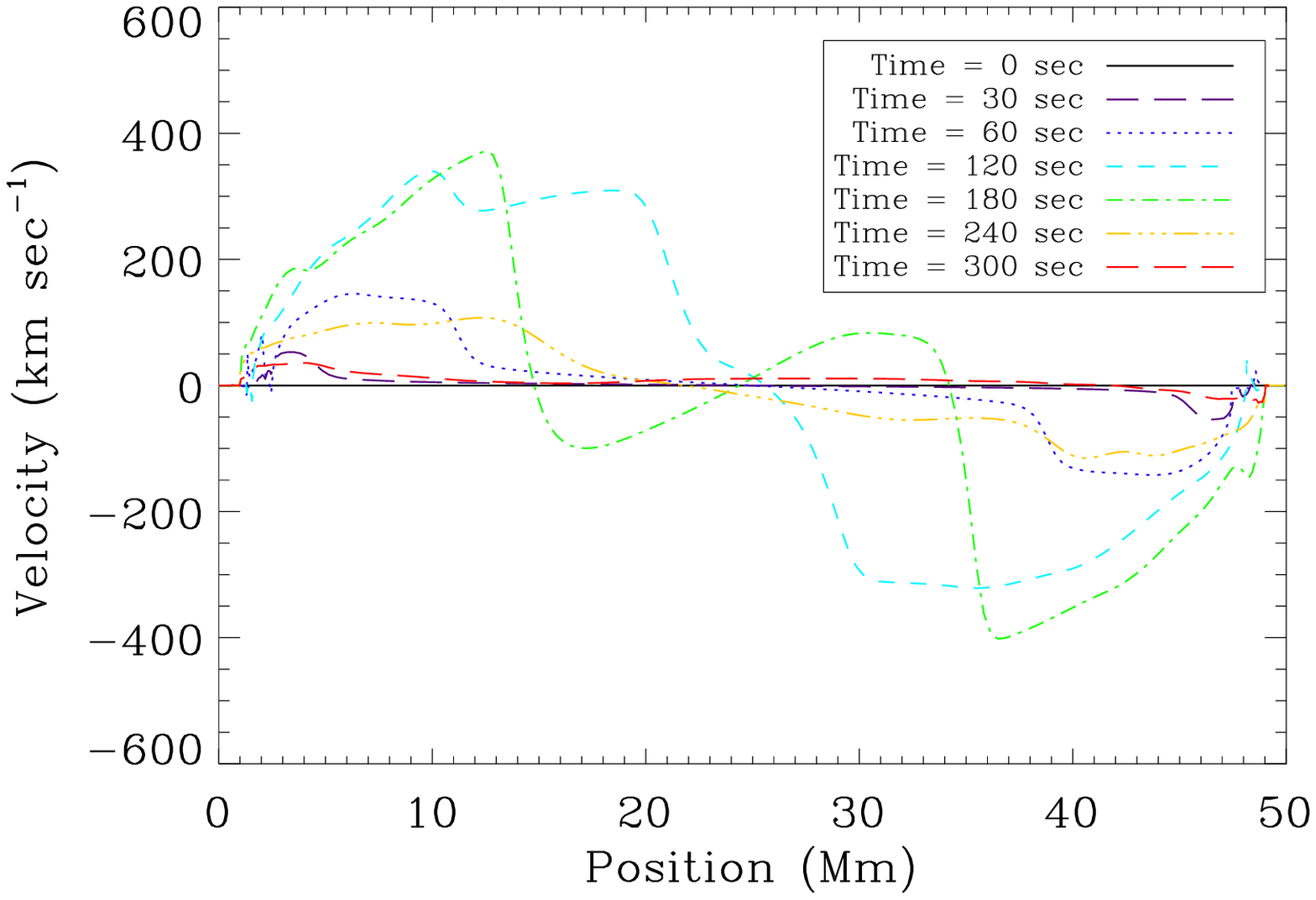}
\end{minipage}
\begin{minipage}[b]{0.32\linewidth}
\centering
\includegraphics[width=2.2in]{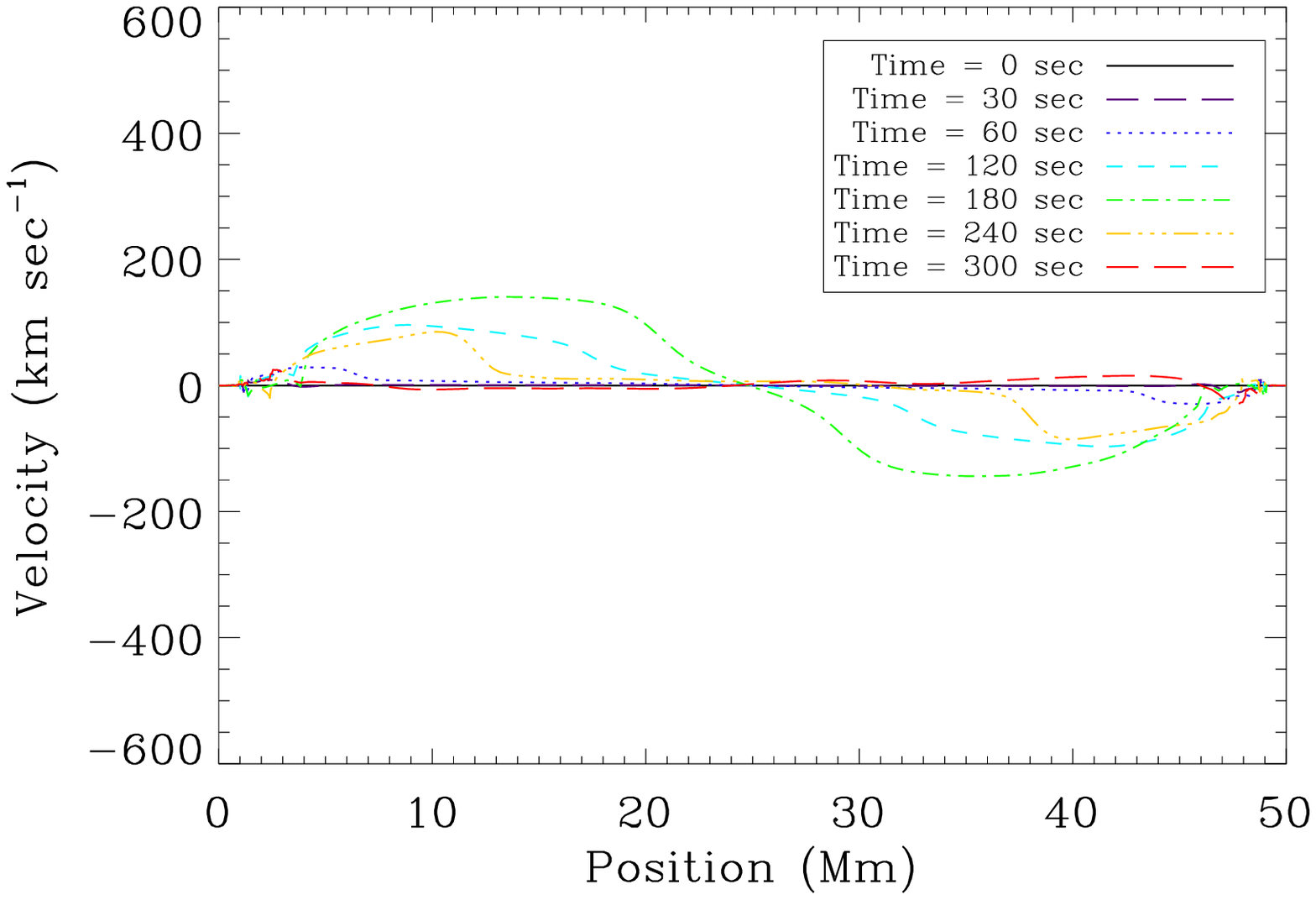}
\end{minipage}
\caption{The electron density (top row), electron temperature (middle row), and bulk flow velocity (bottom row) in Runs 9 (left column), 12 (middle column), and 16 (right column).  The simulations had equal energy fluxes, at the explosive threshold of \citet{fisher1985a}, and electron energies of 5, 20, and 50 keV, respectively.  }
\label{isoprofiles}
\end{figure}

\begin{figure}
\centering
\begin{minipage}[b]{0.32\linewidth}
\centering
\includegraphics[width=2.1in]{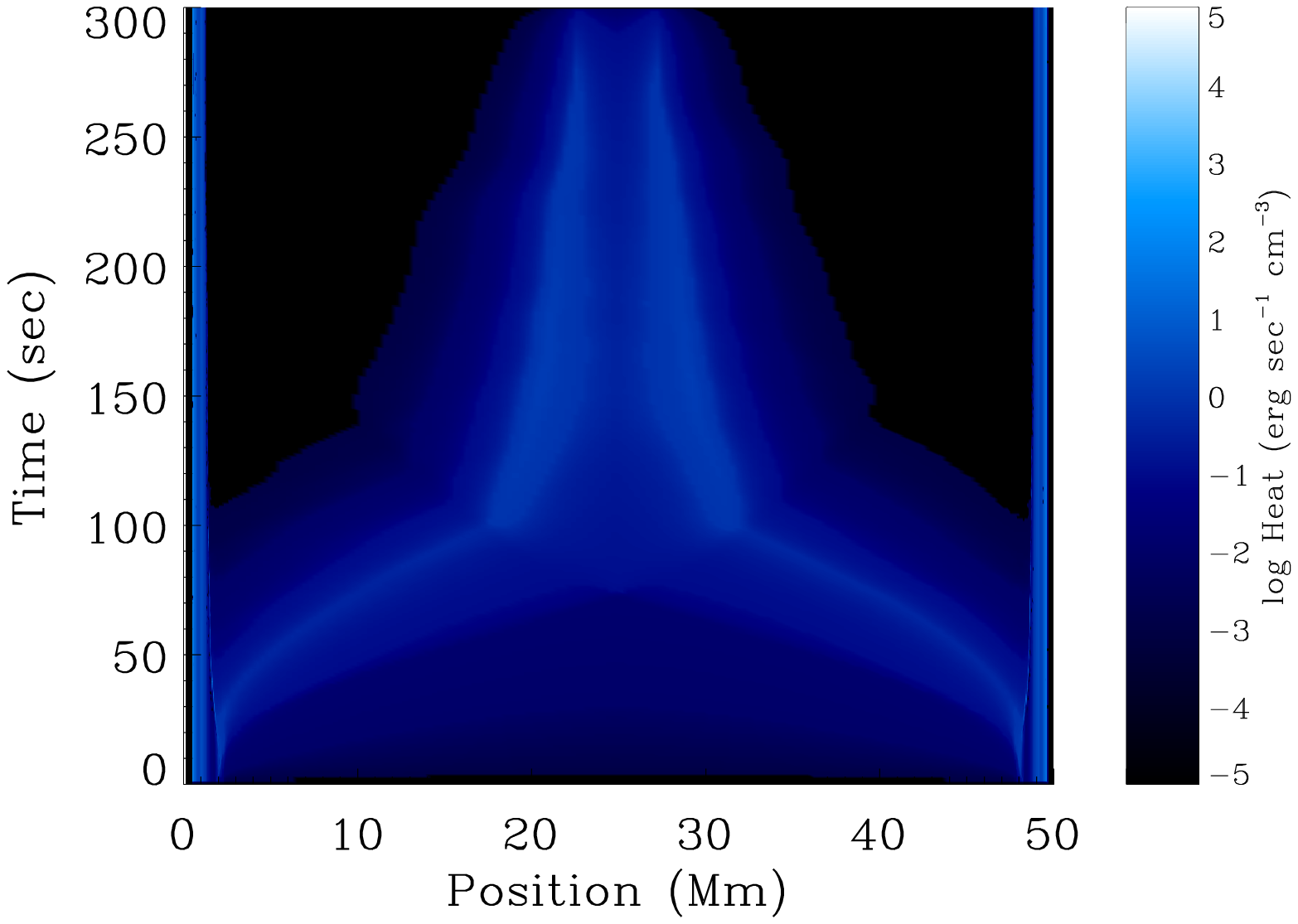}
\end{minipage}
\begin{minipage}[b]{0.32\linewidth}
\centering
\includegraphics[width=2.1in]{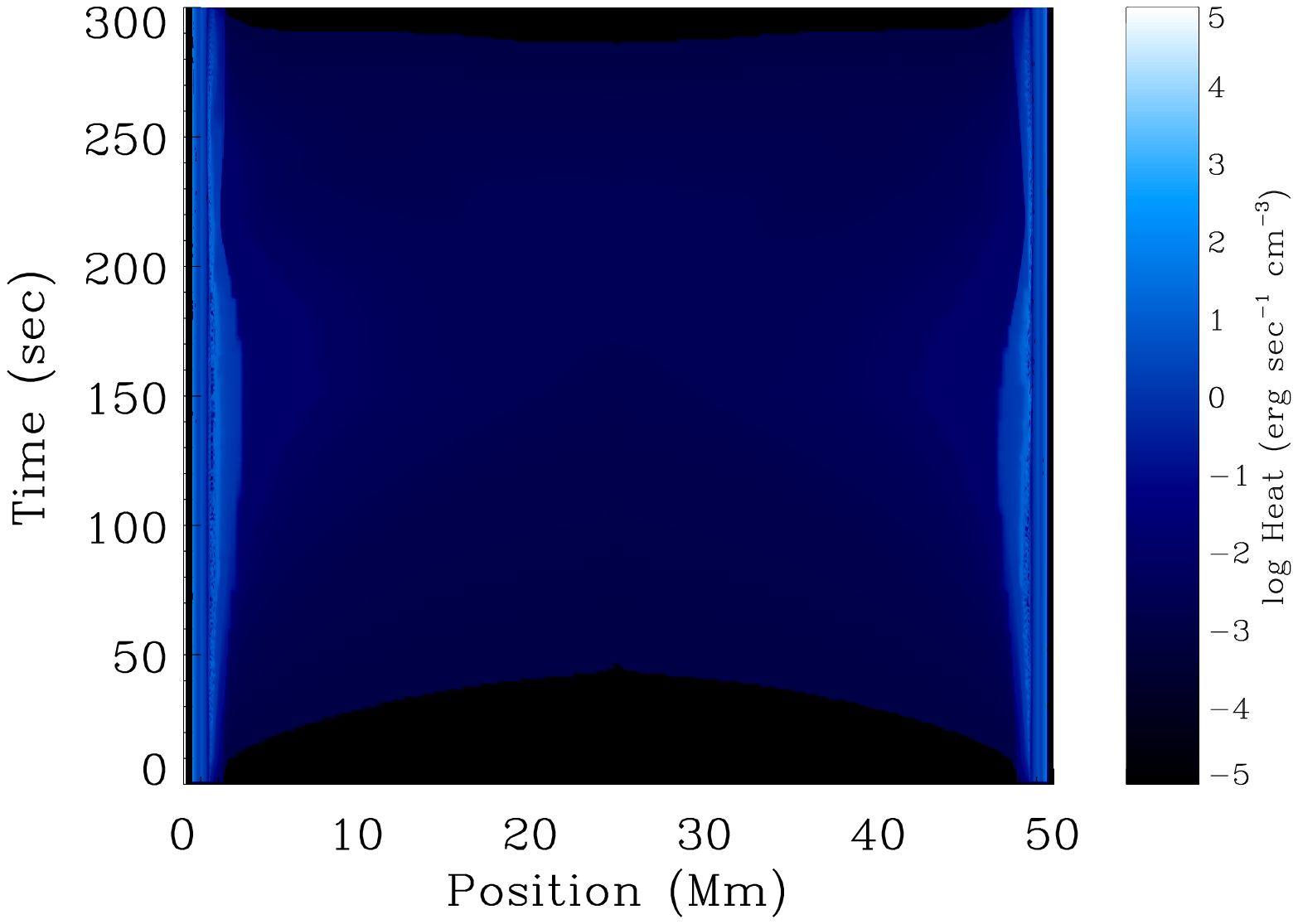}
\end{minipage}
\begin{minipage}[b]{0.32\linewidth}
\centering
\includegraphics[width=2.1in]{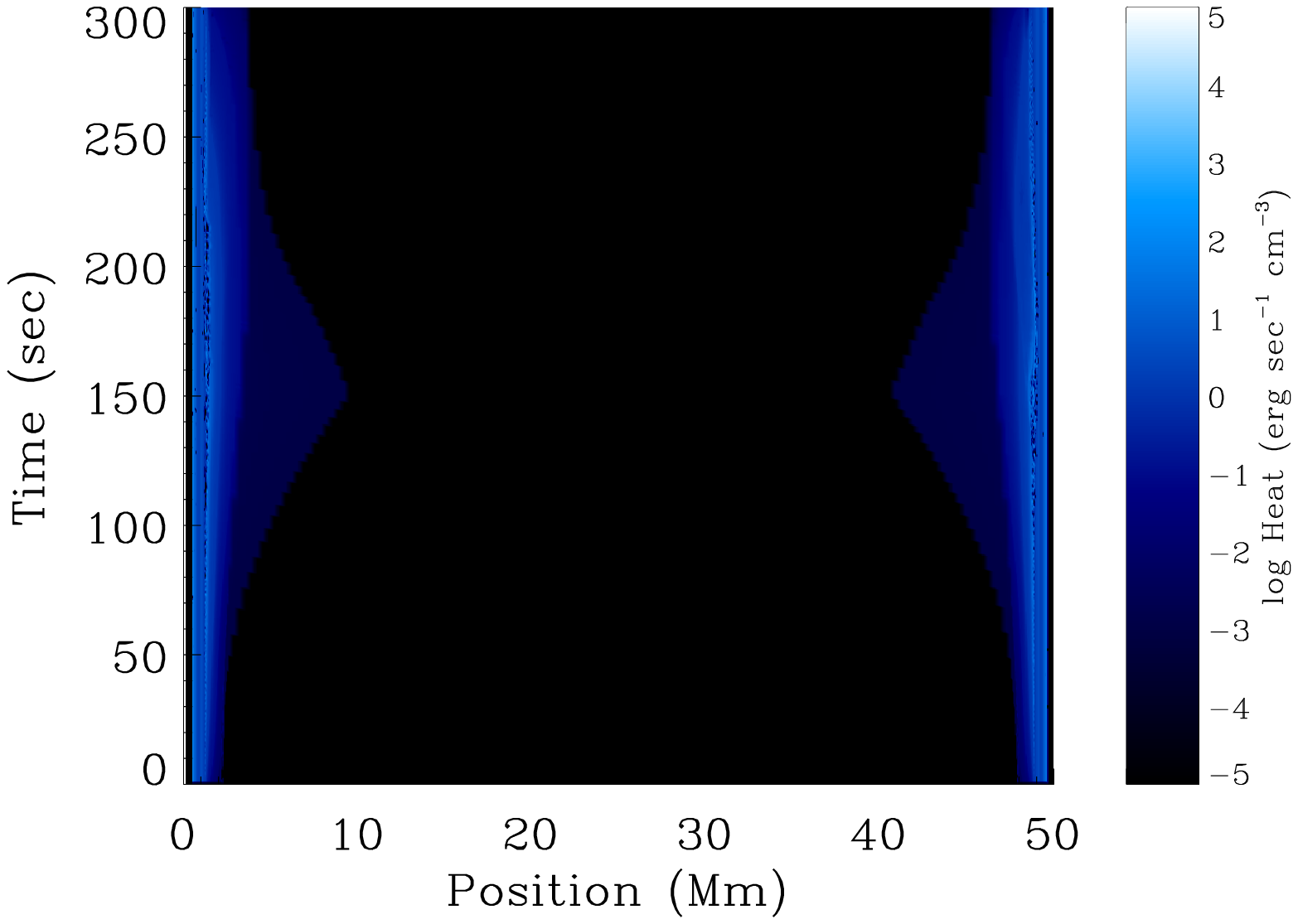}
\end{minipage}
\begin{minipage}[b]{0.32\linewidth}
\centering
\includegraphics[width=2.1in]{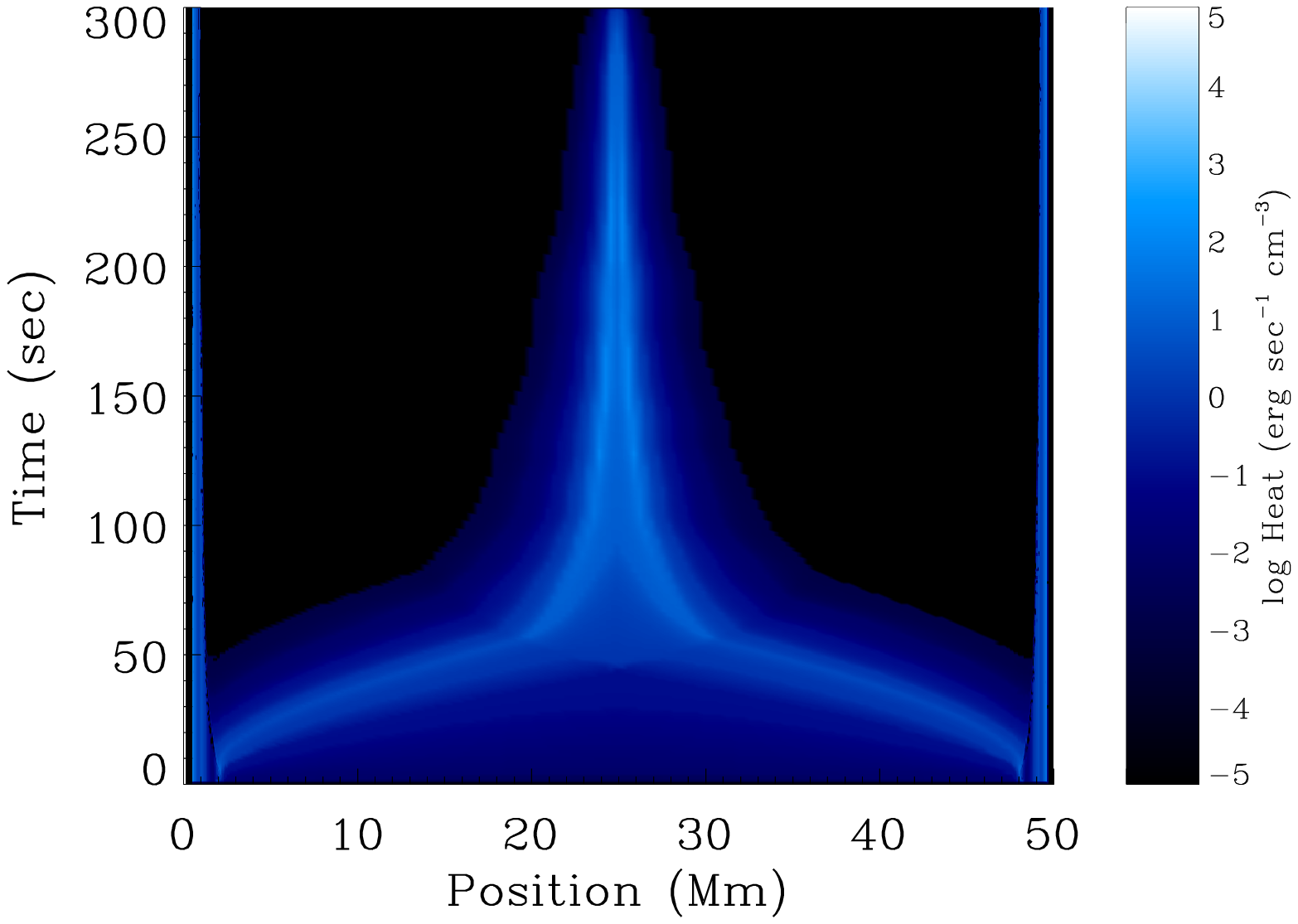}
\end{minipage}
\begin{minipage}[b]{0.32\linewidth}
\centering
\includegraphics[width=2.1in]{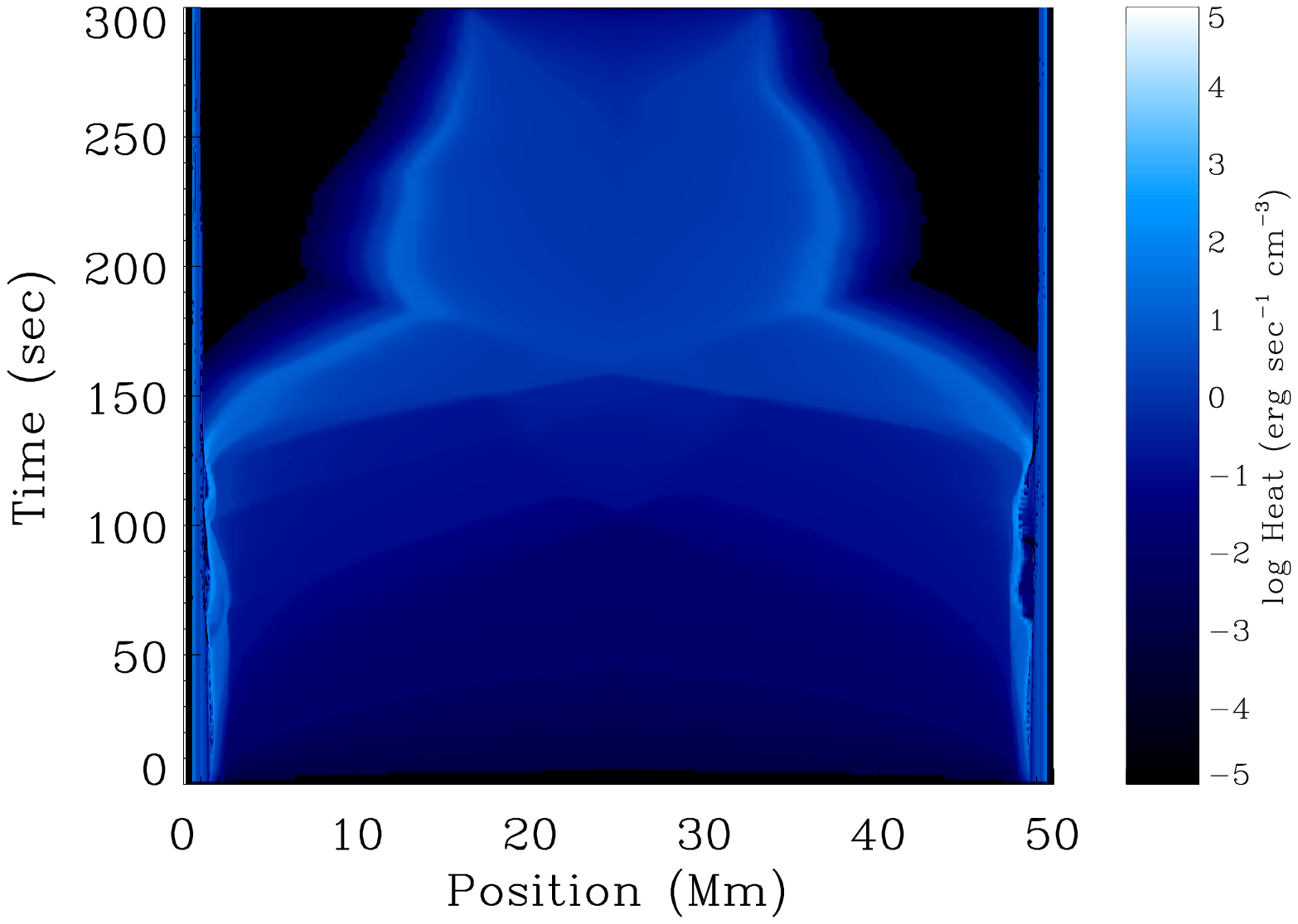}
\end{minipage}
\begin{minipage}[b]{0.32\linewidth}
\centering
\includegraphics[width=2.1in]{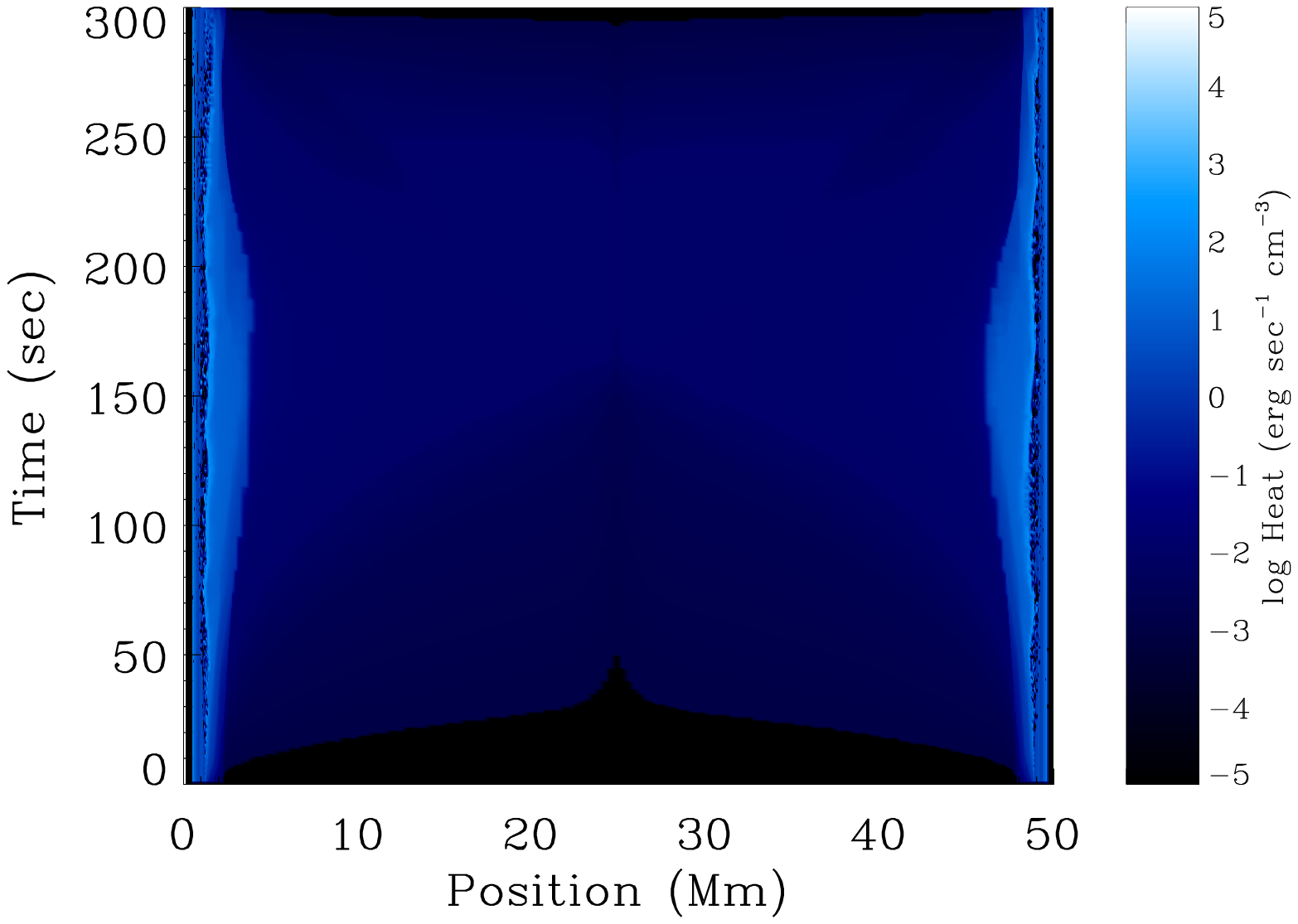}
\end{minipage}
\begin{minipage}[b]{0.32\linewidth}
\centering
\includegraphics[width=2.1in]{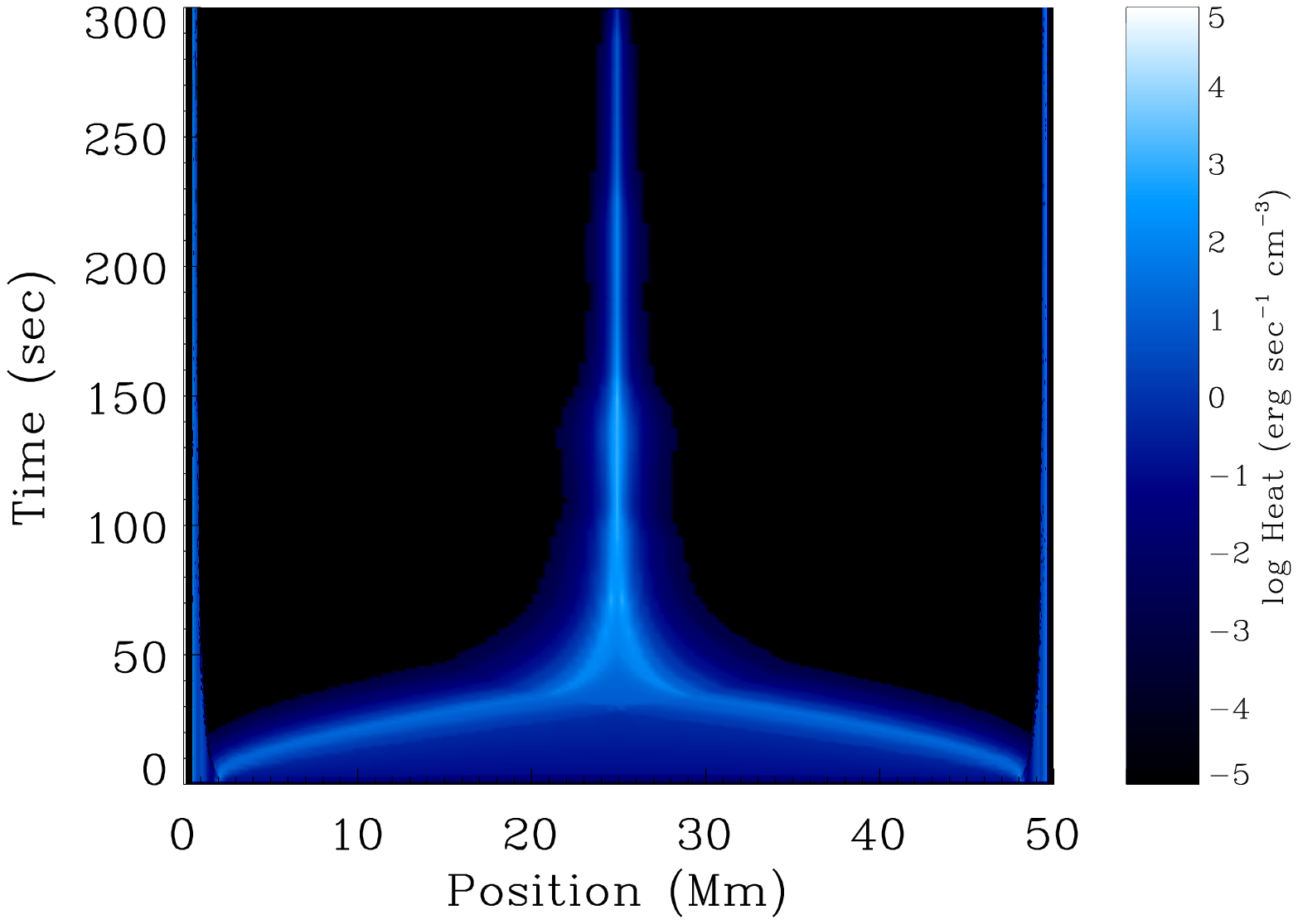}
\end{minipage}
\begin{minipage}[b]{0.32\linewidth}
\centering
\includegraphics[width=2.1in]{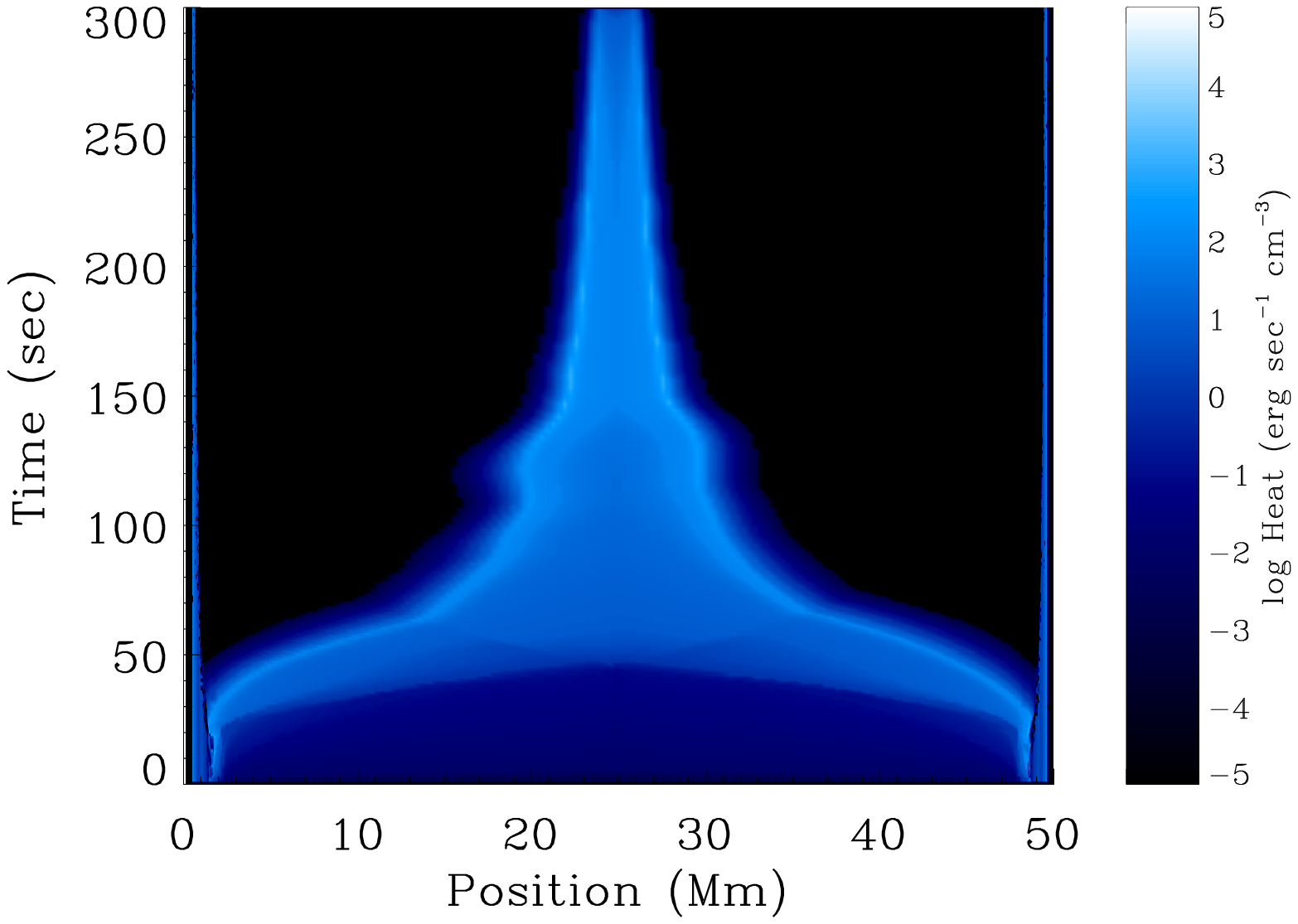}
\end{minipage}
\begin{minipage}[b]{0.32\linewidth}
\centering
\includegraphics[width=2.1in]{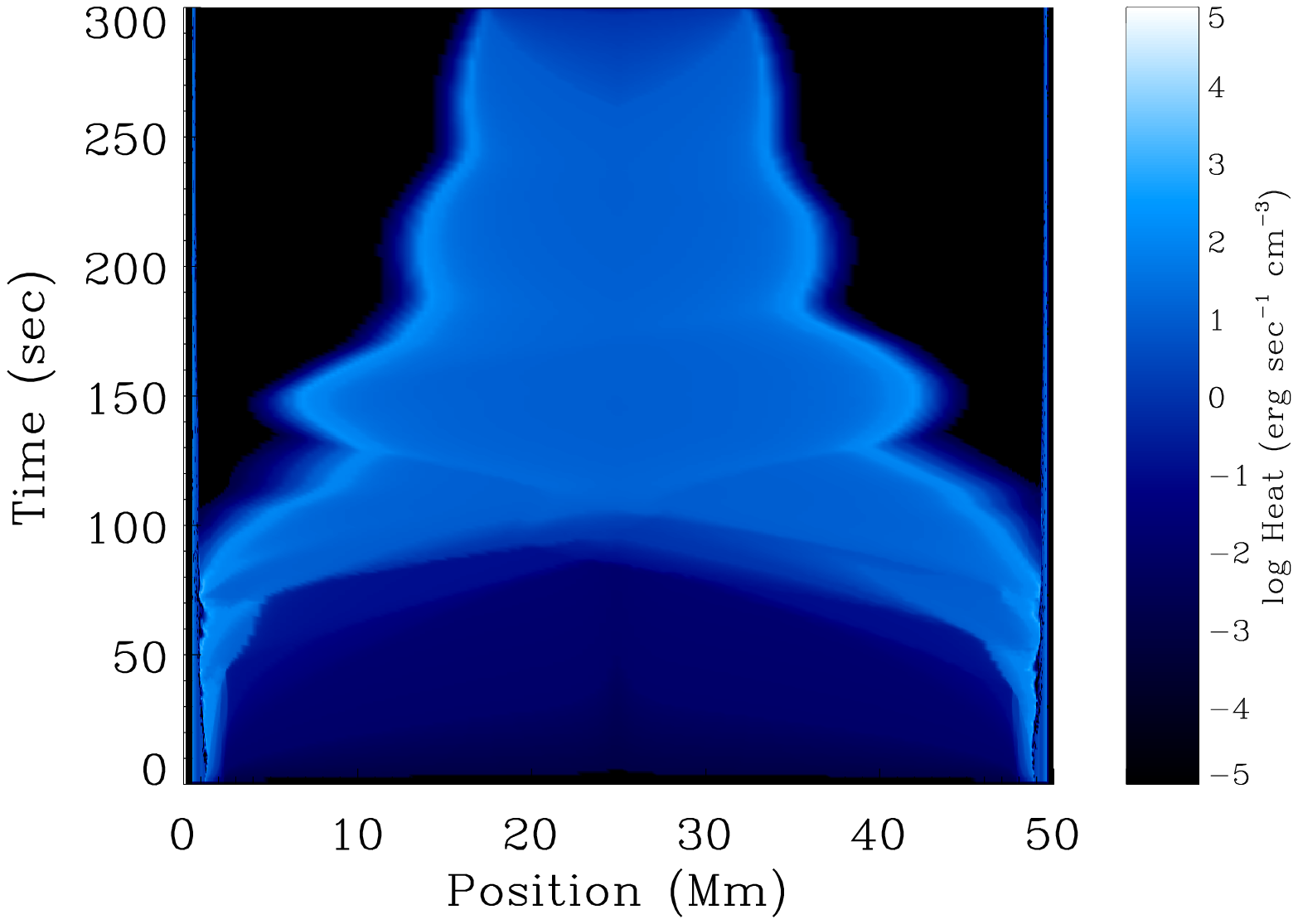}
\end{minipage}
\caption{Energy deposition as a function of time and position in nine of the simulations.  At top, Runs 1, 4, 8, with a maximal beam flux of $10^{9}$ erg sec$^{-1}$ cm$^{-2}$ and electron energies of 5, 20, 50 keV, respectively.  At center, Runs 9, 12, 16, with a maximal beam flux of $10^{10}$ erg sec$^{-1}$ cm$^{-2}$ and electron energies of 5, 20, 50 keV, respectively.  At bottom, Runs 17, 20, 24, with a maximal beam flux of $10^{11}$ erg sec$^{-1}$ cm$^{-2}$ and electron energies of 5, 20, 50 keV, respectively.}
\label{isoenergydep}
\end{figure}

Consider the atmospheric response of Runs 9, 12 and 16, which had beam electron energies of 5, 20, and 50 keV, respectively.  Figure \ref{isoprofiles} shows the electron densities (top row), electron temperatures (middle row), and bulk velocities (bottom row) in the three simulations.    Comparing the density and velocity profiles shows that Run 9 (first column) quickly develops very strong upflows of material in under 30 seconds, even though the maximum of the beam flux occurs at 150 seconds.  Essentially, although most of the energy is deposited in the chromosphere, a significant fraction of electrons are depositing their energy in the corona (see Figure \ref{isoenergydep}), which quickly raises the temperature above 10 MK and drives a strong thermal conduction front.  The combination of chromospheric energy deposition and the conduction front causes a large, explosive evaporation of material back into the chromosphere, leading to velocities up to 576 km sec$^{-1}$ into the corona.  Compare this with Run 12 (middle column), which had much less coronal energy deposition, and thus does not heat up as quickly.  Instead, the electrons in this simulation heat the chromosphere directly, so that evaporation starts later than in Run 9 (although Run 12 reaches a higher maximal flow velocity, 750 km sec$^{-1}$, at around 150 seconds).  At later times, as the corona fills, the electrons no longer stream directly through, so that they start depositing their energy {\it in situ}.  Finally, consider Run 16 (right column), with an electron energy of 50 keV, so that the electrons travel essentially collisionlessly through the corona at all times in the simulation.  The result is that there is no direct heating of the corona, and because higher energy electrons deposit their energy deeper down, the heat capacity is significantly higher so that there will be a much smaller pressure increase.  There is very little evaporation of material, and the density and temperature only rise slightly from their initial values.  

\begin{figure}
\centering
\begin{minipage}[b]{0.32\linewidth}
\centering
\includegraphics[width=2.2in]{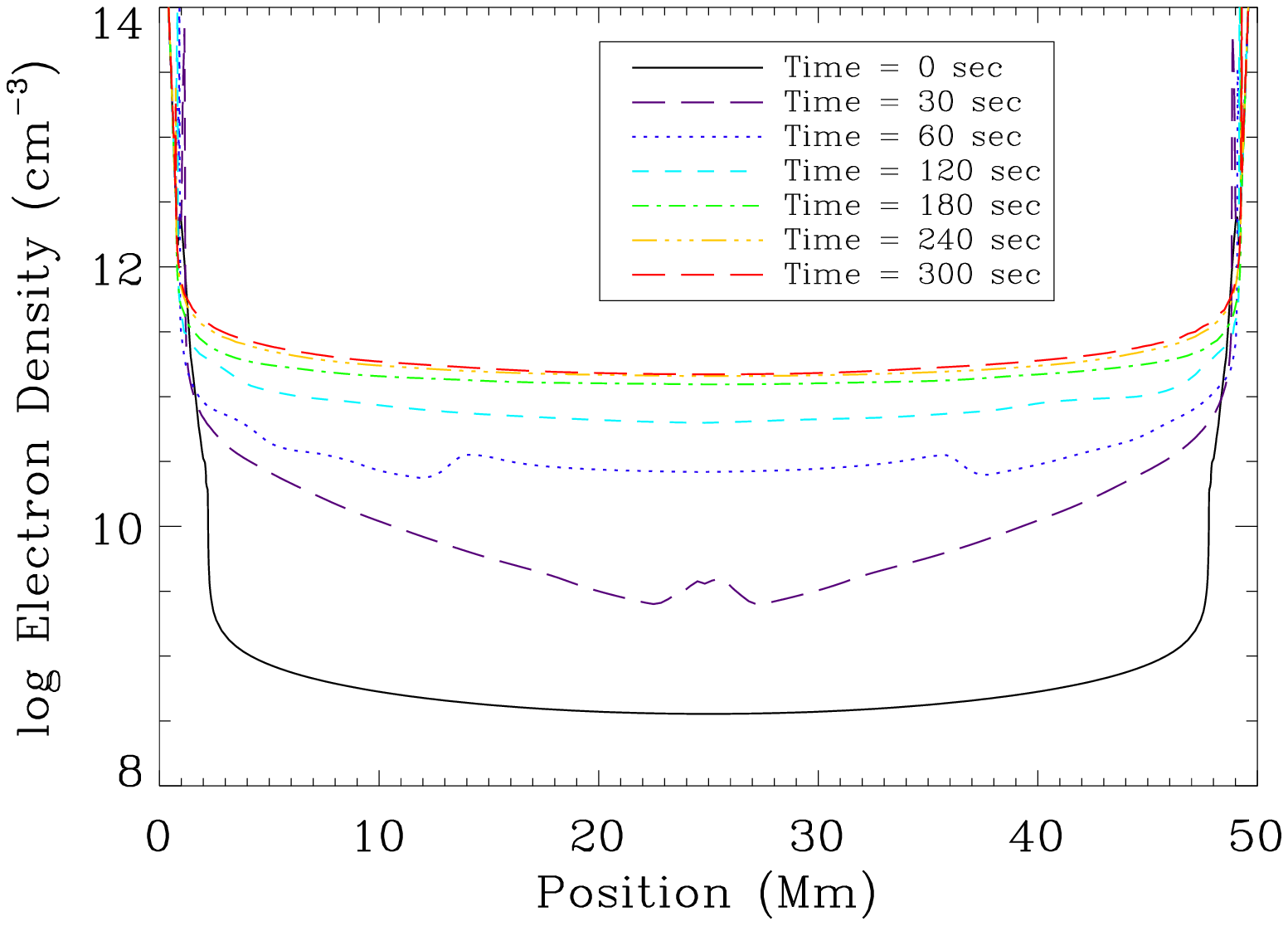}
\end{minipage}
\begin{minipage}[b]{0.32\linewidth}
\centering
\includegraphics[width=2.2in]{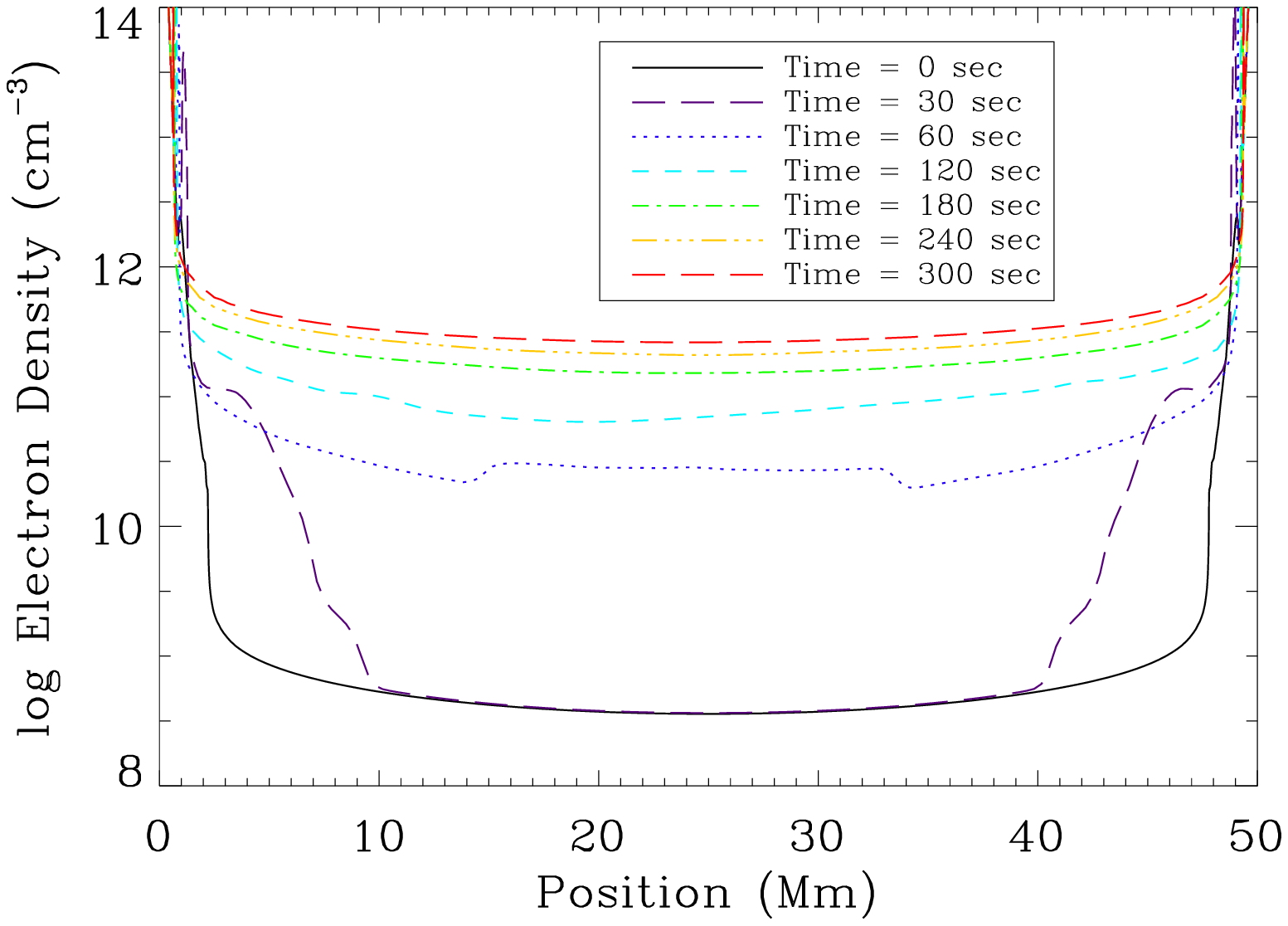}
\end{minipage}
\begin{minipage}[b]{0.32\linewidth}
\centering
\includegraphics[width=2.2in]{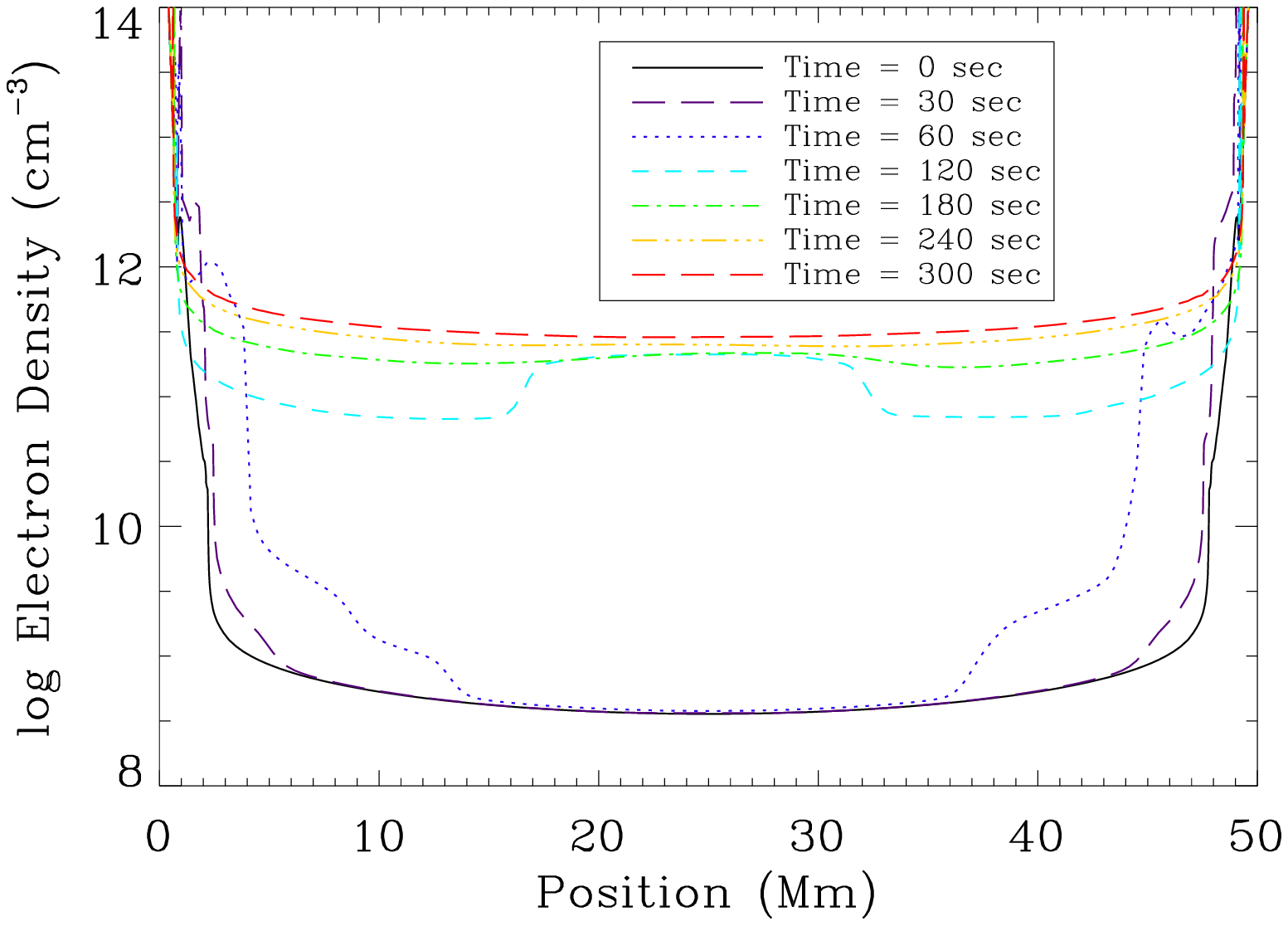}
\end{minipage}
\begin{minipage}[b]{0.32\linewidth}
\centering
\includegraphics[width=2.2in]{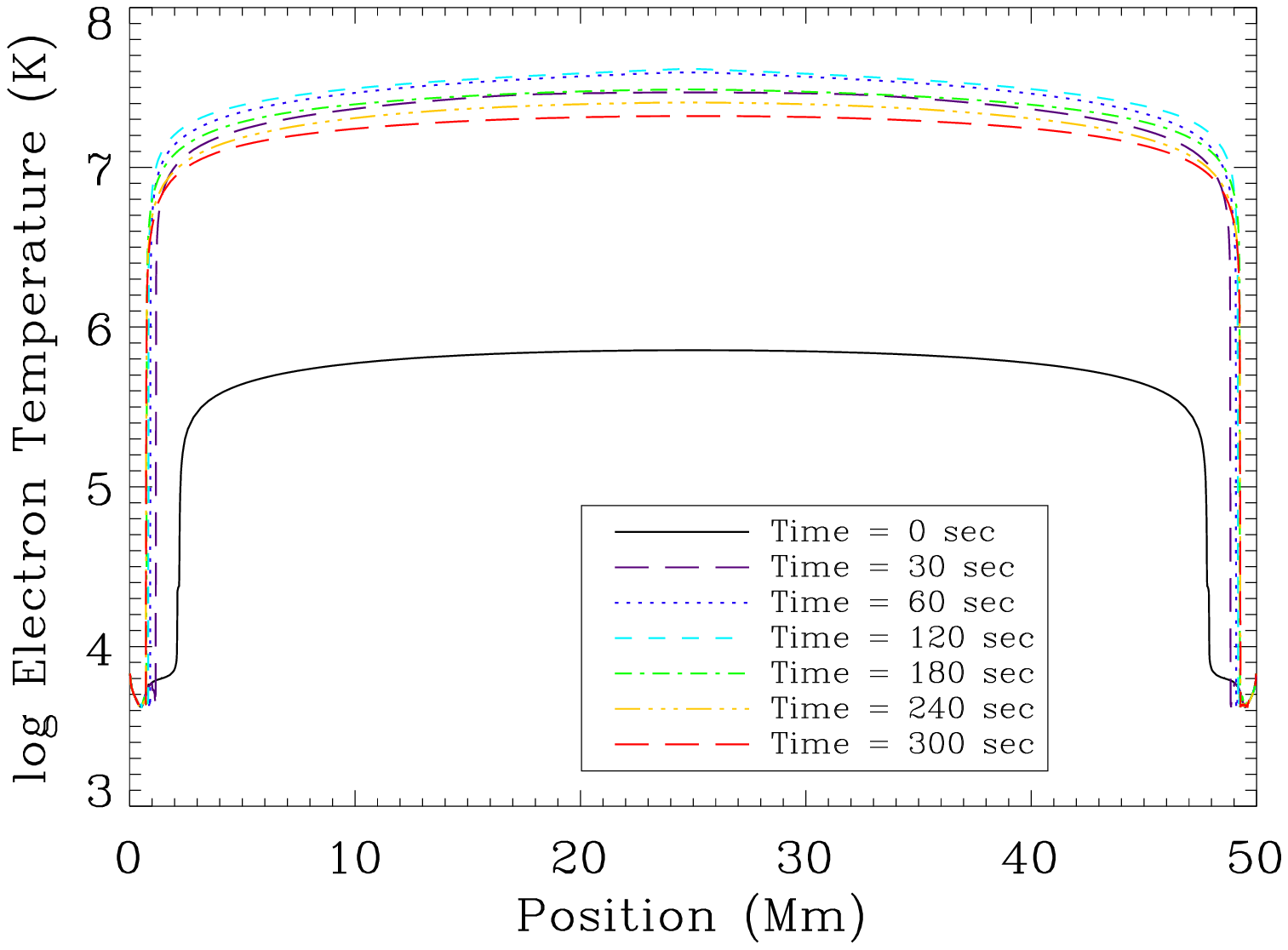}
\end{minipage}
\begin{minipage}[b]{0.32\linewidth}
\centering
\includegraphics[width=2.2in]{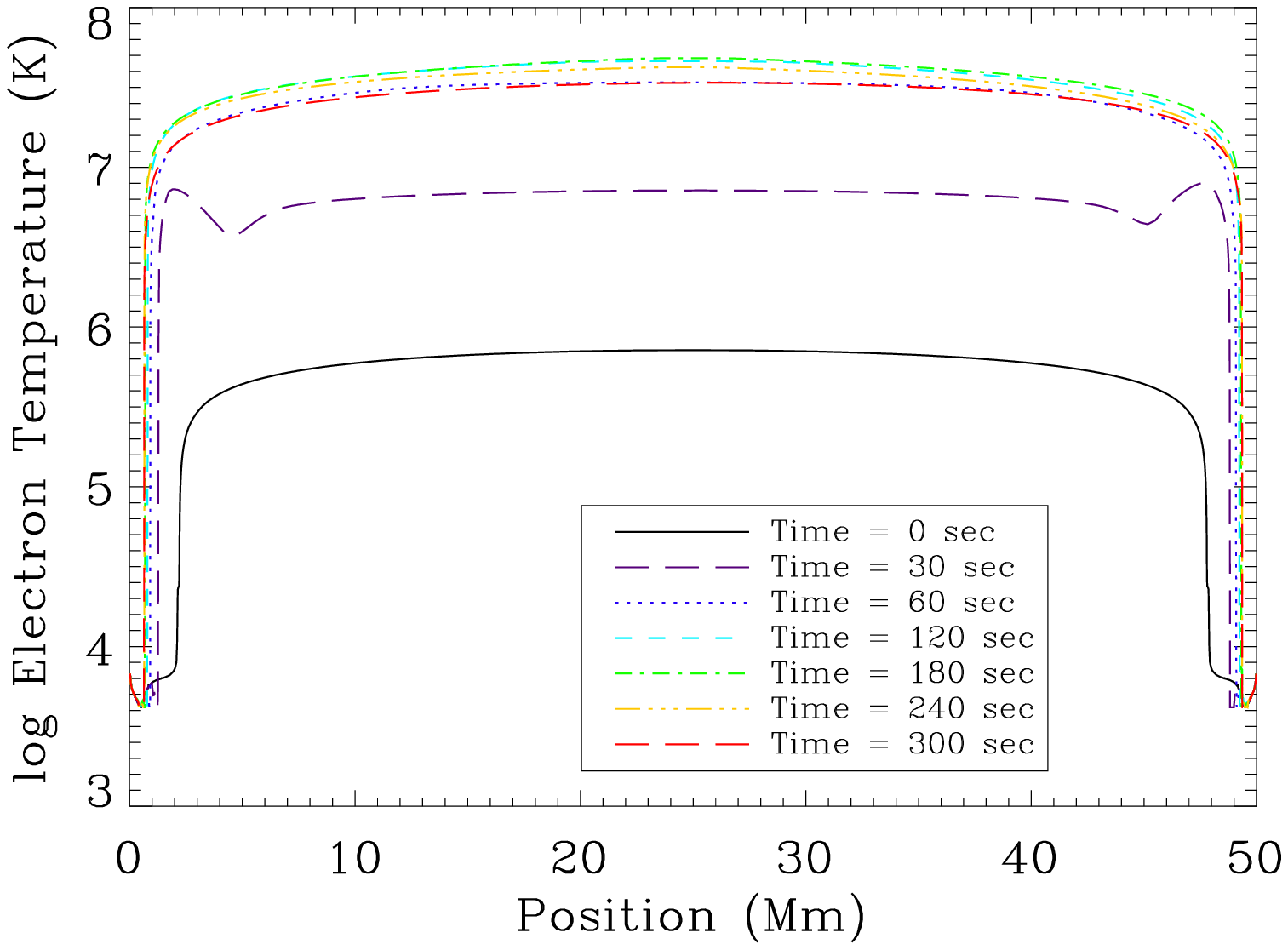}
\end{minipage}
\begin{minipage}[b]{0.32\linewidth}
\centering
\includegraphics[width=2.2in]{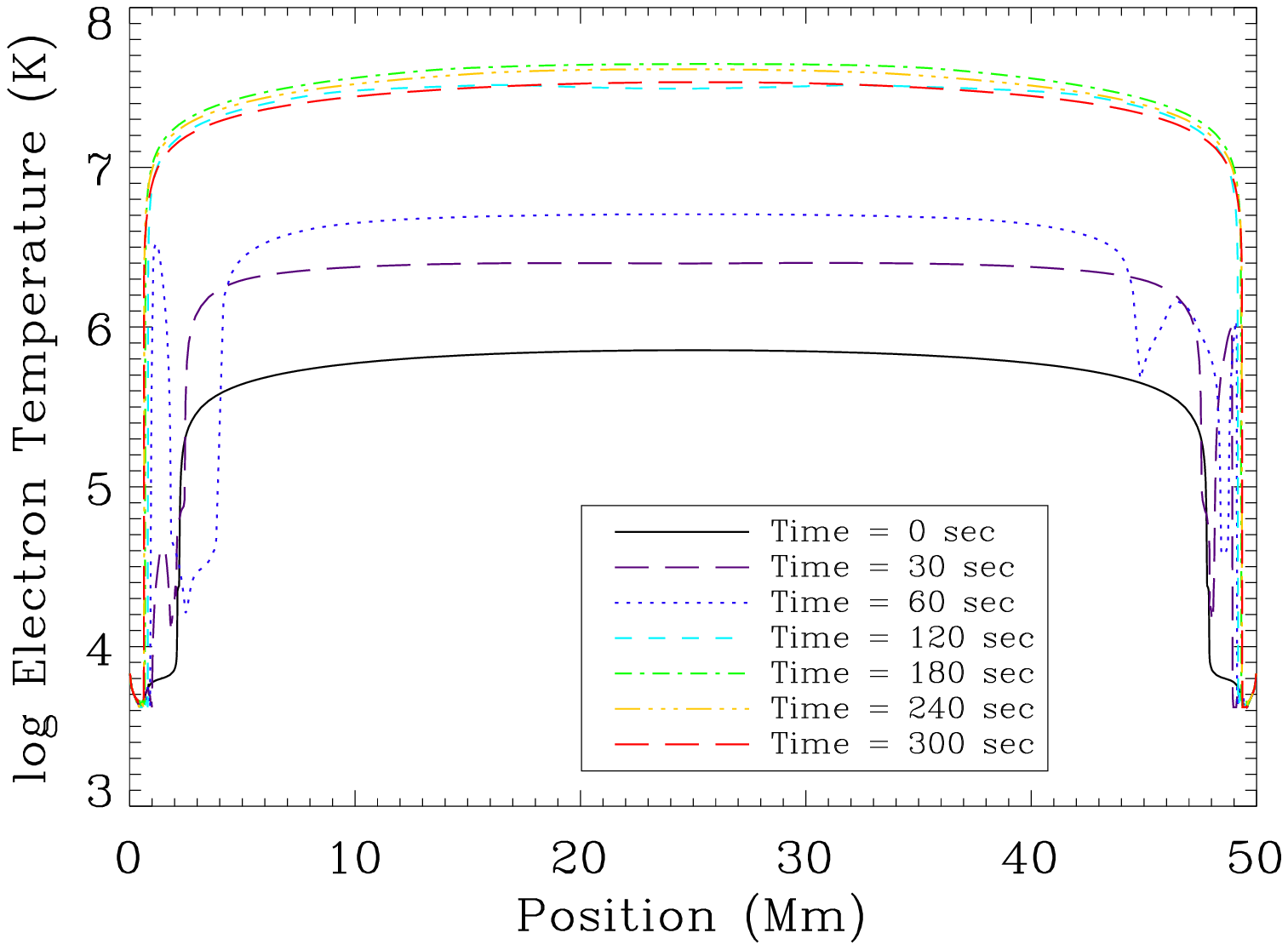}
\end{minipage}
\begin{minipage}[b]{0.32\linewidth}
\centering
\includegraphics[width=2.2in]{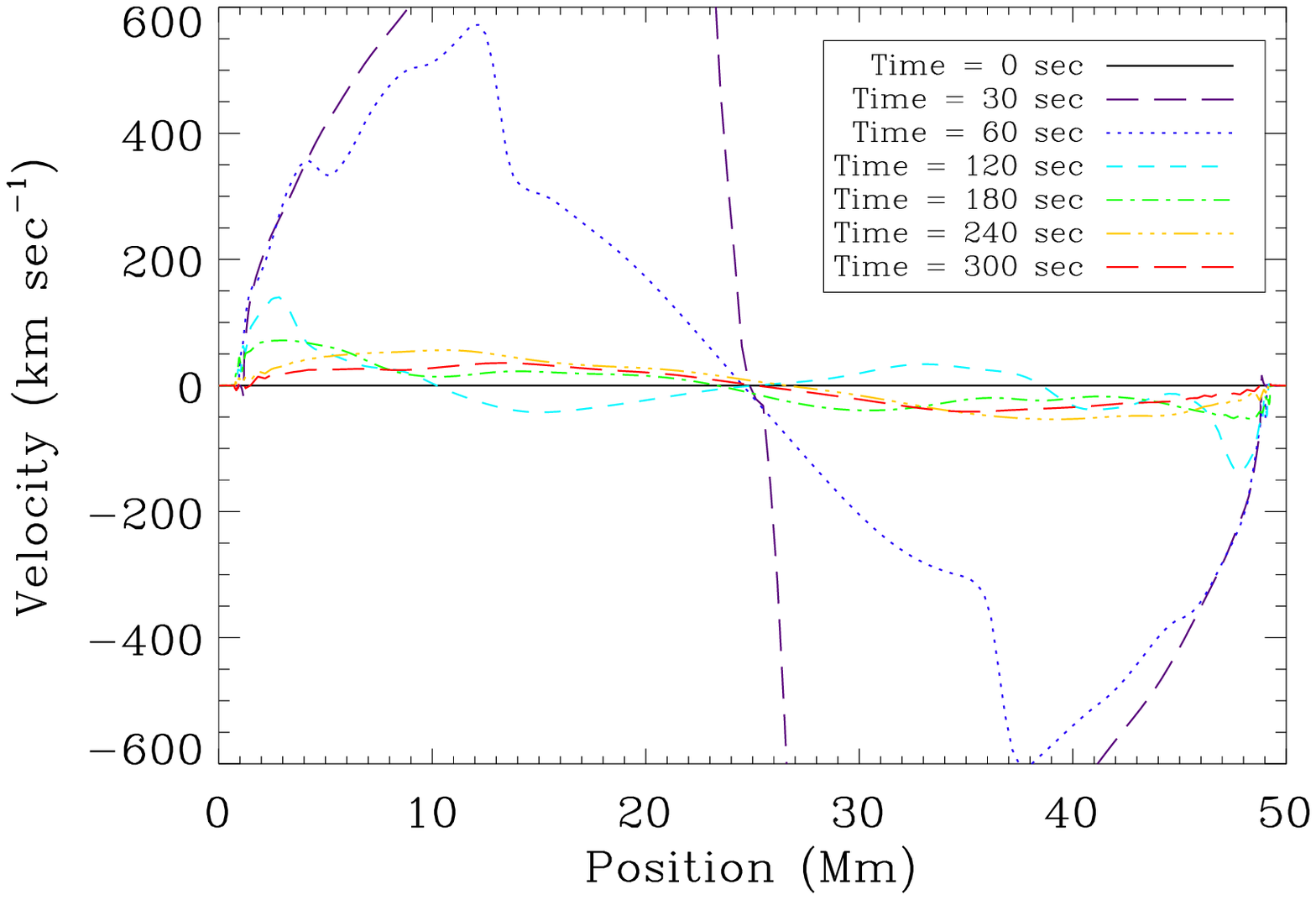}
\end{minipage}
\begin{minipage}[b]{0.32\linewidth}
\centering
\includegraphics[width=2.2in]{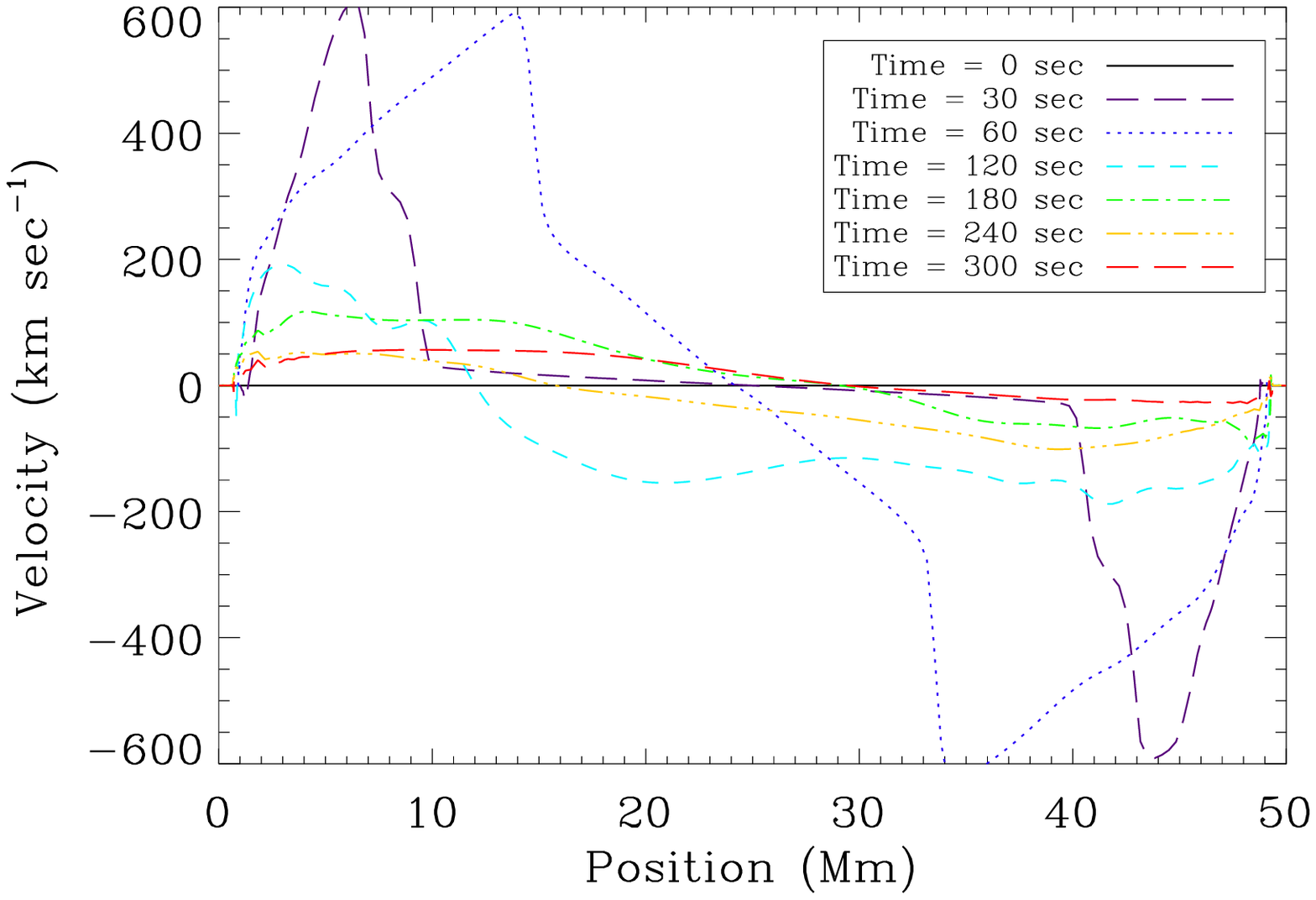}
\end{minipage}
\begin{minipage}[b]{0.32\linewidth}
\centering
\includegraphics[width=2.2in]{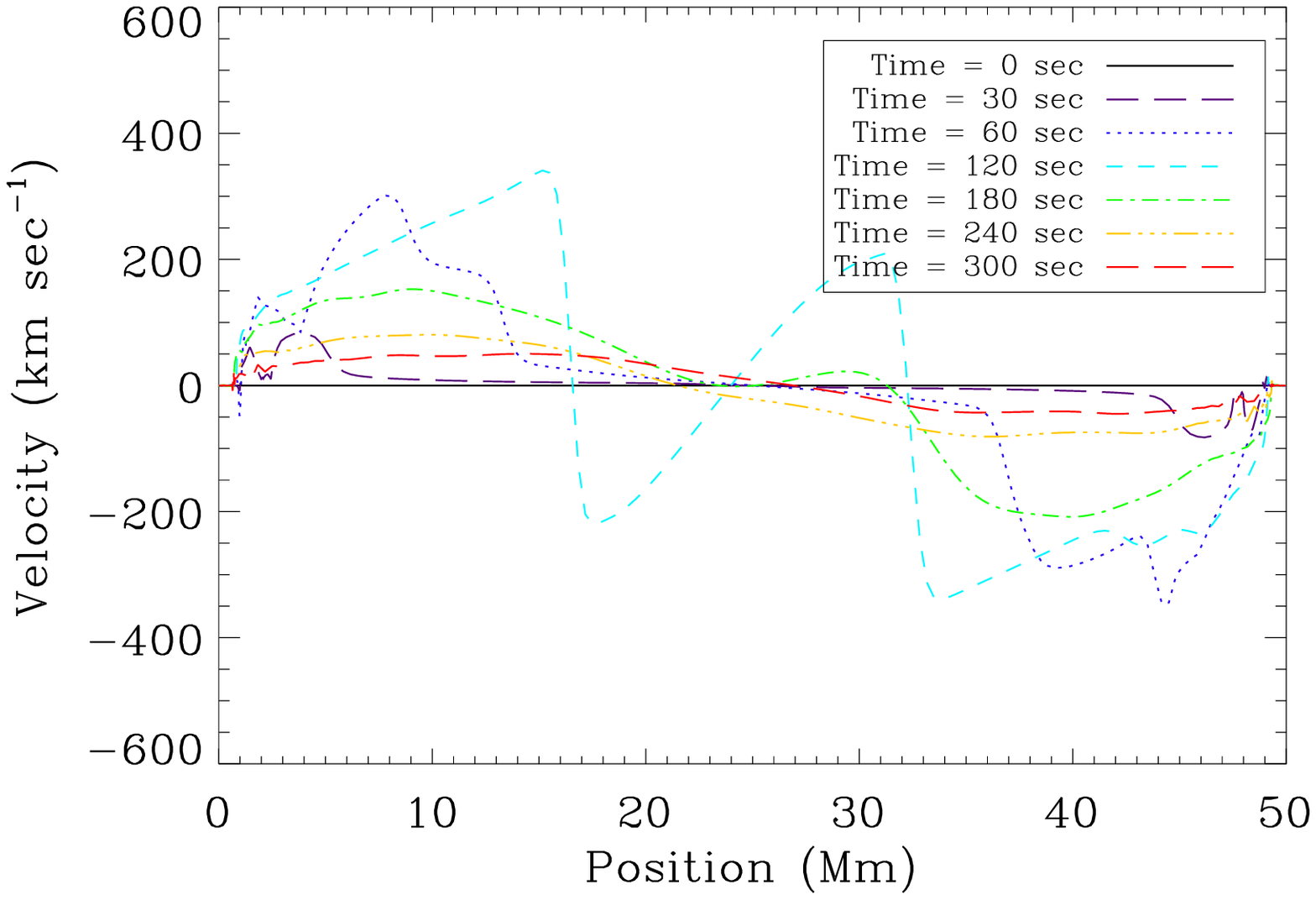}
\end{minipage}
\caption{The electron density (top row), electron temperature (middle row), and bulk flow velocity (bottom row) in Runs 17 (left column), 20 (middle column), and 24 (right column).  The simulations had equal energy fluxes, above the explosive threshold of \citet{fisher1985a}, and electron energies of 5, 20, and 50 keV, respectively.  }
\label{isoprofiles2}
\end{figure}

Figure \ref{isoprofiles2} similarly compares Runs 17, 20, and 24.  At the end of the heating, the three simulations have similar densities (just over $10^{11}$ cm$^{-3}$ at the apex) and similar temperatures ($\approx 50$ MK).  In Run 17, as with Run 9, large upflows develop in under 30 seconds, quickly raising the coronal density and temperature.  The coronal temperature quickly rises above 10 MK, driving a thermal conduction front down the loop, which combined with the chromospheric energy deposition, drives a very strong evaporation of material.  The heating becomes more and more localized toward the apex (see Figure \ref{isoenergydep}) as the flows slow and the density reaches its peak.  In Run 20, the energy is primarily deposited in the chromosphere, which drives an explosive evaporation upwards.  Although the flows take more time to develop than in Run 17, they carry a similar amount of material into the corona, filling and heating it drastically.  Finally, the behavior in Run 24 is completely different from Run 16.  Now, the heat flux deposited in the transition region and chromosphere is large compared to the thermal energy at that depth and the excess energy cannot be radiated away fast enough, so that the pressure rises dramatically (note the chromospheric temperature spikes at 60 seconds).  The flows in this simulation begin around 60 seconds into the simulation, quickly raising the coronal density by nearly three orders of magnitude thereafter.

To elucidate the differences between the beams and the resultant atmospheric response, consider the energy deposition in the simulations.  Figure \ref{isoenergydep} shows the energy deposition in 9 of the simulations: Runs 1, 4, 8 (top, left to right), 9, 12, and 16 (center, left to right), and 17, 20, and 24 (bottom, left to right).  A few properties are readily apparent in these figures.  First of all, the smaller the electron energy, the more the energy deposition becomes localized near the apex.  This is in agreement with the predictions of \citet{nagai1984}, and clearly shows that the highest energy electrons will stream through the corona.  Secondly, a stronger beam evaporates more material, leading to a denser corona and thus shorter mean-free paths of the electrons, so that heating becomes concentrated toward the apex.  Compare the energy deposition in Runs 1, 9 and 17, which quickly become localized near the apex of the loop, although Run 1 is more spread out spatially at all times than Run 9, which is more spread out than Run 17.  Note that, even though Runs 1 and 2 are supposed to be below the explosive evaporation threshold of \citet{fisher1985a}, the bulk velocities exceed 300 km sec$^{-1}$, and the coronal densities and temperatures become nearly an order of magnitude higher than in Runs 3-8.  Since the electrons are low energy, a significant amount of their energy is deposited in the corona, which drives a thermal conduction front, further increasing the pressure in the region of evaporation.   These results indicate the threshold is a function of electron energy, which will be examined in Section \ref{sec:evap}.  

\begin{figure}
\centering
\begin{minipage}[b]{0.32\linewidth}
\centering
\includegraphics[width=2.2in]{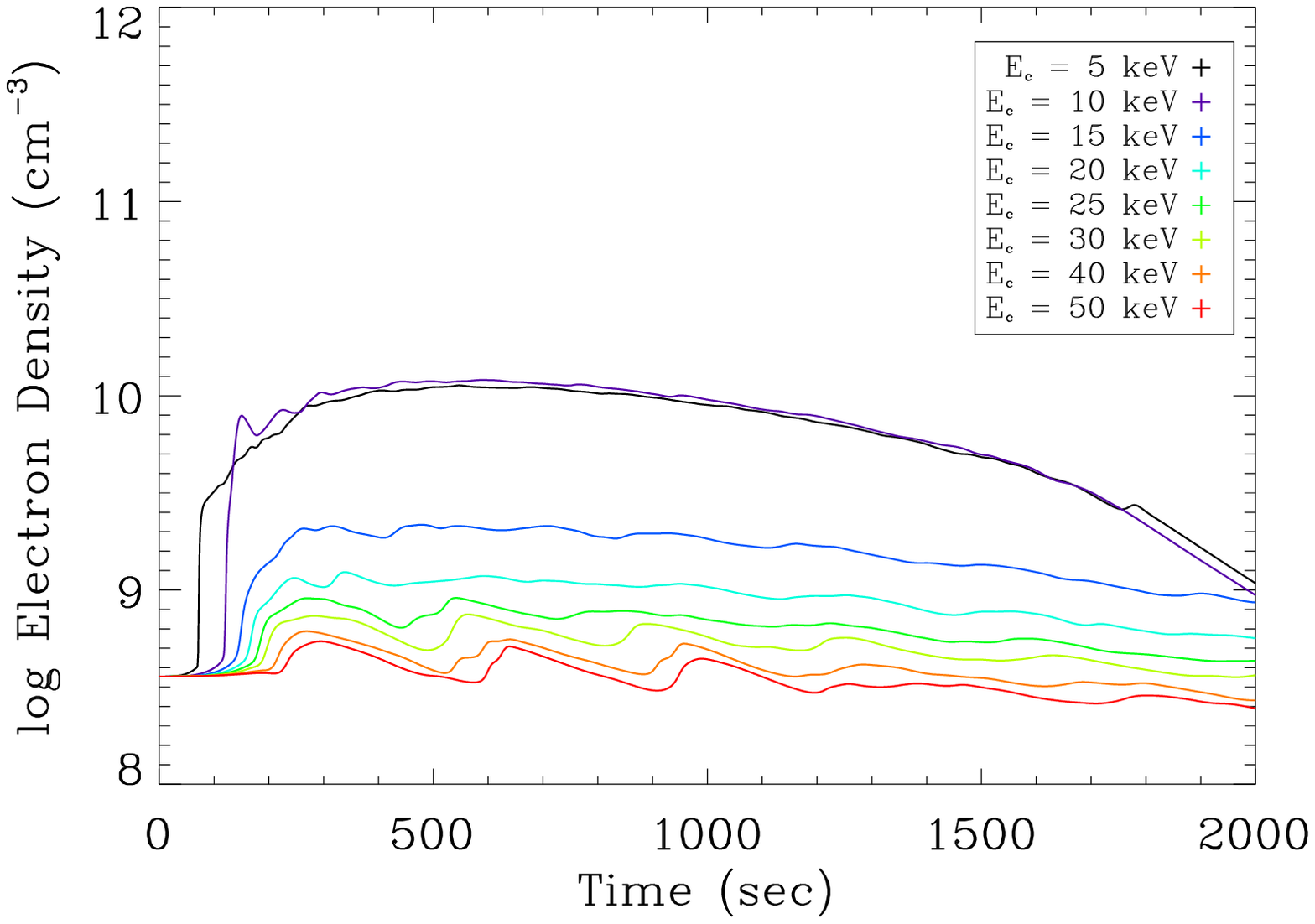}
\end{minipage}
\begin{minipage}[b]{0.32\linewidth}
\centering
\includegraphics[width=2.2in]{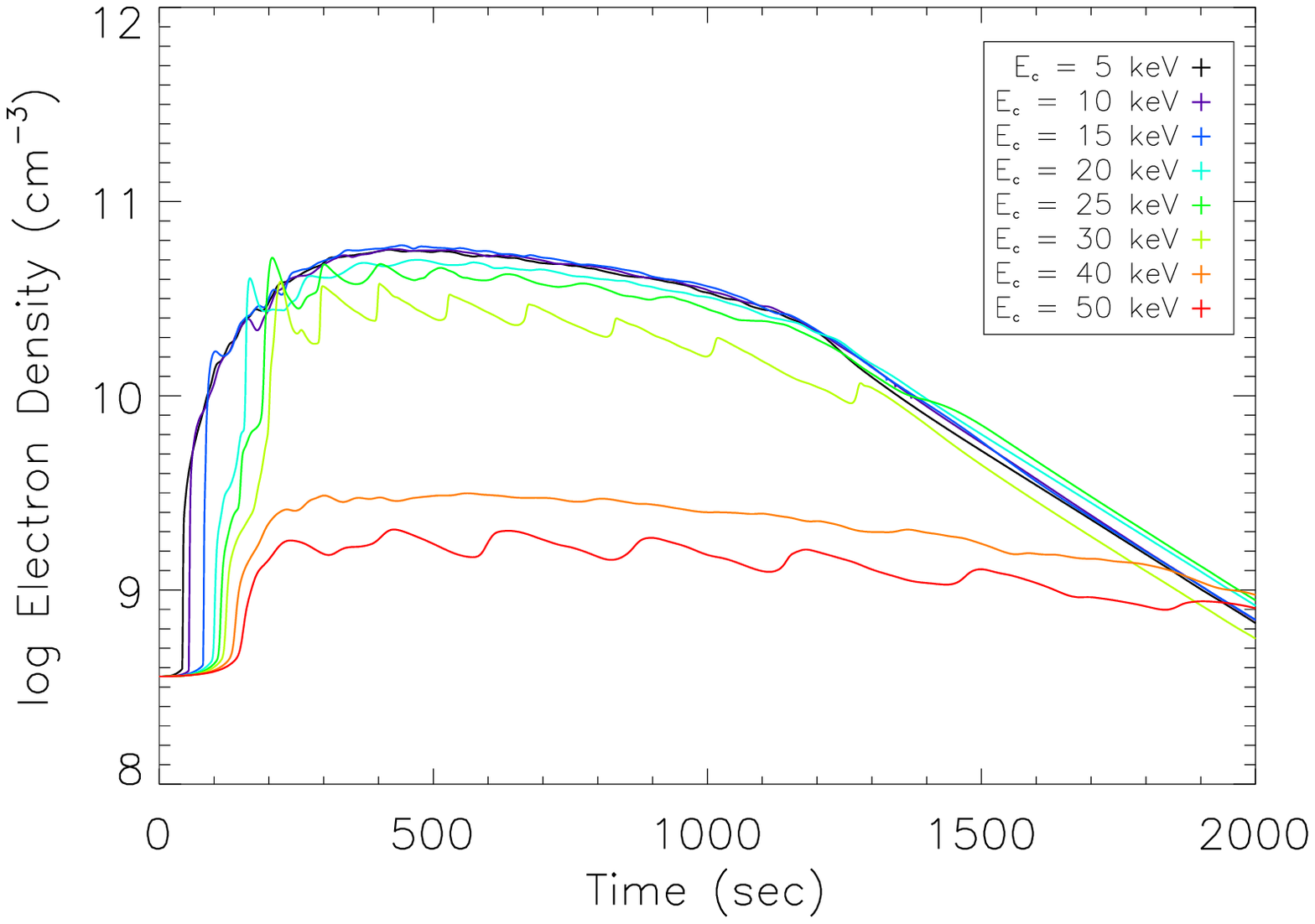}
\end{minipage}
\begin{minipage}[b]{0.32\linewidth}
\centering
\includegraphics[width=2.2in]{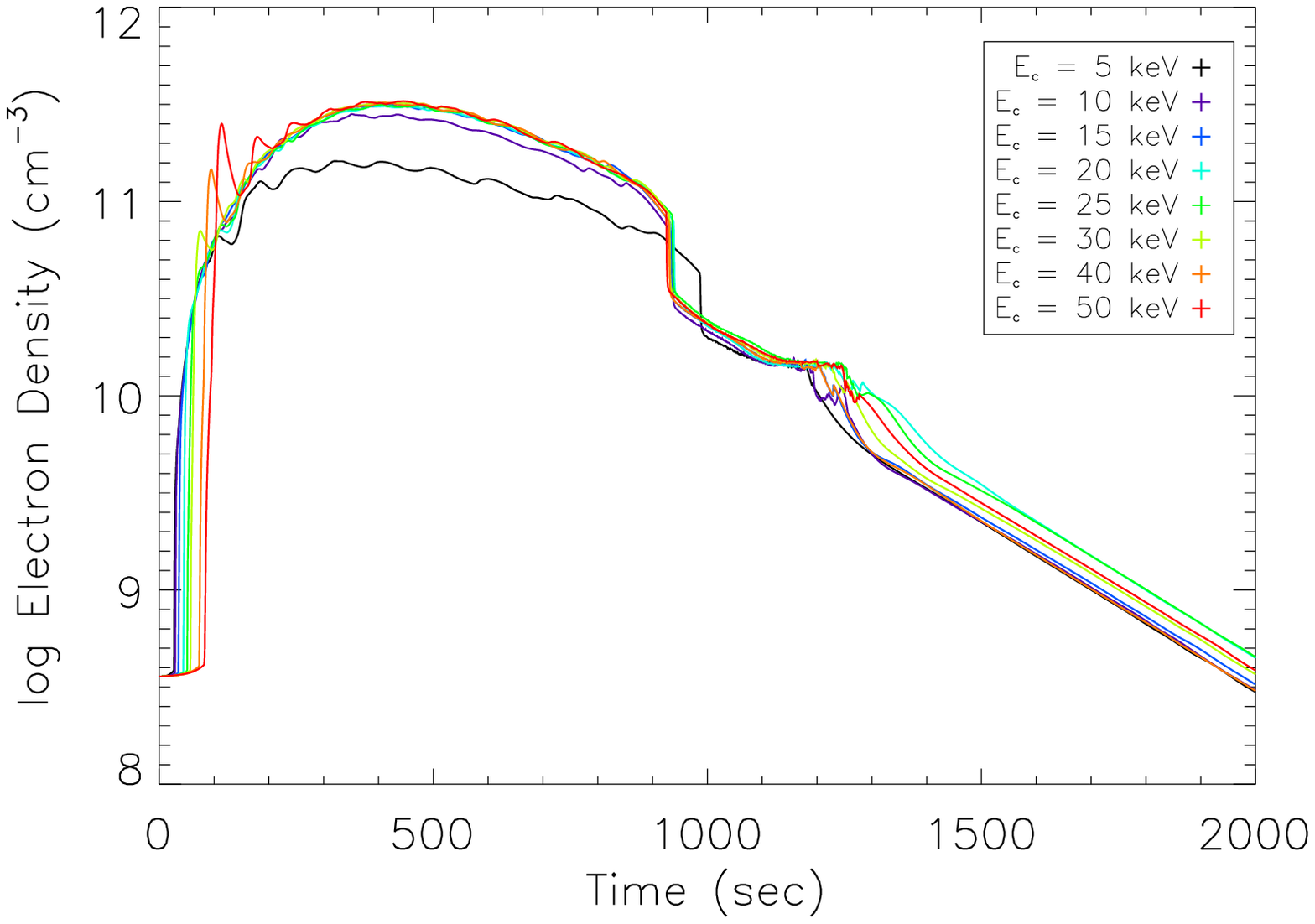}
\end{minipage}
\begin{minipage}[b]{0.32\linewidth}
\centering
\includegraphics[width=2.2in]{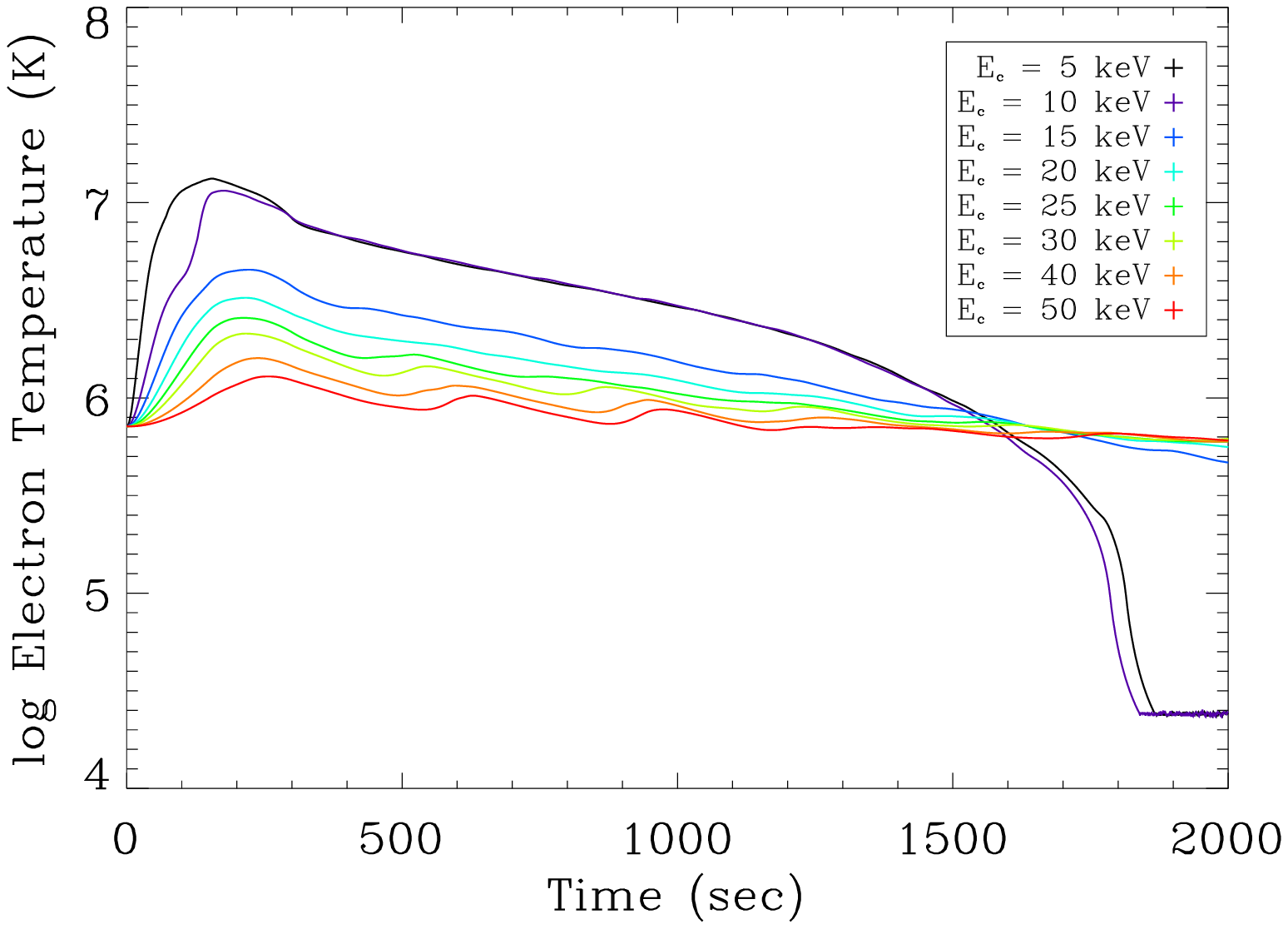}
\end{minipage}
\begin{minipage}[b]{0.32\linewidth}
\centering
\includegraphics[width=2.2in]{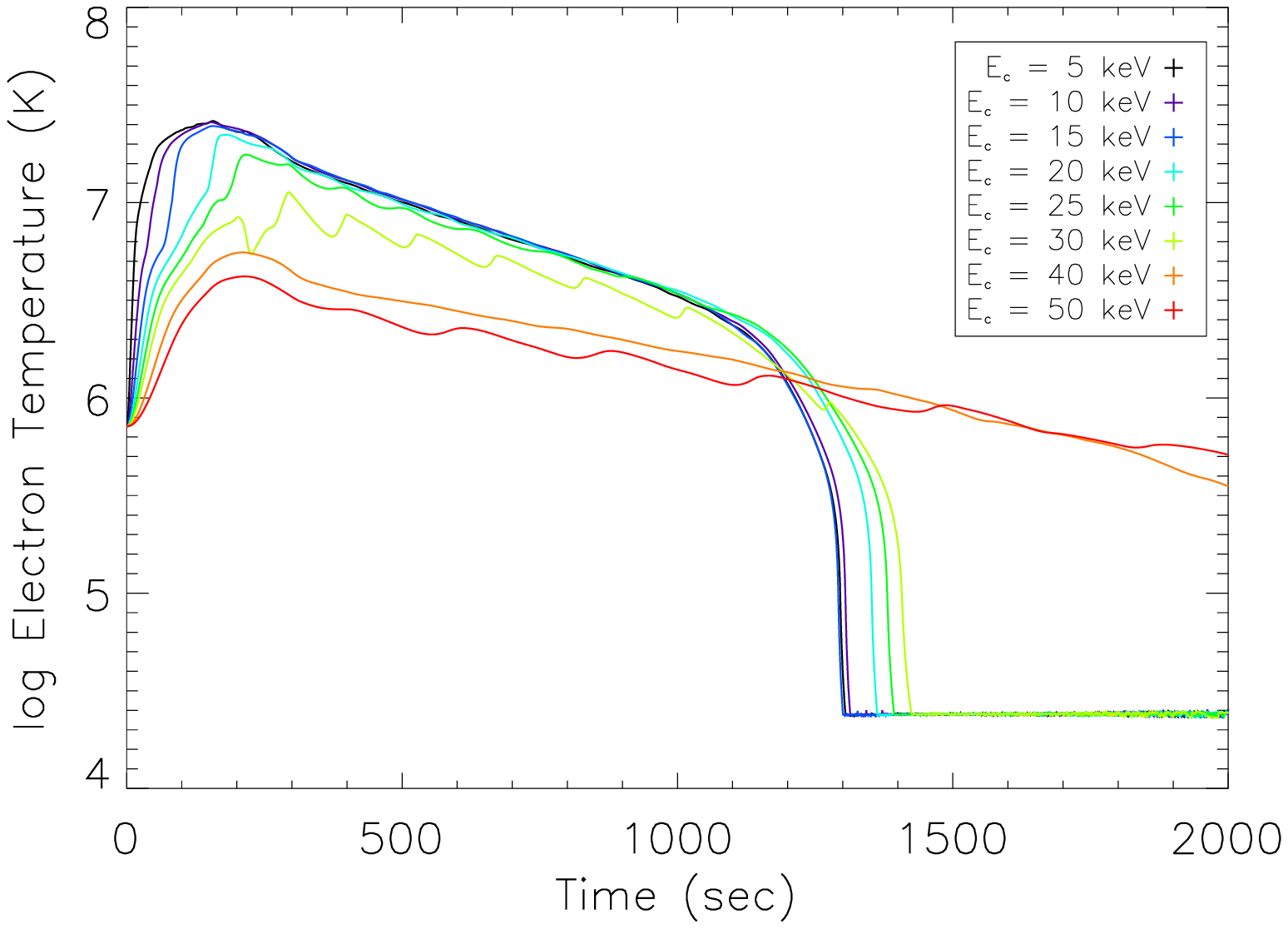}
\end{minipage}
\begin{minipage}[b]{0.32\linewidth}
\centering
\includegraphics[width=2.2in]{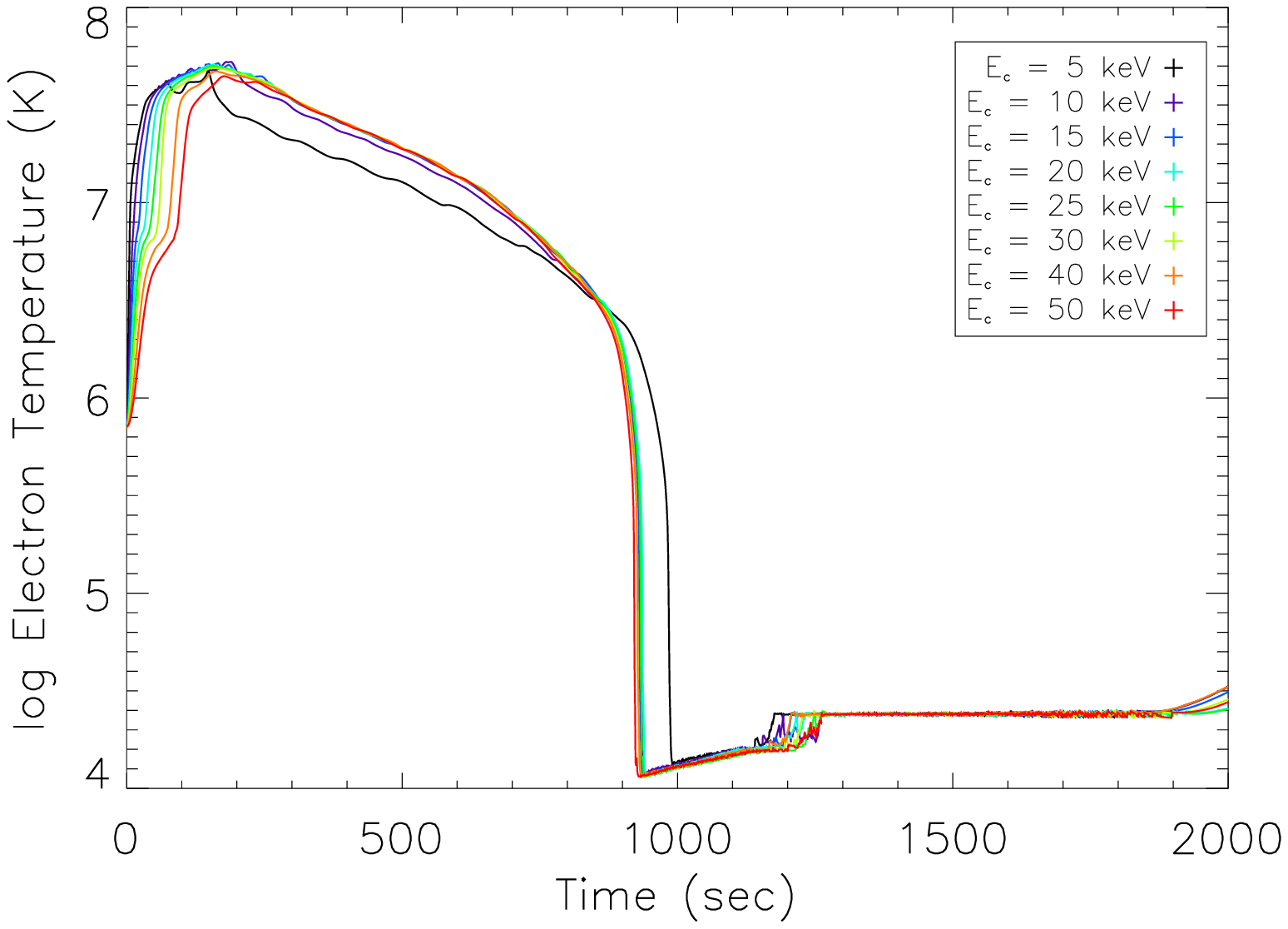}
\end{minipage}
\begin{minipage}[b]{0.32\linewidth}
\centering
\includegraphics[width=2.2in]{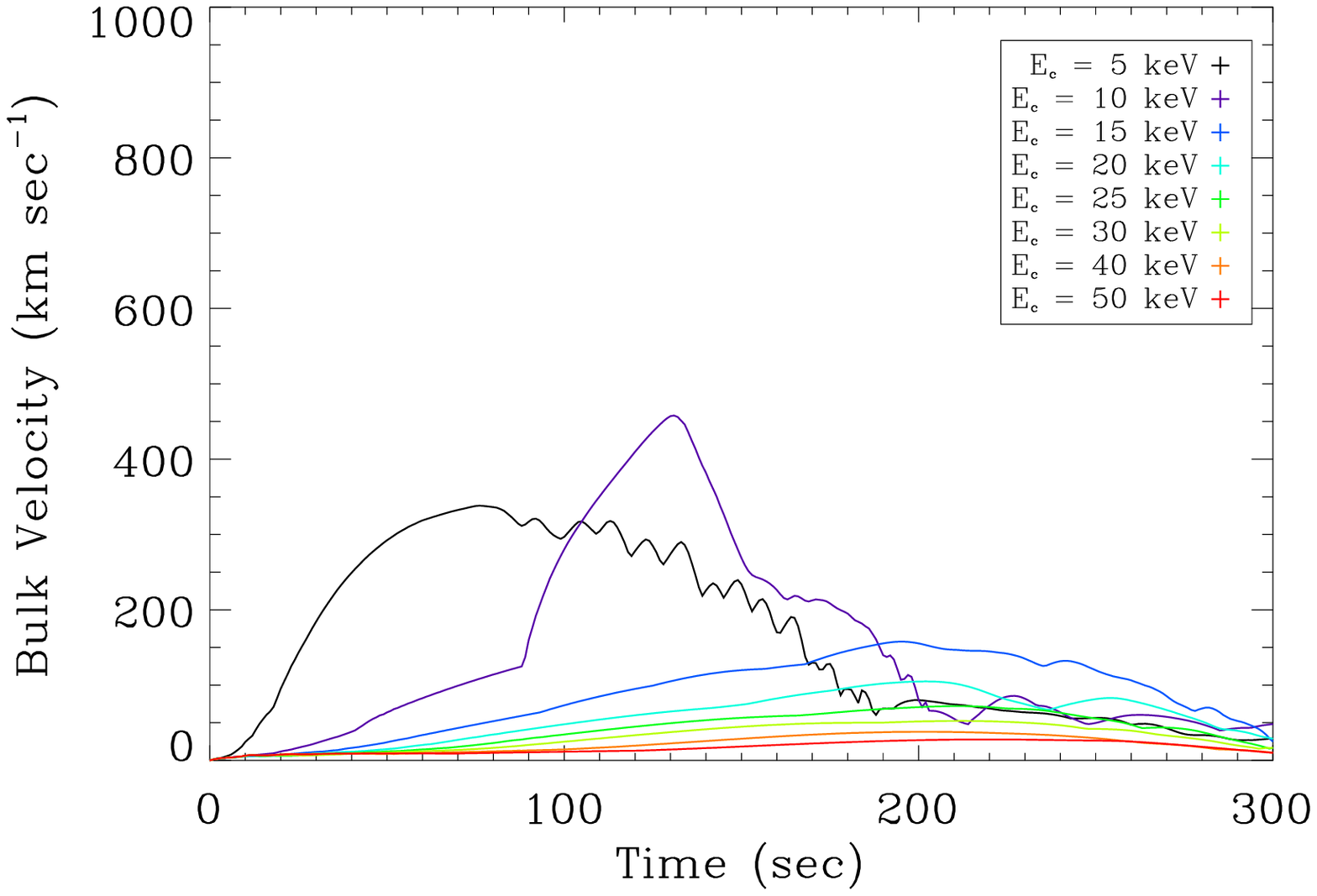}
\end{minipage}
\begin{minipage}[b]{0.32\linewidth}
\centering
\includegraphics[width=2.2in]{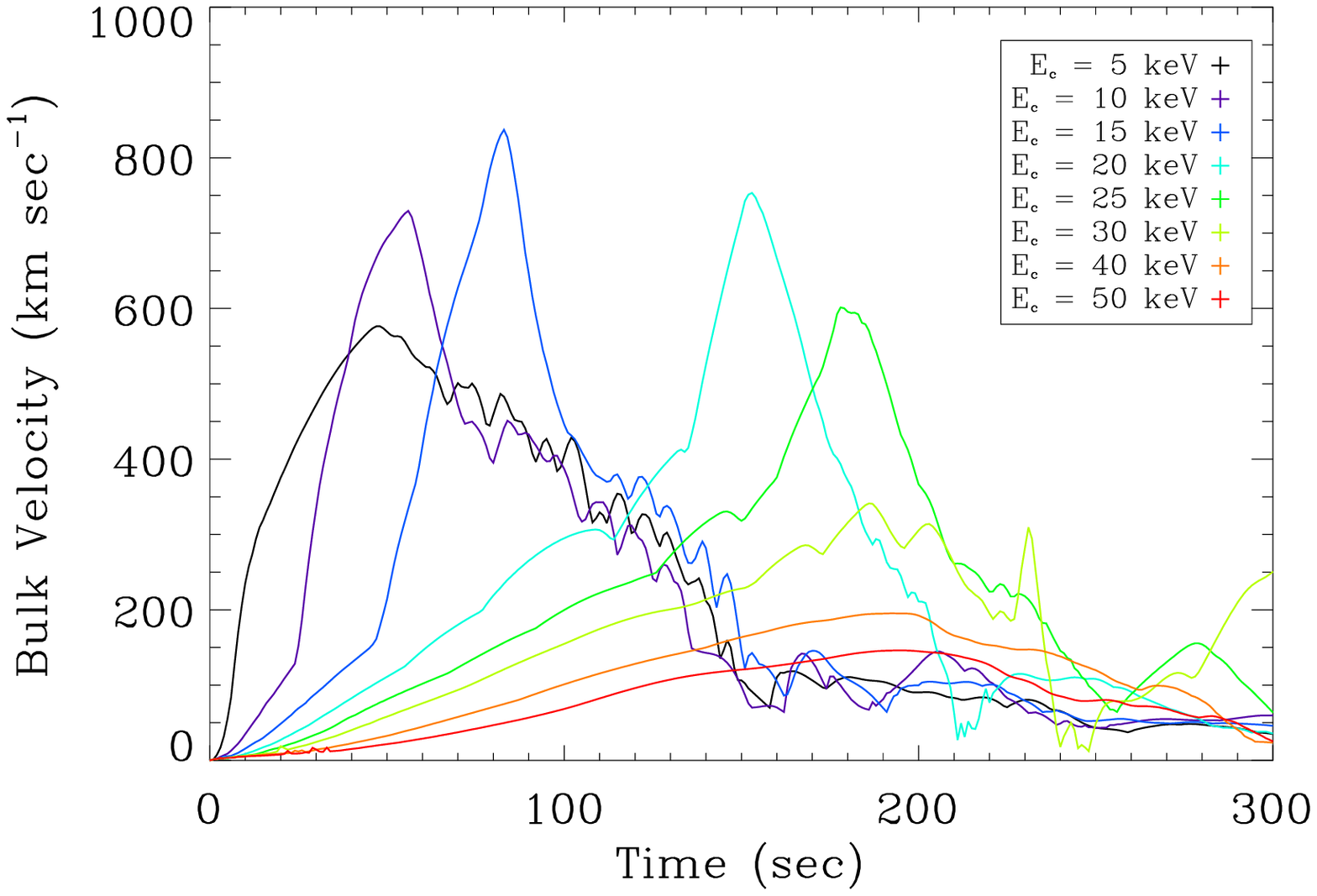}
\end{minipage}
\begin{minipage}[b]{0.32\linewidth}
\centering
\includegraphics[width=2.2in]{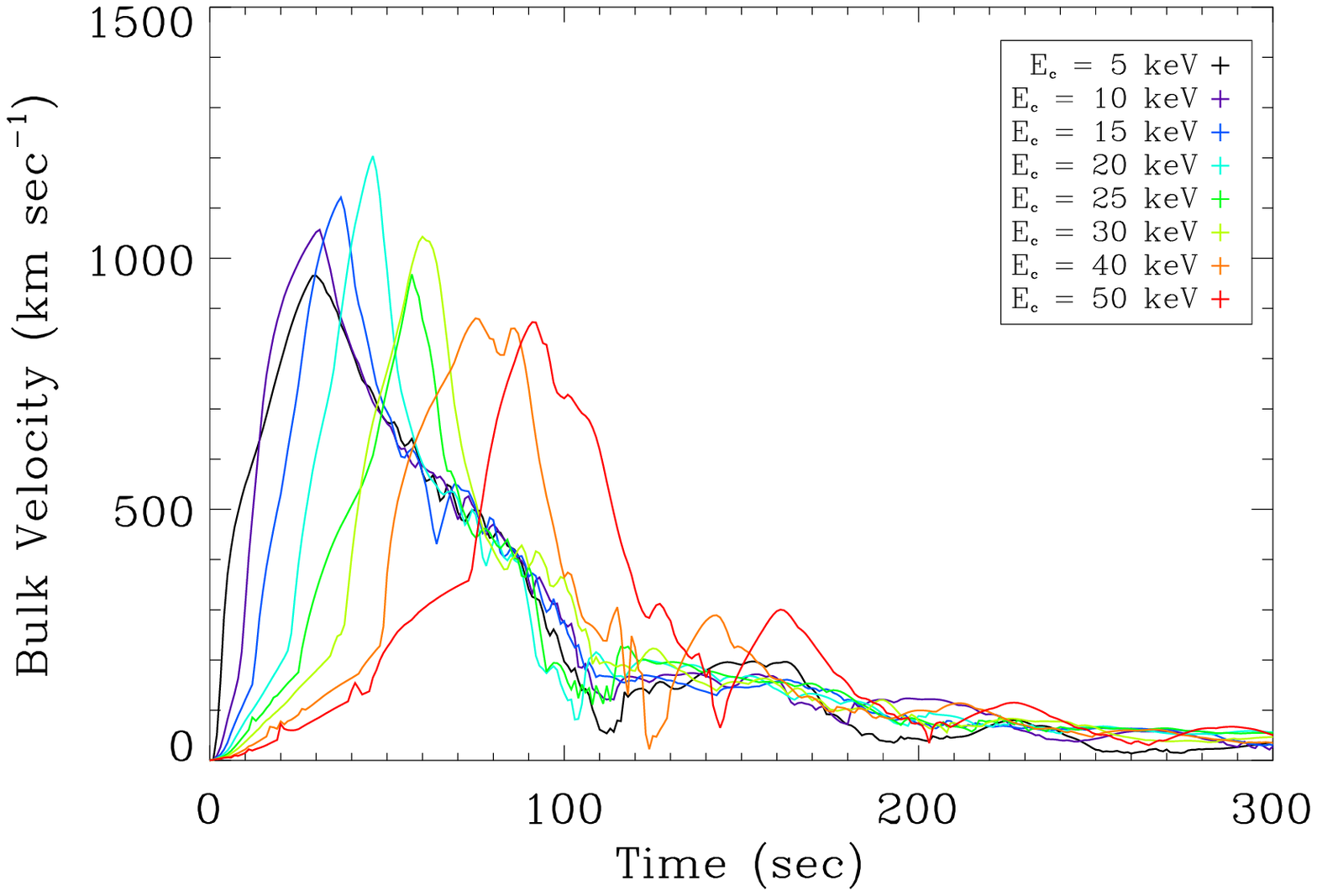}
\end{minipage}
\caption{How the temperature, density, and bulk flow velocity vary for the experiments in Table \ref{isoenertable}.  The apex electron density (top row), apex electron temperature (middle row), and maximum bulk flow velocity (bottom row) in Runs 1-8 (left column), 9-16 (middle column), and 17-24 (right column). }
\label{tempdens}
\end{figure}

There are significant differences for the three groups of simulations.  For those simulations below the canonical explosive evaporation threshold, both the temperature and density are strongly dependent on the energy $E_{\ast}$.  The left column of Figure \ref{tempdens} shows the apex electron temperature, apex electron density, and maximal bulk flow velocity as functions of time for Runs 1-8.  In Runs 1 and 2, the bulk flows develop in a short amount of time, and reach velocities of a few hundred km sec$^{-1}$, while in the other runs, the flows are much slower.  Runs 1 and 2 accordingly reach much higher densities than the other six runs.  The up-flowing material in Runs 1 and 2 brings a significant enthalpy flux into the corona, significantly raising the temperature, compared to Runs 3-8.  It is important to point out that Run 3, with a maximal velocity around 150 km sec$^{-1}$, proceeds gently, despite the speeds significantly exceeding those in the gentle scenarios of \citet{fisher1985b}.

Similarly, the middle column of Figure \ref{tempdens} shows the apex electron temperature, apex electron density, and maximal bulk flow velocity as functions of time for Runs 9-16.  It is clear that lower energy electrons, which deposit their energy higher up in the loop, cause a larger increase in temperature, which then drives a strong conduction front and leads to higher densities.  Higher energy electrons deposit their energy deeper in the chromosphere, where the ambient density is much larger and therefore has a larger heat capacity and stronger radiative losses.  In addition, the time it takes for significant flows to develop is strongly dependent on the electron energy.  There is a clear trend showing that lower energy electrons cause upflows to develop sooner than higher energy electrons (note the times of peak velocity in the plot).  Lower energy electrons heat lower density plasma, which has less inertia and a lower heat capacity, so that the pressure rises and flows develop sooner than for higher energy electrons.  At very late times ($\approx 1200$ sec), the loops in Runs 9-14 catastrophically cool and drain, as they become unable to sustain the radiation and enthalpy losses \citep{cargill2013}.  

The results are completely different in the case where the energy flux $F_{0}$ of the beam is above the canonical explosive evaporation threshold (Runs 17-24).  In these eight simulations, the evaporation is explosive, regardless of the electron energy.  The right column of Figure \ref{tempdens} similarly shows the apex electron density and apex electron temperature as functions of time for these simulations.  In this case, the densities and temperature are very nearly equal in all 8 simulations.  Because the energy flux is extremely large, there is enough energy to heat the chromosphere and cause a large rise in the pressure, even for the high energy electrons that are stopped much deeper in the chromosphere.  As with the previous diagram, lower energy electrons cause upflows to develop sooner than higher energy electrons (once again, compare the times at which the velocity peaks for each electron energy).  Similarly to the previous case, at around 900 seconds, the loops in all 8 simulations catastrophically cool and drain \citep{cargill2013}.  

These results suggest important conclusions.  First, for beams above the canonical explosive evaporation threshold, the final state of the atmosphere is not strongly dependent on the electron energy.  In Runs 17-24, the maximal apex densities and temperatures are nearly identical.  Below the threshold, however, lower energy electrons are more efficient at heating the corona, leading to higher maximal temperatures.  Second, lower energy electrons which deposit their energy higher in the atmosphere drive evaporation into the corona sooner than higher energy electrons.  They do not necessarily, however, drive upflows with higher velocities.  Finally, the explosive evaporation threshold found by \citet{fisher1985a} is dependent on the cut-off energy, a point that they note in their paper but do not examine in detail (they assume a constant 20 keV).  The results here suggest that the threshold could be lower for lower energy cut-offs (see Section \ref{sec:evap}).  

\section{Isoenergetic Beams with a Constant Number Flux}
\label{sec:samenumber}

There is an additional possibility worth considering: to what extent does the number of electrons in the beam matter?  In the previous examples, beams with electron energies $E_{1}$ and $E_{2}$ such that $E_{1} > E_{2}$, with equal total energy will have different numbers of electrons.  In this case, the second beam will have a larger number of lower energy electrons and the first beam will have a smaller number of higher energy electrons.  Since beams composed of lower energy electrons appear to heat loops more efficiently, then what role does the electron number flux play?  This question is examined here.  

To begin, the necessary conditions for two beams to have equal number fluxes are derived.  Assume that there are two isoenergetic beams, each of different electron energy $E_{1}$ and $E_{2}$, such that $E_{1} > E_{2}$.  The number flux $\mathcal{N}$ in each is given by the zeroth moment of the distribution:
\begin{eqnarray}
\mathcal{N} &=& \int_{0}^{\infty} \mathfrak{F}(E_{0},t)\ dE_{0}  \nonumber \\ 
	&=& \frac{F_{0}}{E_{\ast}^{2}}\ \frac{(\delta + 2)\ (\delta - 2)}{2\delta}\ \Bigg[\int_{0}^{E_{\ast}} \frac{E_{0}^{\delta}\ dE_{0}}{E_{c}^{\delta}} +\int_{E_{c}}^{\infty} \frac{E_{0}^{-\delta}\ dE_{0}}{E_{\ast}^{-\delta}} \Bigg] \nonumber \\ 
	&=& \frac{F_{0}}{E_{\ast}}\ \frac{(\delta + 2)\ (\delta -2)}{(\delta + 1)\ (\delta - 1)}\ \ \mbox{[electrons sec$^{-1}$ cm$^{-2}$]}
\label{numfluxeq}
\end{eqnarray}

Then, equate $\mathcal{N}_{1}$ with $\mathcal{N}_{2}$ so that the two beams have the same number of electrons.  Note that since $E_{1} \neq E_{2}$, the spectral index of each differs as well. 
\begin{eqnarray}
\mathcal{N}_{1} &=& \mathcal{N}_{2} \nonumber \\ 
\frac{F_{1}}{E_{1}}\ \frac{(\delta_{1} + 2)\ (\delta_{1} -2)}{(\delta_{1} + 1)\ (\delta_{1} - 1)} &=& \frac{F_{2}}{E_{2}}\ \frac{(\delta_{2} + 2)\ (\delta_{2} -2)}{(\delta_{2} + 1)\ (\delta_{2} - 1)} 
\end{eqnarray}

\noindent where the subscripts 1 and 2 refer to each separate beam.  The condition for equal number flux is thus found to be:
\begin{equation}
\frac{F_{1}}{F_{2}} =  \frac{E_{1}}{E_{2}}\ \frac{(\delta_{2} + 2)\ (\delta_{2} -2)}{(\delta_{2} + 1)\ (\delta_{2} - 1)}\ \frac{(\delta_{1} + 1)\ (\delta_{1} - 1)}{(\delta_{1} + 2)\ (\delta_{1} - 2)}  
\label{samenumberflux}
\end{equation}

\noindent Therefore, given an energy flux for an isoenergetic beam (of energy $E_{1}$ and index $\delta_{1}$), the energy flux carried by a different isoenergetic beam (of energy $E_{2}$ and index $\delta_{2}$) can then be determined such that they have an equal number of electrons.  

8 numerical experiments have been performed and Table \ref{numfluxtable} displays the key results.  Each experiment is assumed to be an isoenergetic beam with 99\% of its electrons within $\pm 2$ keV of an electron energy [5, 10, 15, 20, 25, 30, 40, 50] keV.  The spectral indices are therefore determined by Equation \ref{isoenergeticindex}.  

In the previous section, it was found that the final state of the atmosphere does not depend strongly on electron energy if the energy flux is above the explosive evaporation threshold.  So, the first simulation, with an electron energy of 5 keV, is assumed to have a maximal beam flux $F_{0} = 10^{10}$ erg cm$^{-2}$ sec$^{-1}$, which is at the canonical explosive evaporation threshold (see \citealt{fisher1985b}).  As before, the simulations were performed using loops of length $2L = 50$ Mm, with a cross-sectional area $A = 7.8 \times 10^{16}$ cm$^{2}$.

\begin{table}
\centering
\begin{tabular}{c c c c c c c c}
\hline
Run \# & $E_{\ast}$ & $F_{0,max}$ & $\delta$ & GOES Class & $v_{\mbox{max}}$ & $T_{\mbox{max}}$ & $n_{\mbox{apex,max}}$ \\
 & (keV) & (erg sec$^{-1}$ cm$^{-2}$) &  & (1-8 \AA) & (km sec$^{-1}$) & (MK) & (cm$^{-3}$) \\
\hline
1 & 5 & $1.00 \times 10^{10}$ & 13.7 & C1.7 & 576.3 & 26.2 & $5.65 \times 10^{10}$  \\
2 & 10 & $1.98 \times 10^{10}$ & 25.3 & M3.9 & 825.0 & 31.8 & $9.87 \times 10^{10}$  \\
3 & 15 & $2.96 \times 10^{10}$ & 36.8 & X3.2 & 938.7 & 35.3 & $1.34 \times 10^{11}$  \\
4 & 20 & $3.94 \times 10^{10}$ & 48.3 & X6.4 & 918.6 & 37.2 & $1.62 \times 10^{11}$  \\
5 & 25 & $4.92 \times 10^{10}$ & 59.8 & X14 & 985.5 & 39.4 & $1.92 \times 10^{11}$  \\
6 & 30 & $5.91 \times 10^{10}$ & 71.4 & X27 & 982.7 & 39.7 & $2.09 \times 10^{11}$  \\
7 & 40 & $7.87 \times 10^{10}$ & 94.4 & X66 & 913.9 & 43.1 & $2.66 \times 10^{11}$  \\
8 & 50 & $9.84 \times 10^{10}$ & 117.4 & X150 & 850.3 & 43.5 & $3.20 \times 10^{11}$  \\
\end{tabular}
\caption{Eight numerical experiments assuming an equal number flux $\mathcal{N}$ of electrons for isoenergetic beams at energies $E_{\ast} = $[5, 10, 15, 20, 25, 30, 40, 50] keV.  The spectral indices $\delta$ were calculated using Equation \ref{isoenergeticindex} and the energy fluxes were then calculated using Equation \ref{samenumberflux}, so that the beams have the same number of electrons.  } 
\label{numfluxtable}
\end{table}

\begin{figure}
\begin{minipage}[b]{0.5\linewidth}
\centering
\includegraphics[width=3.25in]{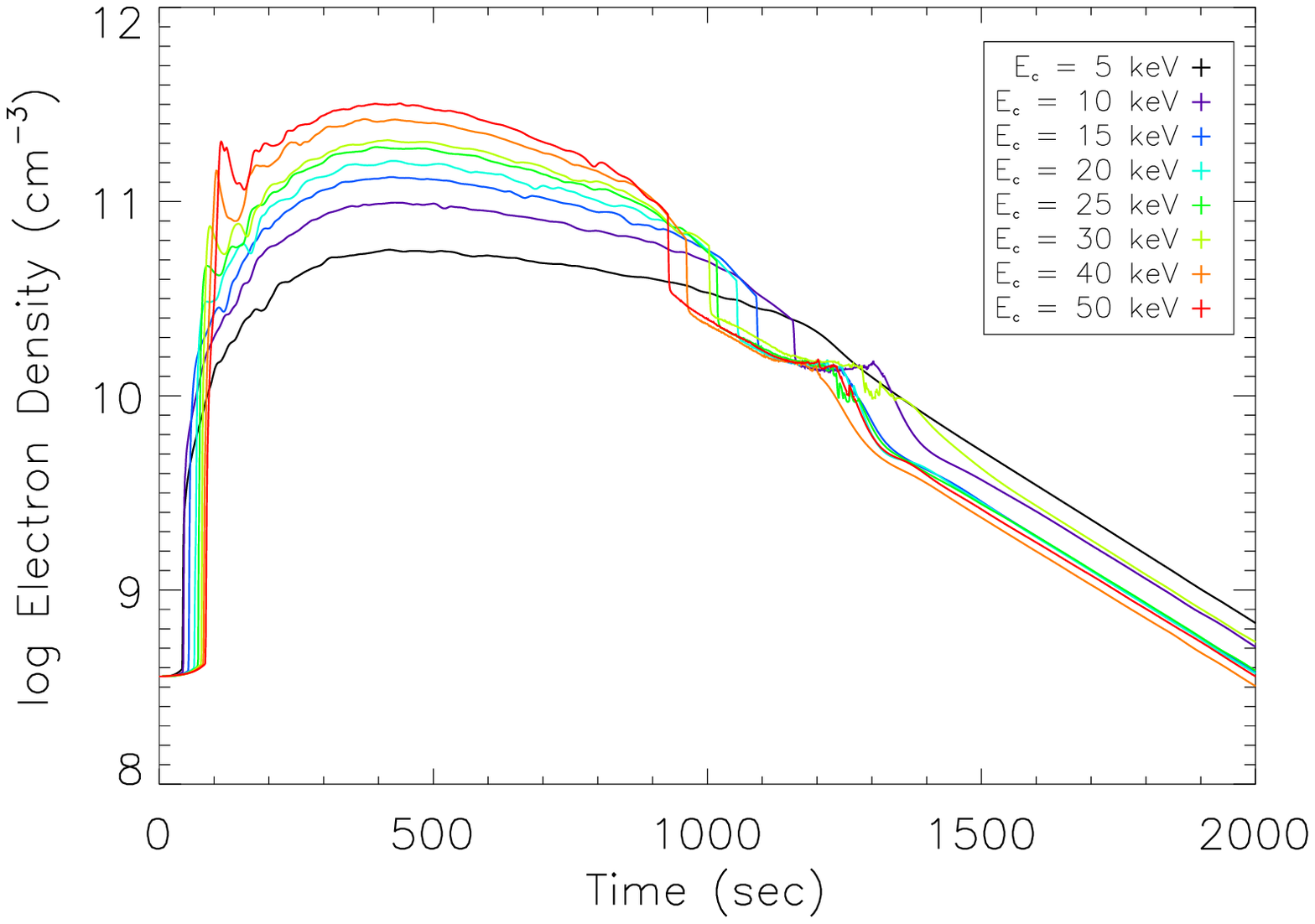}
\end{minipage}
\hspace{0.1in}
\begin{minipage}[b]{0.5\linewidth}
\centering
\includegraphics[width=3.25in]{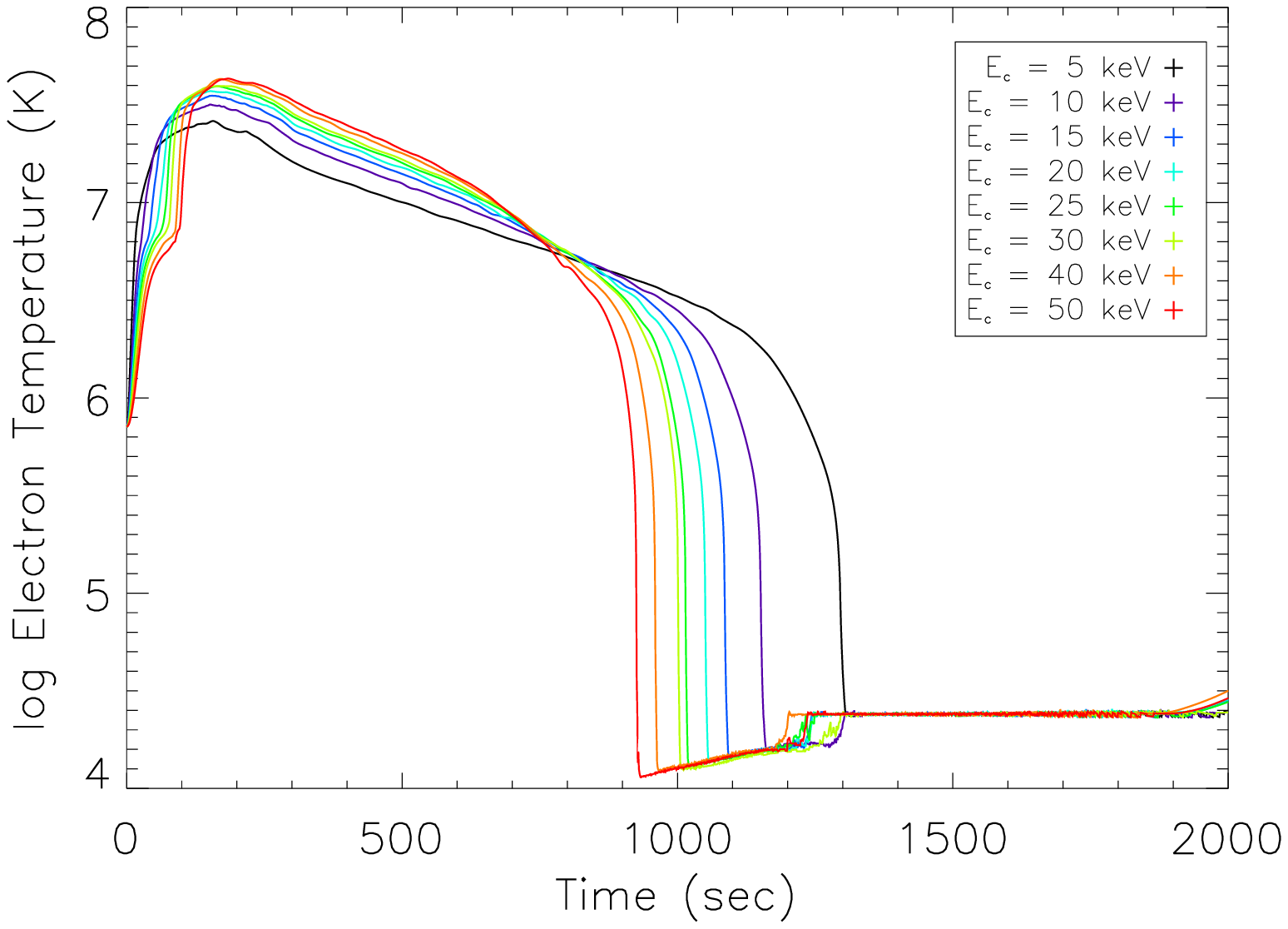}
\end{minipage}
\hspace{0.1in}
\begin{minipage}[b]{\linewidth}
\centering
\includegraphics[width=4.25in]{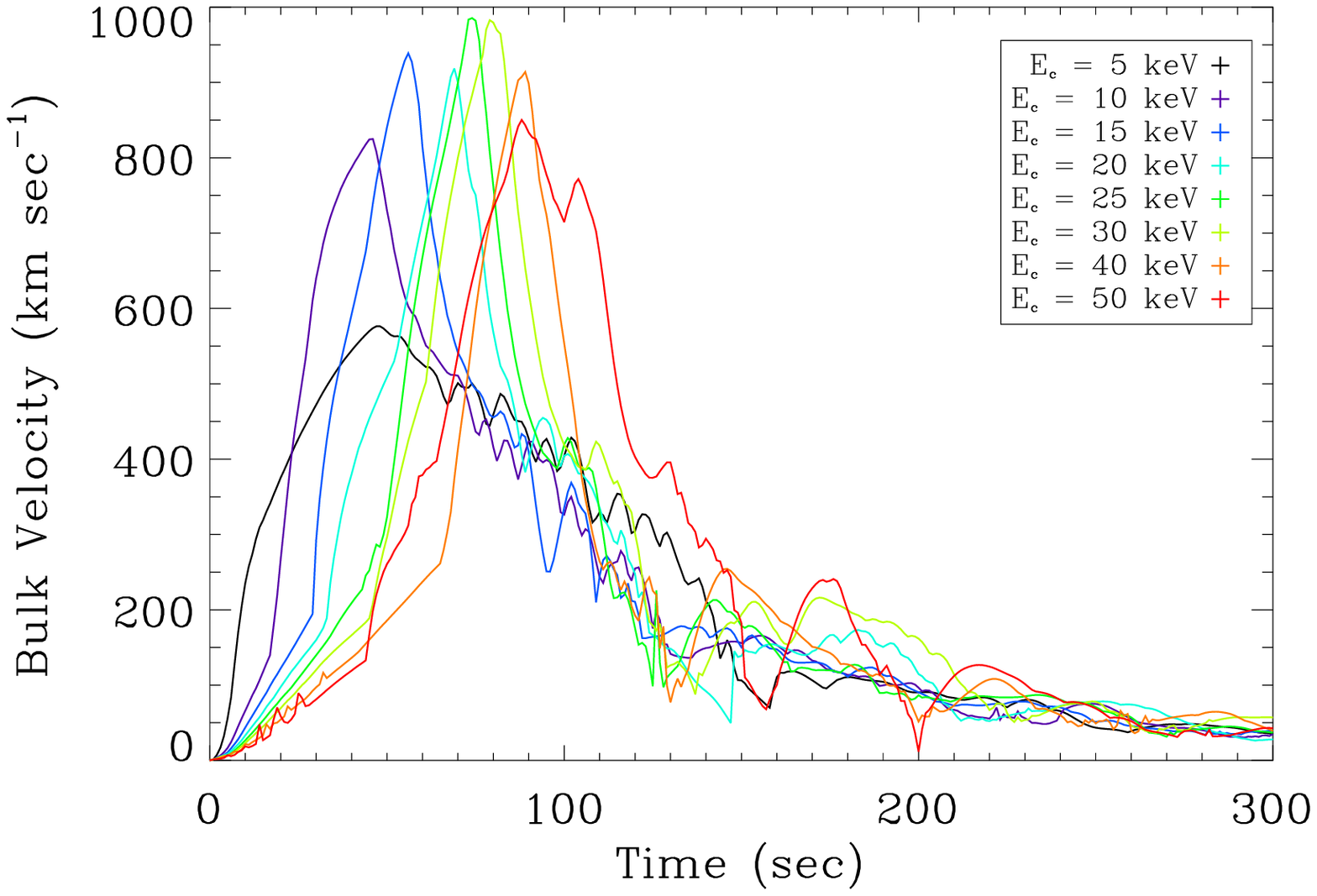}
\end{minipage}
\caption{How the temperature, density, and bulk flow velocity vary for the simulations in Table \ref{numfluxtable}.  {\it Top Left:} The apex electron density in the 8 simulations as a function of time.  {\it Top Right:} The apex electron temperature in the 8 simulations as a function of time.  {\it Bottom:} The maximum of the bulk flow velocity as a function of time (for the first 300 seconds).}
\label{numfluxtempdens}
\end{figure}

Figure \ref{numfluxtempdens} shows the evolution of the density, temperature, and bulk flow velocity as a function of time in the 8 experiments.  The top left hand side of the figure shows the apex electron density in each simulation, while the top right hand side shows the apex electron temperature, and the bottom shows the maximum of the bulk flow velocity.  Several aspects are readily apparent.  Lower energy beams heat and evaporate material earlier, but the loops do not become as dense or hot, and they drain more slowly than those subject to higher energy beams.  

The explanation for this behavior is straightforward: the total beam energy primarily determines the response of the atmosphere.  Lower energy electrons are more efficient at heating the loops since their energy is deposited higher in the atmosphere where it is substantial compared to the local thermal energy and less of the energy will be radiated away immediately (as seen in the previous section).  However, if the number of high and low energy electrons is equal, the high energy electrons will ultimately cause a stronger up-flow of material and a sharper rise in the temperature despite being deposited deeper in the atmosphere since they carry more total energy.  There is a clear and strong correlation between the beam flux and the maximum temperatures and densities.

For 8 beams, with equal number of electrons at different electron energies, there are significant differences between the post-flare atmospheres.  Compare this result with the previous section, where it was found (Runs 17-24, for example) that the post-flare density and temperature are not strongly dependent on the electron energy above the explosive evaporation threshold.  Finally, compare Run 8 of this section with Runs 23 and 24 of the previous section.  The number fluxes (Equation \ref{numfluxeq}), respectively, are $1.23 \times 10^{18}$, $1.24 \times 10^{18}$, and $1.56 \times 10^{18}$ e$^{-}$ sec$^{-1}$ cm$^{-2}$ at their peaks.  Despite the different number fluxes, the maximal bulk flow velocities, temperatures, and densities are all very similar.  The conclusion that can be drawn is that the number flux is only of secondary importance compared to the energy flux carried by the beam.  Even though this may slightly alleviate the problem, the number fluxes in all cases considered here are extremely high and characteristic of the electron number problem \citep{fletcher2008}.

In the limit of very large electron energies, however, this result would not hold.  For example, if the electrons had an extremely large energy, {\it e.g.} 1 MeV, they would stop at a column density of around $10^{23}$ cm$^{-2}$ \citep{nagai1984}.  This column density corresponds to a photospheric depth, where the local thermal energy is too high for there to be significant heating from the electrons without an unrealistically high energy flux.  

\section{On the Explosive Evaporation Threshold}
\label{sec:evap}

\citet{fisher1985a,fisher1985b,fisher1985c} found, using numerical simulations, that beam fluxes above about $10^{10}$ erg sec$^{-1}$ cm$^{-2}$ drive explosive evaporation, with velocities exceeding a few hundred km sec$^{-1}$.  They derive this threshold analytically by equating the radiative loss rate at the top of the chromosphere with the heat deposition, such that if the heating exceeds the losses, the pressure rises and drives evaporation.  However, since the radiative loss rate varies with depth (as the density increases), this threshold must be a function of the cut-off energy.  In this section, the dependence of the explosive evaporation threshold on the electron energy is examined.  

Using isoenergetic beams, 34 simulations have been performed to examine this threshold as a function of the electron energy.  Similar to the work of \citet{fisher1985a}, a beam duration of 5 seconds is assumed with a {\it constant} beam flux (unlike in Section \ref{sec:isosims} where it varied as a function of time) and constant spectral index.  The electron energy is varied between [5, 10, 15, 20, 25, 30, 40, 50] keV, with spectral indices derived from Equation \ref{isoenergeticindex}.  The beam flux is then varied between [$10^{8}$, $5 \times 10^{8}$, $10^{9}$, $5 \times 10^{9}$, $10^{10}$, $5 \times 10^{10}$ , $10^{11}$] erg sec$^{-1}$ cm$^{-2}$.  Table \ref{evapsims} shows the parameters of the simulations performed, along with the maximal bulk flow velocities (km sec$^{-1}$), electron temperatures (MK), and apex electron densities (cm$^{-3}$) that were attained.  

\LTcapwidth=\textwidth

\begin{longtable}{c c c c c c c c c c}
\caption{Simulations across a range of electron energies (5-50 keV), for a range of fluxes (10$^{8}$ - 10$^{11}$ erg sec$^{-1}$ cm$^{-2}$), assuming isoenergetic beams.  The maximal velocities, electron temperatures, and electron apex densities are shown.  \label{evapsims}  } \\

\hline
Run \# & $E_{\ast}$ & $F_{0}$ & $\delta$ &  $v_{\mbox{max}}$ & $T_{\mbox{max}}$ & $n_{\mbox{apex,max}}$ \\
 & (keV) & (erg sec$^{-1}$ cm$^{-2}$) &   & (km sec$^{-1}$) & (MK) & (cm$^{-3}$) \\
\hline
1 & 5 & $1.00 \times 10^{8}$ & 13.7 & 63.6 & 1.06 & $6.11 \times 10^{8}$  \\
2 & 5 & $5.00 \times 10^{8}$ & 13.7 &  255 & 2.69 & $2.18 \times 10^{9}$  \\
3 & 5 & $1.00 \times 10^{9}$ & 13.7 &  342 & 4.60 & $3.13 \times 10^{9}$  \\
4 & 5 & $5.00 \times 10^{9}$ & 13.7 &  596 & 12.1 & $5.42 \times 10^{9}$  \\
5 & 5 & $1.00 \times 10^{10}$ & 13.7 &  737 & 15.5 & $7.93 \times 10^{9}$  \\
6 & 5 & $5.00 \times 10^{10}$ & 13.7 &  1220 & 28.6 & $1.90 \times 10^{10}$  \\
\hline
7 & 10 & $1.00 \times 10^{9}$ & 25.3 &  84.5 & 1.60 & $7.80 \times 10^{8}$  \\
8 & 10 & $5.00 \times 10^{9}$ & 25.3 &  577 & 5.16 & $5.61 \times 10^{9}$  \\
9 & 10 & $1.00 \times 10^{10}$ & 25.3 &  759 & 8.52 & $8.11 \times 10^{9}$  \\
10 & 10 & $5.00 \times 10^{10}$ & 25.3  & 1210 & 20.4 & $1.92 \times 10^{10}$  \\
\hline
11 & 15 & $1.00 \times 10^{9}$ & 36.8 & 40.0 & 1.11 & $5.12 \times 10^{8}$  \\
12 & 15 & $5.00 \times 10^{9}$ & 36.8 & 146 & 2.74 & $1.14 \times 10^{9}$  \\
13 & 15 & $1.00 \times 10^{10}$ & 36.8 & 538 & 4.72 & $1.09 \times 10^{10}$  \\
14 & 15 & $5.00 \times 10^{10}$ & 36.8 & 1190 & 14.6 & $2.07 \times 10^{10}$  \\
\hline
15 & 20 & $1.00 \times 10^{9}$ & 48.3  & 21.8 & 0.93 & $4.24 \times 10^{8}$  \\
16 & 20 & $5.00 \times 10^{9}$ & 48.3  & 97.3 & 1.85 & $7.34 \times 10^{8}$  \\
17 & 20 & $1.00 \times 10^{10}$ & 48.3  & 199 & 3.00 & $1.68 \times 10^{9}$  \\
18 & 20 & $5.00 \times 10^{10}$ & 48.3  & 1070 & 10.7 & $2.81 \times 10^{10}$  \\
\hline
19 & 25 & $1.00 \times 10^{9}$ & 59.8 & 19.8 & 0.85 & $3.91 \times 10^{8}$  \\
20 & 25 & $5.00 \times 10^{9}$ & 59.8 & 67.7 & 1.43 & $6.37 \times 10^{8}$  \\
21 & 25 & $1.00 \times 10^{10}$ & 59.8 & 113 & 2.17 & $1.05 \times 10^{9}$  \\
22 & 25 & $5.00 \times 10^{10}$ & 59.8 & 886 & 8.90 & $3.84 \times 10^{10}$  \\
\hline
23 & 30 & $1.00 \times 10^{9}$ & 71.4 & 17.5 & 0.81 & $3.75 \times 10^{8}$  \\
24 & 30 & $5.00 \times 10^{9}$ & 71.4 & 49.4 & 1.21 & $5.54 \times 10^{8}$  \\
25 & 30 & $1.00 \times 10^{10}$ & 71.4 & 88.3 & 1.71 & $7.68 \times 10^{8}$  \\
26 & 30 & $5.00 \times 10^{10}$ & 71.4 & 667 & 6.26 & $3.39 \times 10^{10}$  \\
\hline
27 & 40 & $1.00 \times 10^{9}$ & 94.4 & 15.7 & 0.76 & $3.63 \times 10^{8}$  \\
28 & 40 & $5.00 \times 10^{9}$ & 94.4 & 26.3 & 0.98 & $4.59 \times 10^{8}$  \\
29 & 40 & $1.00 \times 10^{10}$ & 94.4 & 54.2 & 1.26 & $5.78 \times 10^{8}$  \\
30 & 40 & $5.00 \times 10^{10}$ & 94.4 & 198 & 3.49 & $1.91 \times 10^{9}$  \\
31 & 40 & $1.00 \times 10^{11}$ & 94.4 & 573 & 6.29 & $9.03 \times 10^{10}$  \\
\hline
32 & 50 & $1.00 \times 10^{10}$ & 117.4 & 36.5 & 1.04 & $4.92 \times 10^{8}$  \\
33 & 50 & $5.00 \times 10^{10}$ & 117.4 & 136 & 2.53 & $1.17 \times 10^{9}$  \\
34 & 50 & $1.00 \times 10^{11}$ & 117.4 & 328 & 4.16 & $7.94 \times 10^{9}$  \\

\end{longtable}

\begin{figure}
\centering
\includegraphics[width=3.5in]{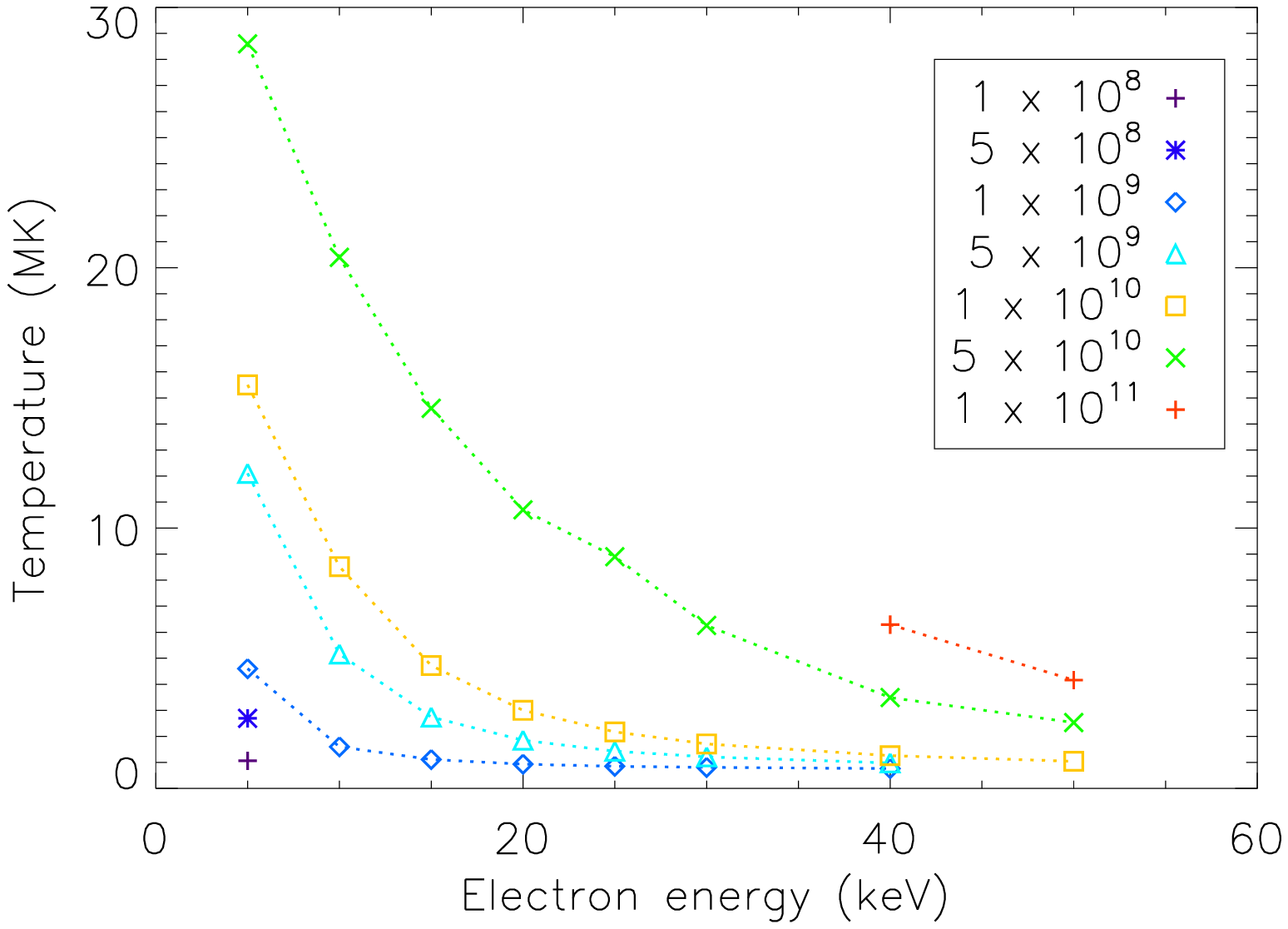}
\includegraphics[width=3.5in]{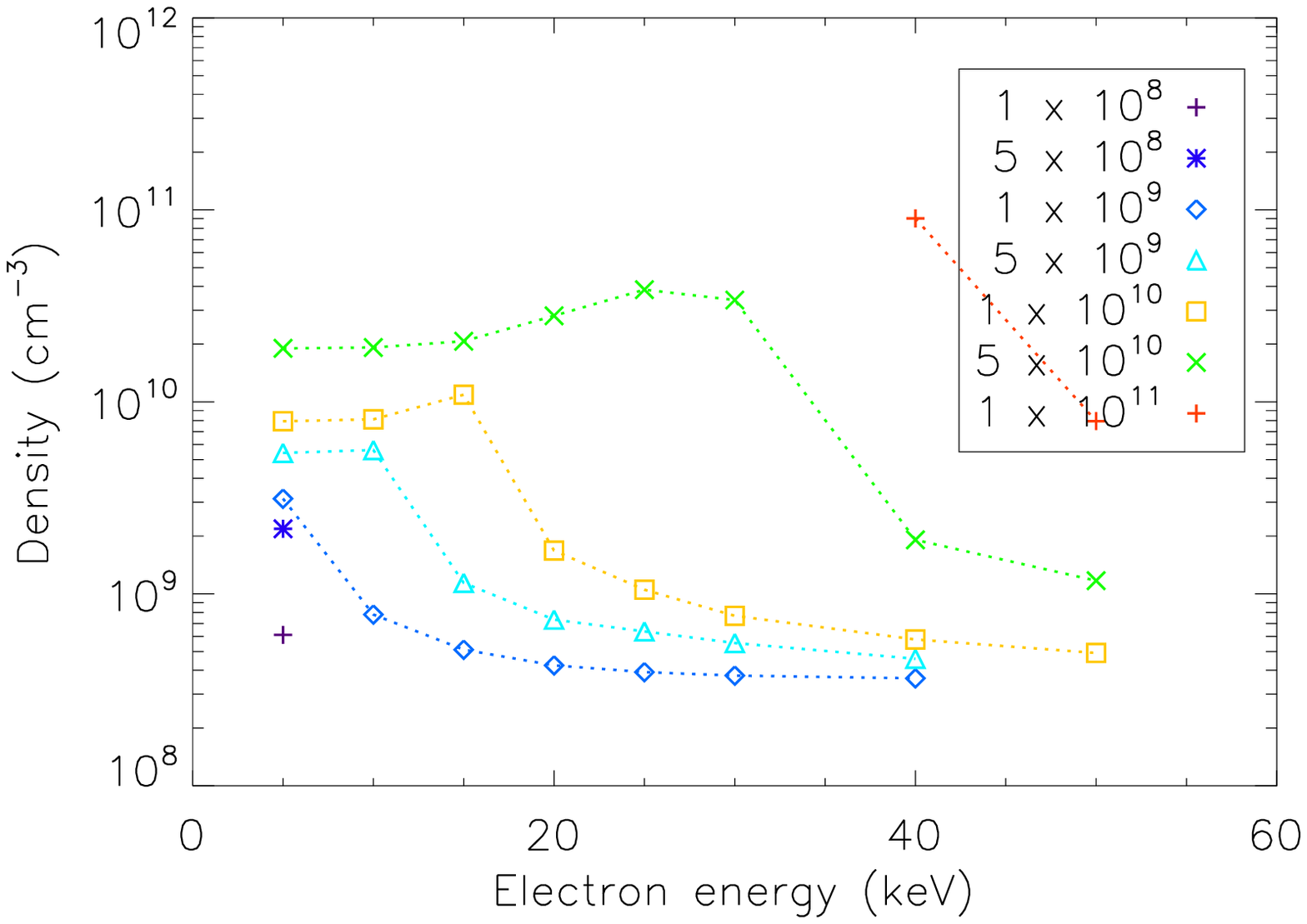}
\includegraphics[width=3.5in]{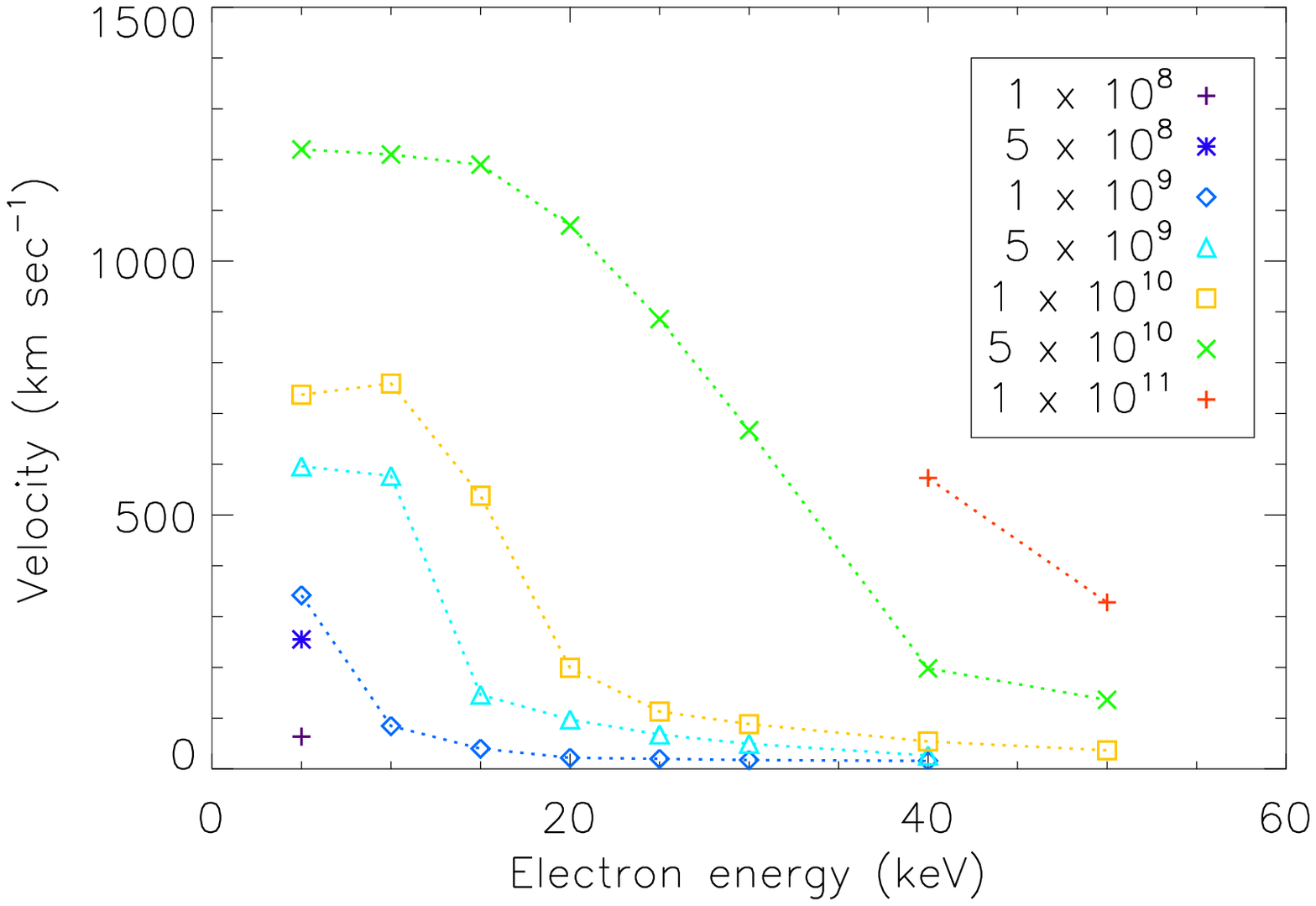}
\caption{The maximal density, temperature, and bulk flow velocity attained in each simulation in Table \ref{evapsims}, as a function of electron energy and beam flux (different colors, as labeled).  {\it Top:} The maximal electron temperature in each simulation.  At every beam flux, lower energy electrons lead to higher temperatures.  {\it Middle:} The maximal electron density at the apex of the loop.  {\it Bottom:} The maximal bulk flow velocity in each simulation.}
\label{evapplots}
\end{figure}

Several things can be learned from these simulations.  Figure \ref{evapplots} shows the temperatures (MK), densities (cm$^{-3}$), and velocities (km sec$^{-1}$) attained in these simulations as a function of energy $E_{\ast}$, for the various beam fluxes in Table \ref{evapsims}.  First, consider the temperature plot.  At every beam flux, lower electron energies lead to monotonically higher temperatures attained in the simulations.  This once again confirms the notion that lower energy electrons are more efficient at heating flaring loops.

The density plot shows something different.  At a given beam flux ({\it e.g.} $5 \times 10^{10}$ erg sec$^{-1}$ cm$^{-2}$), the simulations with the highest energy electrons become only slightly denser than the initial conditions ($\approx 4 \times 10^{8}$ cm$^{-3}$), while the others increase in density by over 2 orders of magnitude (in this case, beams with electron energy $< 40$ keV).  However, note that for all of the simulations below this threshold, the maximal apex density is approximately the same ($\approx 2 \times 10^{10}$ cm$^{-3}$).  Since the total non-thermal energy content was the same in each simulation, this suggests that the amount of upflowing material does not depend strongly on the electron energy, below 40 keV.  Similarly at $5 \times 10^{9}$ erg sec$^{-1}$ cm$^{-2}$, for 15 keV and above, the density only increases slightly from the initial values, and there are correspondingly small velocities.  At 5 and 10 keV, though, the density reaches about $5 \times 10^{9}$ cm$^{-3}$, at speeds of just under 600 km sec$^{-1}$ (for both simulations).  Importantly, these results imply that the lower energy electrons are more efficient than higher energy electrons at providing the corona with hot, dense plasma.

Finally, consider the velocity plot, where a similar trend is found.  At a given beam flux, the highest velocities are attained by beams with the lowest energy electrons, in general.  However, as might be expected from the density case, below a certain threshold, the velocities are approximately equal.  For example, at $5 \times 10^{10}$ erg sec$^{-1}$ cm$^{-2}$, the maximal velocities at 5, 10, and 15 keV are all around 1200 km sec$^{-1}$.  By considering the pressure balances in the chromosphere, \citet{fisher1984,fisher1985b} derive the maximal speed attained by explosive evaporation to be around 2.35 $c_{s}$ (the sound speed).  Assuming an ideal gas, this gives:
\begin{equation}
2.35 c_{s} = 2.35 \sqrt{\frac{(5/3) k_{B} T}{ m_{i}}} \approx (2.76 \times 10^{4})\ T^{0.5}\ \mbox{[cm sec$^{-1}$]}
\end{equation}

\noindent for $k_{B}$ the Boltzmann constant and $m_{i}$ the ion mass (assumed to be hydrogen).  For example, in the case of Run 6, for a temperature of around 20 MK, the speed works out to around 1170 km sec$^{-1}$.  Similarly, for Run 9, at a temperature of about 8 MK, the equation gives a speed of 780 km sec$^{-1}$.  The maximal speeds at each energy $E_{\ast}$ in Table \ref{evapsims} are in agreement with the speed derived by \citet{fisher1985b}.  

These results confirm the existence of two regimes of underlying physics: gentle and explosive evaporation.  In the explosive evaporation case, the momentum transport through the solar atmosphere strongly depends on both the beam flux and the electron energy.  The maximal velocity of upflowing material is around $2.35 c_{s}$, as derived by \citet{fisher1985b}.  However, the threshold between gentle and explosive evaporation {\it also depends} on the electron energy and can be estimated from these results.  Calling supersonic flows explosive (see below for a more precise definition), and fitting a line in log-log space to the average of the upper and lower limits of the threshold, the following relation is found here:  
\begin{equation}
\log_{10}{F} = 6.99 + 2.43 \log_{10}{E_{\ast}}
\end{equation}

\begin{figure}
\centering
\includegraphics[width=4.5in]{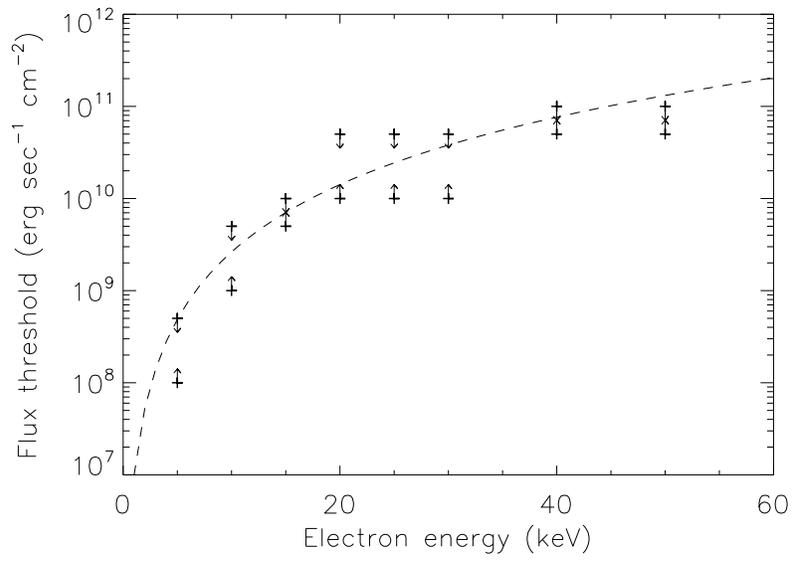}
\caption{The energy flux threshold for explosive evaporation as a function of the electron energy.  At each electron energy, the upper and lower limits attained by the simulations in Table \ref{evapsims} are plotted.  A linear fit in log-log space to the average of the limits has been over-plotted, for comparison.}
\label{evapthresh}
\end{figure}

Figure \ref{evapthresh} shows the upper and lower limits obtained by the simulations in this section, listed in Table \ref{evapsims}, along with the fit to the data.  In agreement with \citet{fisher1985b}, the threshold at 20 keV is  $\gtrsim 10^{10}$ erg sec$^{-1}$ cm$^{-2}$.  

This result can be derived analytically.  Following the definition of explosive evaporation found in \citet{fisher1985b}, the evaporation is explosive if the heating time scale is less than the hydrodynamic expansion time scale:
\begin{equation}
\frac{3 k_{B} T}{Q} \lesssim \frac{L_{0}}{c_{s}}
\end{equation}

\noindent for $Q = F / N$ the heating rate (erg sec$^{-1}$), $L_{0}$ the length of the heated region, and $c_{s}$ the sound speed after heating.  Following \citet{emslie1978}, electrons with energy $E_{\ast}$ stop at a column density $N = \frac{\mu_{0} E_{\ast}^{2}}{2 \pi e^{4} (2 + \beta/2) \gamma}$, where the variables are defined in that reference.  Substituting for $Q$ and $c_{s}$ and solving:
\begin{equation}
\frac{F}{E_{\ast}^{2}} \gtrsim \frac{3 \sqrt{5} (k_{B} T)^{3/2} \mu_{0}}{2 \pi e^{4} (2+\frac{\beta}{2}) \gamma L_{0} \sqrt{3 m_{i}}}
\end{equation}

\noindent Thus the threshold energy flux changes with the electron energy squared:
\begin{equation}
\log{F} \propto \log{E_{\ast}^{2}}
\end{equation}

The threshold of explosive evaporation depends quadratically on the electron energy and linearly on the energy flux.  The numerical results do not agree precisely because of the sparse energy fluxes examined here.  For example, at 20, 25, and 30 keV, the threshold is in the same range (between 1 and $5 \times 10^{10}$ erg sec$^{-1}$ cm$^{-2}$), although it is clear from Figure \ref{evapplots} that a smaller electron energy results in larger velocities.

\section{Conclusions}
\label{sec:isoconclusions}

In this work, the deposition of energy in the solar atmosphere by non-thermal electrons as the driving mechanism for solar flares has been examined.  In order to study the transport of mass, momentum, and energy through the solar atmosphere, knowledge of the properties of the energy deposition and the detailed response of the atmosphere are crucial.  Observations have revealed accelerated electrons ranging in energy from a few keV to well over 100 keV ({\it e.g.} \citealt{holman2003}; \citealt{warmuth2009}; \citealt{ireland2013}).  How the atmosphere responds to heating by an electron beam depends strongly on the properties of the beam, and thus upon the electrons that comprise the beam.  To this end, isoenergetic beams (that is, beams composed of electrons all at the same energy) have been used to understand the relative importance of the different components of the beam.  

The isoenergetic assumption is artificial because measured electron spectra in solar flares are generally found to be a sharp power-law in form.  However, most of the energy in the beam is concentrated near the low-energy cut-off due to the sharpness of the spectra.  The problem is thus simplified to assume that all of the electrons are found at one energy, which allows us to isolate the effects of energy deposition by electrons of varying energies.  After deriving the necessary equations to model isoenergetic beams in Section \ref{sec:isoenergetic}, simulations were carried out assuming otherwise realistic parameters in Section \ref{sec:isosims}.  The electron beams were assumed to last for 5 minutes, with the energy pulse rising and falling for 150 seconds, reaching a peak energy flux consistent with observed quantities.  The electron energies comprised a wide-range of energies, from 5 to 50 keV, well within observed bounds.  The loops were assumed to be 50 Mm in length, consistent with measurements of active region structures.  

These simulations show several important features of electron beam heating:

\noindent 1. Above the explosive evaporation threshold, the response of the atmosphere does not strongly depend on the electron energy.  Although properties of the impulsive phase may still differ, during the gradual phase the densities and temperatures in the corona are fairly independent of the electron energy.  

\noindent 2. Lower energy electrons are significantly more efficient at heating the atmosphere.  Because their energy is deposited higher up (towards the top of the chromosphere and transition region) than higher energy electrons, the deposited energy is comparable to the local thermal energy and less of the energy is lost through the efficient radiation deeper in the chromosphere.  

\noindent 3. Lower energy electrons drive up-flows sooner than higher energy electrons, although not necessarily with higher velocities.  This may be because the higher coronal heating drives a thermal conduction front, contributing energy in addition to the chromospheric energy deposition, and/or because the lower density material has less inertia.  

Since isoenergetic beams at different electron energies will carry different numbers of electrons, the importance of number flux was examined.  It was found that the number flux is relatively unimportant compared to the total energy being carried by the beam.  The energy flux carried by the beam dominates the response of the atmosphere, regardless of the number of electrons in that beam (within sensible bounds, though).

Finally, the results of Section \ref{sec:isosims} indicated that the explosive evaporation threshold depends upon the energy of the electrons comprising the beam.  Physically, since higher energy electrons deposit their energy deeper down (due to having a much longer mean-free path) where the heat capacity and radiative losses are much higher, a beam with higher energy electrons will require significantly more total energy to drive explosive evaporation.  As explained by \citet{fisher1985a,fisher1985b,fisher1985c}, explosive evaporation is driven by a local over-pressure in the chromosphere, which forces material both upwards into the corona (evaporation) as well as deeper into the chromosphere (condensation).  

Accordingly, in Section \ref{sec:evap} many simulations, using parameters similar to the investigations of \citet{fisher1985a} were performed.  The simulations ranged from electron energies of 5 to 50 keV and beam fluxes from $10^{8}$ to $10^{11}$ erg sec$^{-1}$ cm$^{-2}$.  These demonstrated important results for heating driven by a thick-target model.  

\noindent 1. The mass of up-flowing material depends strongly upon both the electron energy and the beam flux.  Low-energy electrons are more efficient at driving evaporation of hot, dense plasma into the corona than higher energy electrons.  The amount of material transported through the solar atmosphere is limited once the beam is well above the explosive threshold, because the material is transported by the flows, which are limited in their speeds.  We find that the more total energy that is carried by the beam, the more mass that is evaporated into the corona.  

\noindent 2. The momentum of up-flowing material also depends strongly upon both the electron energy and the beam flux.  The speeds at which material up-flows depend strongly on the electron energy, with lower energy electrons more efficiently causing a drastic increase in the chromospheric pressure and thus the flow speeds.  However, above the explosive evaporation threshold, the speed of up-flows is limited to $2.35 c_{s}$ \citep{fisher1985b}, which holds regardless of the electron energy or flux.  The simulation results were found to be in good agreement with this speed.  Due to the limitation in the speed of the bulk flows, the amount of material is limited as well.  

\noindent 3. The energy transported likewise depends upon both the electron energy and the beam flux.  Lower energy electrons are more efficient at driving up-flows and thus providing hot, dense plasma to the corona, at all beam fluxes.  Lower energy electrons are also more efficient at heating the atmosphere, thus raising the temperatures more drastically.   A larger beam flux more efficiently drives a stronger up-flow, similarly providing the corona with hot, dense plasma, at all electron energies.

From these three results, we conclude that the threshold between explosive evaporation and gentle evaporation depends strongly upon the electron energy, as well as the beam flux (Figure \ref{evapthresh}).  We show analytically that the threshold depends quadratically upon the electron energy and linearly with the beam flux.  Lower energy electrons require significantly less energy to drive explosive evaporation, due to depositing energy higher in the atmosphere, which is assisted by thermal conduction.  

These results have implications on events smaller than flares, as well.  Electrons around 5 keV can drive explosive evaporation with very little total non-thermal energy.  In observed microflares, X-ray brightenings with energy release about $10^{-6}$ times that of a large flare \citep{hannah2011}, observations point to cut-off energies below 7 keV \citep{phillips2004}.  Even with non-thermal energies significantly smaller than solar flares, this leads to the possibility that explosive evaporation could occur in these microflares.  For example, \citet{chifor2008} found recurring EUV jets associated with microflares, with speeds up to $150$ km sec$^{-1}$, which they attributed to chromospheric evaporation due to recurring magnetic reconnection.  Likewise, this may have further implications for nanoflares heated by a particle beam ({\it e.g.} \citealt{testa2014}).  

There are still many features of energy deposition by electron beams that need to be examined in detail.  It is necessary to understand the dependence of material flow speeds on the loop length, the duration of heating, the shape of the beam pulse, and the pitch angle distribution of the electrons.  A wider range of simulations can explore all of these features directly and systematically.  The insight provided here, however, has allowed for a clearer interpretation of the physics underpinning the heating and evolution of observed flares.  In particular, we have gained a deeper understanding of the beam properties, the transport of mass, momentum, and energy through the solar atmosphere, and the interplay between the heating mechanisms and the atmospheric response.

\end{document}